\documentclass[aps,prd,reprint,nofootinbib,longbibliography,superscriptaddress,floatfix]{revtex4-2}
\usepackage[T1]{fontenc}
\usepackage[utf8]{inputenc}
\usepackage{graphicx,amsmath,amssymb,amsthm,booktabs,array}
\usepackage{microtype}
\usepackage[hidelinks]{hyperref}
\newcommand{\ud}{\mathrm{d}}
\newcommand{\GeV}{\mathrm{GeV}}

\hypersetup{pdftitle={Required spectral contributions in glueball QCD sum rules beyond single-pole fits},pdfauthor={Yong-Qi Kang and Zi-Yan Yang}}
\relax{}
\begin{document}
\title{Required spectral contributions in glueball QCD sum rules beyond single-pole fits}
\author{Yong-Qi Kang}
\affiliation{School of Robotics Engineering, Guangzhou City University of Technology, Guangzhou 510800, China}
\author{Zi-Yan Yang}
\email[Contact author: ]{yangzh@gcu.edu.cn}
\affiliation{School of Robotics Engineering, Guangzhou City University of Technology, Guangzhou 510800, China}
\begin{abstract}
We provide a continuously verified diagnostic that distinguishes a compatible single-pole fit from a mass region required by QCD sum-rule moments. For specified glueball sum-rule inputs, the moments can require contributions in broad mass bands even when narrower candidate bands are empty. We compare fourteen two- and three-gluon currents at central fixed-scale inputs, undeformed continua and 1\% matching tolerance. All twelve nonempty inputs permit an empty candidate band, yet nine still require a broader contribution. This separates two statements that a pole fit conflates: a successful pole fit does not make its mass region necessary, and excluding every single pole does not remove the information that a broader region must contribute. The five diagnostic classes of the survey organize these outcomes by whether a pole is compatible and whether the 30\% band is required or can be empty; two inputs admit no nonnegative spectrum. Four of the nine positive bounds persist on subwindows satisfying the fixed-scale input criteria. For the $GG,0^{-+}$ current, closest to the nominal $X(2370)$ mass, the 20\% band containing $X(2370)$ can be empty while the 30\% band requires about 44.2\% of each spectrum's own low-energy Borel integral. Determining the physical nature of $X(2370)$ requires further theoretical and experimental progress.
\end{abstract}
\maketitle
\raggedbottom
\section{Introduction}
Gluon self-interactions make glueballs a distinctive probe of nonperturbative QCD. Pure-gauge lattice calculations predict a spectrum of such states~\cite{morningstar1999}, while mixing with quark configurations and hadronic decay channels complicates their identification in full QCD~\cite{jiang2023,chen2023review}. Masses, production mechanisms and decay patterns therefore provide complementary information. Radiative heavy-quarkonium decays are an important experimental setting for this search.

The $X(2370)$ illustrates the difficulty of identifying a glueball from these observables. BESIII first observed it in $J/\psi\to\gamma\pi^+\pi^-\eta'$~\cite{besiii2011}; a subsequent amplitude analysis established $J^{PC}=0^{-+}$ in $\gamma K_S^0K_S^0\eta'$~\cite{besiii2024}. A recent search found no significant $X(2370)\to K^*(892)^0\bar K^0+\mathrm{c.c.}$ signal. Combining this suppression with earlier production and decay information, BESIII argued for a dominant pseudoscalar-glueball component~\cite{besiii2026}.

Several approaches support a glueball assignment under their respective assumptions. Current--field and Fierz analyses describe three-pseudoscalar decay ratios~\cite{wang2026duality,tan2026decays}; an anomaly-based effective model describes the measured channels and proposes further decay tests~\cite{giacosa2026}. A constituent-gluon model finds compatible masses and widths~\cite{buisseret2026}. Radiative-production calculations emphasize the sensitivity to glueball--quarkonium mixing and interference~\cite{zhu2026}. Competing descriptions remain possible: radial-hybrid templates~\cite{lu2026}, a comparison of hybrid, tetraquark and three-gluon assignments~\cite{zhang2026nature}, and distinct $\Sigma\bar\Sigma$~\cite{ben2026} and $\phi h_1(1380)$~\cite{khemchandani2026} molecular mechanisms can accommodate parts of the available information. A coupled-channel reanalysis finds that the $K_S^0K_S^0\pi^0$ distributions do not yet require an additional distinct $X(2370)$ contribution within its amplitude model~\cite{dong2026}. These analyses use different dynamical assumptions and can accommodate parts of the available data, while the physical nature of $X(2370)$ remains unsettled. 

Existing QCD sum-rule analyses typically infer a mass from a single-pole fit and then interpret that mass region as hosting a state. This inference assumes that a successful pole fit requires a contribution in the fitted mass region. Whether the moments themselves enforce that contribution, once the single-pole ansatz is relaxed, is a separate question. We study this question in QCD sum rules~\cite{svz1979,gubler2019} for the fourteen nonzero two- and three-gluon currents of Ref.~\cite{chen2021}. Among their pseudoscalar channels, the $GG,0^{-+}$ mass result lies closest to the nominal $X(2370)$ mass. The $GG$ and $GGG$ labels describe interpolating operators, rather than constituent compositions of physical states. Sum rules relate the current correlator to its operator product expansion (OPE)~\cite{nielsen2010}. In the usual pole-plus-continuum description, the Borel moment ratio estimates the pole mass squared and the zeroth moment fixes its squared coupling~\cite{wang2026review,narison2023status}. For several states or a continuous distribution, the ratio instead gives a Borel-weighted mean mass squared. Different allocations of spectral contribution can produce nearly the same mean throughout the window. Effective-threshold studies also show how continuum choices can bias pole-based extractions despite Borel stability~\cite{lucha2007,lucha2009}. The $X(2370)$ decay data motivate this study but do not enter its moment constraints; we do not test the state's glueball assignment.

At fixed input, we minimize the fraction of a candidate spectrum's own low-energy Borel integral that comes from a chosen mass band. We allow all nonnegative spectra matching the two OPE moments throughout the Borel window within a specified relative tolerance. Contributions may come from several poles, a continuous distribution, or both. A positive lower bound rules out spectra with too little contribution in the band; an attained zero supplies an empty-band alternative. We compare this bound with the best single-pole mismatch at the same input and tolerance.

Inverse-problem~\cite{liumeeda2020,li2021glueball,li2024excited} and maximum-entropy~\cite{gubleroka2010} methods reconstruct spectral shapes. Positivity inequalities constrain sum-rule consistency and hadronic parameters~\cite{benmerrouche1995,narison2023laplace}; more recent work bounds spectral integrals through Lagrange duality~\cite{lawrence2024} and semidefinite kernel inequalities~\cite{abbott2026,mutzel2026}. Glueball scattering bootstrap studies constrain the scattering amplitude and its couplings~\cite{bocchia2026}. We use the established positivity and duality framework for two-point glueball sum rules. It allows us to compare pole compatibility and band necessity for each current without selecting one low-energy spectral shape. Both the inequalities and compatible spectra are checked over the continuous Borel window.

At central fixed-scale inputs and 1\% matching, twelve of the fourteen complete-window inputs admit nonnegative spectra. All twelve allow an empty candidate band, while nine require a broader contribution. Four of these nine currents retain positive bounds in independent calculations on subwindows satisfying the fixed-scale input criteria; the other five remain complete-window diagnostics. Two of these currents show why pole fits and band bounds give different information. At a deformed-continuum setting, the three-gluon $0^{-+}$ current $\widetilde{\eta}_0$ admits both a pole in the test band and a smooth spectrum entirely outside it. In the detailed three-gluon $2^{++}$ calculation for $\eta_2$, every single pole is excluded, yet the broad band must supply at least 22.2\% of each compatible spectrum's own low-energy Borel integral. A mass region can therefore remain optional despite a successful pole fit, while a contribution in a broad mass band can still be required when every single pole is excluded.

The next step is to determine how much of this spectral information survives input variation. For $\widetilde{\eta}_0$, we vary the condensates and continuum threshold jointly, with one parameter set generating both moments across the window. The broad-band requirement survives throughout these continuous parameter ranges for an undeformed continuum. We then compute the minimum total relative moment discrepancy that permits an empty band. This threshold specifies the input accuracy needed to sustain the required contribution; assessing the physical OPE remainder and continuum uncertainty is a separate task.

This paper is organized as follows. Section~\ref{sec:model} defines the two diagnostics. Section~\ref{sec:applications} validates them with known spectra, classifies the current inputs and examines two members in detail. Section~\ref{sec:input-dependence} tests input dependence and determines empty-band discrepancy thresholds. Section~\ref{sec:scope} discusses the five classes, spectral-model tests and input accuracy. The appendices give current definitions, OPE coefficients, derivations, continuous verification and individual-current plots.

\section{Spectral contributions from Borel moment constraints}\label{sec:model}\label{sec:bounds}
To distinguish a fitted mass from a contribution required by the moments, we use two complementary diagnostics. The minimum single-pole mismatch, $\epsilon_{\rm pole}$, tests whether one pole can describe the moments. The minimum band fraction, $\inf F_{\mathcal B}$, tests how much a mass region must contribute when general nonnegative spectra are allowed. We first relate these quantities to current couplings and Borel moments, then specify the compatible spectral class and the bounds used to evaluate them.

\subsection{Current couplings and Borel moments}
The two-point function of a gluonic current measures its coupling to states with the current's quantum numbers. Its spectral density resolves that contribution by invariant mass $\sqrt{s}$. We write $\ud\rho(s)=\rho(s)\ud s$ for a continuous density and use the same measure notation for delta-function poles. This lets us treat narrow states and continuous hadronic contributions in one sum rule.

For a scalar or pseudoscalar current $J$, with two-point correlator $\Pi$ and spectral density \mbox{$\rho=\pi^{-1}\operatorname{Im}\Pi$}, a zero-width state $H$ contributes
\begin{equation*}
\begin{aligned}
 \ud\rho_{\rm pole}(s)&=A\,\delta(s-m_H^2)\,\ud s,\\
 A&=|\langle0|J(0)|H\rangle|^2.
\end{aligned}
\end{equation*}
The residue is the squared current--state coupling. Its relation to a decay constant follows from the current convention; Section~\ref{sec:borel-input} specifies the convention for our pseudoscalar example. A spinful current requires the corresponding polarization projection, given for the tensor example in Appendix~\ref{app:tensor-input}. Integrating over a mass band collects the contributions of all states in it. The Borel weight $e^{-\tau s}$ then suppresses higher masses, so the integral probes both the couplings and their distribution in mass~\cite{hansen2019,lawrence2024}.

For a nonzero nonnegative spectrum below the continuum threshold $s_0$, the Borel moments are
\begin{equation}\label{eq:borel-mean}
\begin{aligned}
 L_k[\rho](\tau)&=\int_0^{s_0}s^k e^{-\tau s}\,\ud\rho(s),\\
 \frac{L_1[\rho](\tau)}{L_0[\rho](\tau)}&=\langle s\rangle_\tau.
\end{aligned}
\end{equation}
Here $\tau=1/M_B^2$ is the inverse Borel mass squared. The zeroth moment gives the low-energy contribution to the Borel-transformed correlator. The first moment weights the same states by an extra factor of $s$, making it more sensitive to higher masses. Their ratio is a weighted mean mass squared: a narrow state of mass $m_n$ and residue $A_n$ enters this mean with weight $A_n e^{-\tau m_n^2}$. Thus a state with a large coupling can dominate the ratio even when other states contribute.

For a single pole $\ud\rho=A\delta(s-s_p)\ud s$, the residue cancels and the ratio equals $s_p$ at every $\tau$~\cite{svz1979,gubler2019}. For several states or a continuous spectrum, $\sqrt{L_1/L_0}$ is instead a weighted root-mean-square mass. It can lie between populated regions, even where the spectrum itself is empty. A moment-ratio mass therefore gives a useful measure of where the correlator receives its contribution; identifying it with a resonance mass uses the additional single-pole assumption.

Increasing $\tau$ shifts the relative weight toward lower masses. The identity $\partial_\tau\langle s\rangle_\tau=-\operatorname{Var}_\tau(s)$ quantifies that shift: the slope measures the spread in mass squared sampled by the Borel kernel (Appendix~\ref{app:moment-ratio}). An exactly constant ratio for a fixed nonnegative measure forces one support point. A nearly flat ratio can also arise from nearby states or from one dominant contribution with weaker components. Borel stability alone therefore leaves their separation unresolved.

We test that remaining freedom by asking whether the same moments can be reproduced with the proposed mass band empty. Both the overall contribution $L_0$ and the mass weighting encoded by $L_1$ enter this test. On the OPE side, $T_1/T_0$ is the target ratio; its interpretation as a spectral mean requires a compatible nonnegative measure. At finite tolerance the candidate and target ratios are only approximately equal.

\subsection{OPE inputs and compatible spectra}\label{sec:positive-model}\label{sec:framework}
\begin{figure*}[t]
\centering\includegraphics[width=\linewidth]{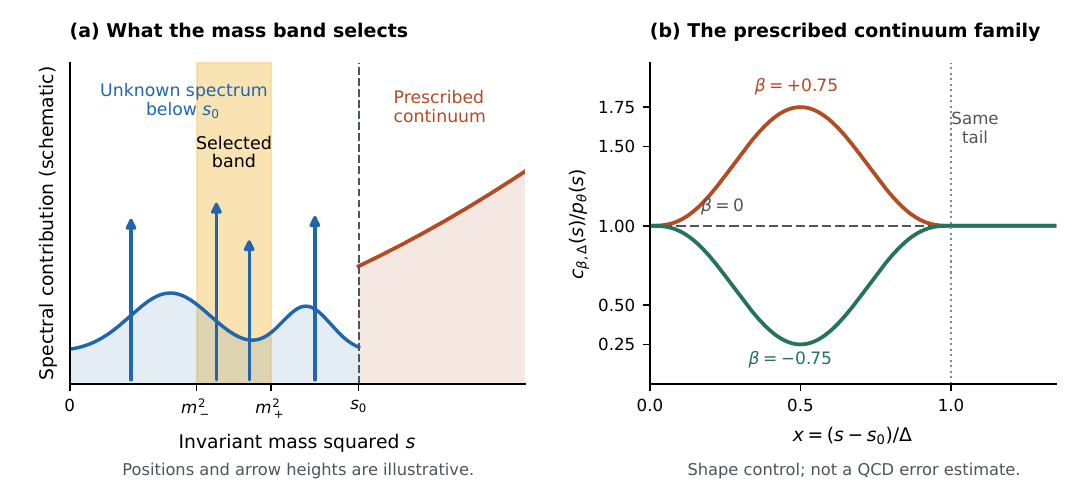}
\caption{The mass band selects all contributions in an interval; the continuum prescription specifies the high-energy part to subtract. (a) The unknown nonnegative spectrum below $s_0$ can contain several poles and a continuous density. The shaded $[m_-^2,m_+^2]$ band includes both. Curves, pole positions and arrow heights are schematic. (b) The relative continuum changes used in the pseudoscalar tests are $c_{\beta,\Delta}/p_\theta=1+\beta b_\Delta$, with $b_\Delta=64x^3(1-x)^3$ for $x=(s-s_0)/\Delta\in[0,1]$ and zero elsewhere. The displayed $\beta=0,\pm0.75$ choices return to the same baseline beyond $s_0+\Delta$; $\beta$ is the fractional change at the deformation center [Eq.~\eqref{eq:new-continuum}].}
\label{fig:continuum-schematic}
\end{figure*}

To test alternatives to a single pole, we allow any finite nonnegative measure $\ud\rho$ on $[0,s_0]$. The threshold divides the unknown low-energy spectrum from a prescribed continuum above $s_0$. Figure~\ref{fig:continuum-schematic} shows the two parts of this model. A selected mass band collects every contribution inside it, including several poles and continuous weight. The prescribed high-energy continuum is subtracted from the OPE moments before testing the low-energy spectrum. Below the threshold we impose no smoothness restriction.

For the fixed-scale baseline, let $O_k(\tau;\theta)$ be the OPE moment before continuum subtraction, with input parameters $\theta$. A nonnegative continuum $c(s;\theta,\gamma)$, with shape parameters $\gamma$, gives the low-energy target
\begin{equation}\label{eq:positive-target}
\begin{aligned}
 T_k(\tau;\theta,\gamma)={}&O_k(\tau;\theta)\\
 &-\int_{s_0}^{\infty}s^k e^{-\tau s}c(s;\theta,\gamma)\,\ud s.
\end{aligned}
\end{equation}
We compare this target with the candidate moment $L_k[\rho](\tau)$ defined in Eq.~\eqref{eq:borel-mean}. The zeroth moment constrains the total Borel-weighted contribution, and the first constrains its distribution in mass.

The matching tolerance $\epsilon$ is the largest permitted relative discrepancy between either candidate moment and its target. For positive targets, the signed relative moment residual is
\begin{equation}\label{eq:relative-residual}
 \mathcal R_k[\rho](\tau)=\frac{L_k[\rho](\tau)}{T_k(\tau)}-1,
 \qquad k=0,1.
\end{equation}
A positive residual means that the candidate moment exceeds the target; a negative residual means that it falls below it. For $0\leq\epsilon<1$ and $0<a<b$, the compatible class $\mathcal A_\epsilon(\theta,\gamma)$ consists of nonnegative measures on $[0,s_0]$ for which $|\mathcal R_k[\rho](\tau)|\leq\epsilon$ throughout $I=[a,b]$. Equivalently,
\begin{equation}\label{eq:positive-class}
\begin{aligned}
(1-\epsilon)T_k(\tau)&\leq L_k[\rho](\tau)\leq(1+\epsilon)T_k(\tau),\\
 &\qquad k=0,1,\quad \tau\in I=[a,b].
\end{aligned}
\end{equation}
We also call these spectra feasible. One spectrum and one set of input parameters must satisfy both moment envelopes throughout the inverse-Borel window. Residual plots display $100\mathcal R_k$ in percent; zero denotes exact matching and the $\pm1\%$ strip denotes the adopted tolerance.

We adopt $\epsilon=1\%$ as a common, stringent benchmark for comparing spectral assumptions. It allows a relative discrepancy of at most 1\% in each moment at every Borel point. This matching tolerance is a comparison setting, not an estimate of the physical OPE error or a confidence level. Convergence tests constrain retained terms; the tolerance scan in Figure~\ref{fig:tensor-resolution} and the critical discrepancies in Section~\ref{sec:correlated-inputs} separately test sensitivity to the allowed mismatch. Only orders $k=0,1$ are constrained.

At fixed renormalization scale, condensates and continuum prescription, the targets obey $T_1=-\partial_\tau T_0$, and each candidate obeys $L_1=-\partial_\tau L_0$. A finite error bound on a function need not control its derivative, so both moment constraints are imposed explicitly. We denote this fixed-coupling prescription by F. The running-coupling prescription control R sets $\mu^2=1/\tau$ and differentiates the same coupling-dependent $T_0$ totally, holding condensates and the current normalization convention fixed. The derivative acts on the coupling in both integrated and nonintegrated terms. A full renormalization-group treatment would additionally require consistent operator evolution and mixing. Here F and R test sensitivity to the stated derivative choices, using the same spectral kernels and bound method. Their paired comparison in Section~\ref{sec:robustness-controls} tests the dependence on the OPE targets.

All reported targets are strictly positive (Appendices~\ref{app:target-positivity}, \ref{app:tensor-input}, \ref{app:chen-inputs}, \ref{app:robustness} and~\ref{app:correlated-inputs}). The relative residuals are thus well defined, and every feasible measure is nonzero. We establish feasibility separately by a spectrum, and exclusion by an inequality.

\begin{samepage}
\subsection{The Borel contribution of a mass band}\label{sec:band-fraction}
We measure a mass region's contribution by its share of the candidate spectrum's low-energy Borel integral. For a closed mass band $\mathcal B=[m_-,m_+]\subset[0,\sqrt{s_0}]$ and a chosen positive weight parameter $\tau_*$, define
\begin{equation}
 F_{\mathcal B}[\rho]=
 \frac{\int_{m_-^2}^{m_+^2}e^{-\tau_*s}\,\ud\rho(s)}{L_0[\rho](\tau_*)},
 \qquad \tau_*>0.
 \label{eq:positive-fraction}
\end{equation}
The fraction $F_{\mathcal B}$ measures how much of the low-energy Borel sum rule is carried by the selected mass region. For narrow states with residues $A_n$, the numerator contains $\sum_{m_n\in\mathcal B} A_n e^{-\tau_*m_n^2}$, together with any continuous contribution in the band. Both band endpoints are included. The denominator collects the same contributions over the whole region below $s_0$. Thus $F_{\mathcal B}=0.2$ means that 20\% of this current's low-energy Borel integral comes from the band.
\par\end{samepage}

In the pseudoscalar convention used below, $A_n=|f_n|^2$, where $f_n$ is the state's decay constant. The fraction therefore constrains a sum of weighted squared decay constants and continuous contributions. If one isolated narrow state exhausts the band, its contribution is $|f_H|^2 e^{-\tau_*m_H^2}=F_{\mathcal B}L_0(\tau_*)$. Extracting that decay constant additionally requires the state's mass and the absolute normalization $L_0$. Our bounds leave the number of states and the division of the band contribution among them free. They characterize the response of a specified current, rather than a count of hadrons or a probability for their internal composition.

A spectrum supported entirely in the band has $F_{\mathcal B}=1$, whether it contains one pole, several poles or a continuous density. Since the Borel weight is positive, $F_{\mathcal B}=0$ means that this spectrum leaves the band empty. An attained zero minimum establishes the existence of such an alternative. A strictly positive lower bound requires every compatible spectrum to contribute there, while leaving its state content open.

We normalize by the candidate's own $L_0(\tau_*)$, so the fraction retains this interpretation even when the candidate and target moments differ. A common rescaling of the current cancels between numerator and denominator. The band and $\tau_*$ are fixed before searching; changing $\tau_*$ changes the balance between lower- and higher-mass contributions. We use $\tau_*=a$ except for the detailed tensor example, whose weight is specified in Section~\ref{sec:tensor-localization}. The low-energy fraction used to choose a Borel window addresses a different partition: it compares the whole region below $s_0$ with the full OPE integral. Here we resolve the contribution within that low-energy region.

\subsection{Minimum single-pole mismatch}
The complementary test restricts the low-energy spectrum to one pole and measures its best possible agreement with the input. At fixed input, optimize one residue $A$ and one position $s_p$ over the full support and minimize the largest residual over both moments and the entire window:
\begin{equation}\label{eq:pole-mismatch}
\epsilon_{\rm pole}=\inf_{\substack{A>0\\0\leq s_p\leq s_0}}
 \max_{k=0,1}\sup_{\tau\in I}
 \left|\frac{A s_p^k e^{-\tau s_p}}{T_k(\tau)}-1\right|.
\end{equation}
The tolerance $\epsilon$ is specified before the test; $\epsilon_{\rm pole}$ is the resulting best achievable mismatch. A larger value means that even the best single pole needs a greater relaxation of the constraints. A certified lower bound above 1\% excludes every single pole at our chosen tolerance. A verified pole within 1\% demonstrates compatibility of that ansatz. Several-state or continuous spectra may remain feasible when the pole fails; their allowed band contributions are determined by the general bounds. Adding higher-moment conditions while retaining the existing targets and constraints cannot enlarge the feasible class or restore a pole already excluded by $k=0,1$.
\relax{}

\subsection{Bounds on the band fraction}
We determine the allowed band fractions by combining explicit spectra with inequalities valid for the entire spectral class. A spectrum verified against every constraint gives an attainable fraction. If it leaves the band empty, the minimum is zero. A positive lower bound requires every compatible spectrum to contribute to the band.

At fixed inputs, verified spectra and their mixtures span an attainable inner interval. Bounds valid for the whole class provide an outer interval enclosing all feasible values:
\begin{equation}\label{eq:interval-logic}
 I_{\rm in}\ \subseteq\ \{F[\rho]:\rho\ \text{feasible}\}\ \subseteq\ I_{\rm out}.
\end{equation}
Here $F$ denotes the spectral integral or fraction being bounded. A normalized fraction varies continuously along a mixture, even though it is generally nonlinear in the mixing parameter. The allowed fraction range describes spectral freedom at the stated input and tolerance.

A bracket for the minimum instead measures how accurately we have determined the infimum of that range. Its lower endpoint bounds that minimum from below, and a verified spectrum reaches or improves on its upper endpoint. Other compatible fractions can be much larger. All zero minima reported for the band fraction are attained by explicit empty-band spectra.

To obtain a bound, we compare a linear combination of moment kernels with the weight selecting the band. Integrating this inequality against any nonnegative spectrum bounds its band fraction (Appendix~\ref{app:bounds}). Optimization supplies the combination coefficients, or dual multipliers.

We check both the proposed inequality and the candidate spectra over their full continuous domains using interval arithmetic, which encloses rounding and integration errors between grid points~\cite{jansson2007,johansson2017,johansson2018}. A verified spectrum serves as a witness to compatibility. The saved inputs, coefficients and verification record form a numerical certificate that can be checked independently of the optimization. Together, $\epsilon_{\rm pole}$ and $\inf F_{\mathcal B}$ test whether one pole can describe the moments and how much a mass region must contribute at the same input. Section~\ref{sec:applications} applies this pair of diagnostics to the glueball currents.
\relax{}
\section{Spectral constraints for glueball currents}\label{sec:applications}\label{sec:borel-application}
We compare single-pole compatibility and band contributions across the currents, using a common tolerance and each current's own inputs. After a check with synthetic spectra, the complete-window classification identifies which bands must contribute. We then study two of these currents, $\widetilde{\eta}_0$ and $\eta_2$, at separately specified inputs to show how pole compatibility and required band contributions can differ.

\subsection{Controlled tests with synthetic spectra}\label{sec:known-truth}
\begin{figure*}[t]
\centering\includegraphics[width=\linewidth]{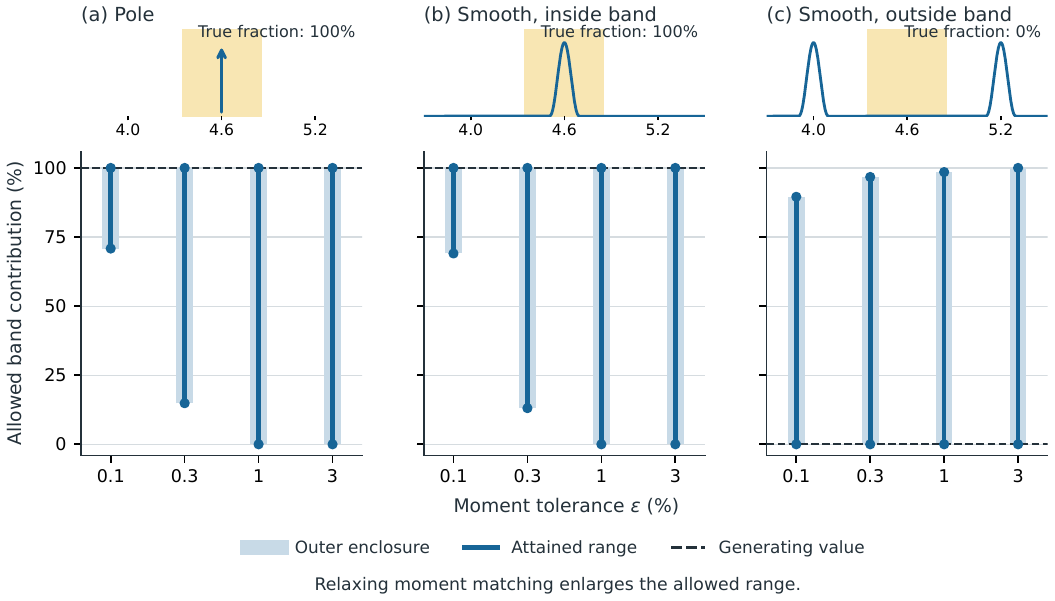}
\caption{Three synthetic spectra at four tolerances give twelve controlled tests, independent of the fourteen-current QCD survey. The lower-panel ordinate is $100F_{\mathcal B}$ in percent, with $F_{\mathcal B}$ defined in Eq.~\eqref{eq:positive-fraction}. The upper sketches show the generators against the gold $[4.35,4.85]\,\GeV$ band; their horizontal coordinate is mass in GeV. They are a pole at 4.60 GeV, one smooth packet there, and equal packets at 4.00 and 5.20 GeV. Each packet has 200 MeV full support width. In the lower panels, the pale outer enclosure contains every compatible fraction, the dark segment spans verified attainable values, and the dashed line gives the generating fraction. These twelve synthetic cases share $M_B^2\in[5.7,7.2]\,\GeV^2$, $s_0=33\,\GeV^2$, $\tau_*=(7.2\,\GeV^2)^{-1}$ and the deterministic admixture $\delta_{\rm mix}=2\times10^{-4}$ defined in Eq.~\eqref{eq:mock-offset}. Each target mixes the generating moments with those of the auxiliary 2.50 GeV pole. Each gluonic current in the later survey instead uses its own threshold and Borel window. Outer and attained endpoints differ by less than 0.010 percentage point before rounding. At 1\%, both initially localized inputs also allow an empty band (Appendix~\ref{app:positive-evidence}).}
\label{fig:known-truth}
\end{figure*}

Before using QCD inputs, we check the bounds in twelve synthetic cases: three generating spectra, each tested at four matching tolerances. These are controlled numerical tests, separate from the fourteen gluonic currents studied in Section~\ref{sec:chen-comparison}. Figure~\ref{fig:known-truth} tests two features: the interval should retain the known answer, and its width should reveal how much freedom the moments leave to alternative spectra. Each vertical bar spans allowed band contributions, rather than repeated estimates of one fitted value.

Within this synthetic comparison, all three generators use $s_0=33\,\GeV^2$, $M_B^2\in[5.7,7.2]\,\GeV^2$, $\tau_*=(7.2\,\GeV^2)^{-1}$ and $\mathcal B=[4.35,4.85]\,\GeV$. The generating normalized Borel measures $\ud\eta=e^{-\tau_*s}\ud\rho/L_0(\tau_*)$ are a pole at 4.60 GeV, a smooth distribution centered at 4.60 GeV, and equal smooth distributions at 4.00 and 5.20 GeV. Each smooth component, called a packet, has a nonnegative $C^2$ density in $m=\sqrt{s}$ supported on a 200 MeV interval. This full support width is a construction parameter rather than a hadronic decay width. The first two spectra are inside the band and the third is outside, giving fractions $1,1,0$. A common known nonnegative continuum is added above $s_0$ and subtracted exactly (Appendix~\ref{app:known-input}).

The supplied target contains the fixed admixture $\delta_{\rm mix}=2\times10^{-4}$:
\begin{equation}
 T_k=(1-\delta_{\rm mix})L_k^{\rm true}+\delta_{\rm mix}L_k^{\rm aux},
 \label{eq:mock-offset}
\end{equation}
The auxiliary spectrum is a 2.50 GeV pole, and both components have $L_0(\tau_*)=1$. This deterministic offset preserves $T_1=-\partial_\tau T_0$. Spectra, input, window, band and offset are held fixed while the tolerance changes. All three generating spectra satisfy the full-window constraints at the smallest tolerance, $\epsilon=0.1\%$. The optimization covers all nonnegative measures in Eq.~\eqref{eq:positive-class}, including shapes and supports different from the generators.

At 0.1\% tolerance, the pole and single-packet inputs require at least 70.82\% and 69.04\% in the band, respectively (Figure~\ref{fig:known-truth}). Their true fractions of one remain allowed. At 1\%, verified empty-band alternatives make both ranges exactly $[0,1]$. Thus relaxing only the matching tolerance can remove a band requirement.

For the two-packet input, the zero-fraction generator is feasible at every tested tolerance. Already at 0.1\%, the upper bound is about 89.56\%, so the same input permits substantial band weight. All 12 cases have verified attainable endpoints and outer bounds (Appendix~\ref{app:positive-evidence}). Changing $\tau_*$ within the window preserves positivity of the two 0.1\% lower bounds and the exact $[0,1]$ ranges at 1\%; Appendix~\ref{app:weight-transfer} gives conservative reweighting bounds.

The bounds thus retain the known band fractions while exposing compatible alternatives, providing a controlled check before the QCD applications.
\relax{}

\subsection{Currents, inputs and mass bands}\label{sec:current-protocol}
A common protocol makes the diagnostics comparable across currents while retaining a separate threshold and Borel window for each. The nonzero local currents of Ref.~\cite{chen2021} comprise four two-gluon and ten three-gluon operators, including six with $C=-$. Their explicit definitions and OPE coefficients are collected in Appendix~\ref{app:chen-inputs}. The spin-one $C=+$ constructions vanish identically and are not members of this set. For higher spin we retain the same parity-projected invariant as the input calculation.

We retain the condensate ranges used in Ref.~\cite{chen2021}. The gluon-condensate interval agrees with the sum-rule average reviewed in Ref.~\cite{narison2018quarkonia}. In a common notation,
\begin{equation}\label{eq:common-condensates}
\begin{aligned}
 G&=\langle\alpha_s GG\rangle\in[0.060,0.067]\,\GeV^4,\\
 G_2&=\langle g_s^2GG\rangle=4\pi G,\qquad H=\langle g_s^3G^3\rangle=Gr,\\
 r&\in[7.2,9.2]\,\GeV^2 .
\end{aligned}
\end{equation}
The central values are $G=0.0635\,\GeV^4$ and $r=8.2\,\GeV^2$. Using the standard relation between the gauge coupling and $\alpha_s$~\cite{pdg2020}, we retain the one-loop coupling prescription and scale parameter of Ref.~\cite{chen2021},
\begin{equation}\label{eq:coupling-law}
 \alpha_s(\mu^2)=\frac{4\pi}{11\log(\mu^2/\Lambda_{\rm QCD}^2)},
 \qquad \Lambda_{\rm QCD}=0.300\,\GeV .
\end{equation}
The retained moments, including the nonintegrated terms, are given in Appendix~\ref{app:chen-inputs}, Eq.~\eqref{eq:survey-target} and Table~\ref{tab:ope-coefficients}.

Each current retains the threshold and full Borel window of Ref.~\cite{chen2021}. We fix central condensates and $\mu^2$ at the window midpoint, retain all OPE terms specified in that calculation, including the nonintegrated terms, and constrain both moments throughout the window. The weight is the inverse upper Borel endpoint. Alongside the candidate band $\mathcal B_C$ defined by the asymmetric mass interval in Ref.~\cite{chen2021}, we test $[0.8M_C,1.2M_C]$ and $[0.7M_C,1.3M_C]$, intersected with the low-energy support. These are uniform resolution choices around each central mass $M_C$ reported in Ref.~\cite{chen2021} (Appendix~\ref{app:chen-inputs}). We abbreviate the corresponding fractions as $F_C$, $F_{20}$ and $F_{30}$.

The complete-window comparison uses the undeformed continuum, $c=p_J$ above each threshold, and 1\% matching. We separately audit the adopted OPE-convergence and low-energy-fraction criteria, abbreviated as the input criteria. We use the 5\%, 10\% and 40\% limits of Ref.~\cite{chen2021}, with the channel-dependent numerators defined in Eq.~\eqref{eq:channel-input-criteria}. Four currents admit qualifying subwindows under the fixed-scale, finite-$s_0$ convention; the other ten do not within the printed source windows. We assess spectral feasibility separately from these input criteria. The four selected-window tests in Table~\ref{tab:chen-gated} form a subsequent control, not additional currents.

\subsection{Complete-window classification}\label{sec:chen-comparison}
We first classify the currents at central, fixed-scale inputs on their complete windows. For each current, the pole and band tests use the same moments and tolerance; the input criteria are audited separately.

\begin{figure*}[t]
\centering\includegraphics[width=\linewidth]{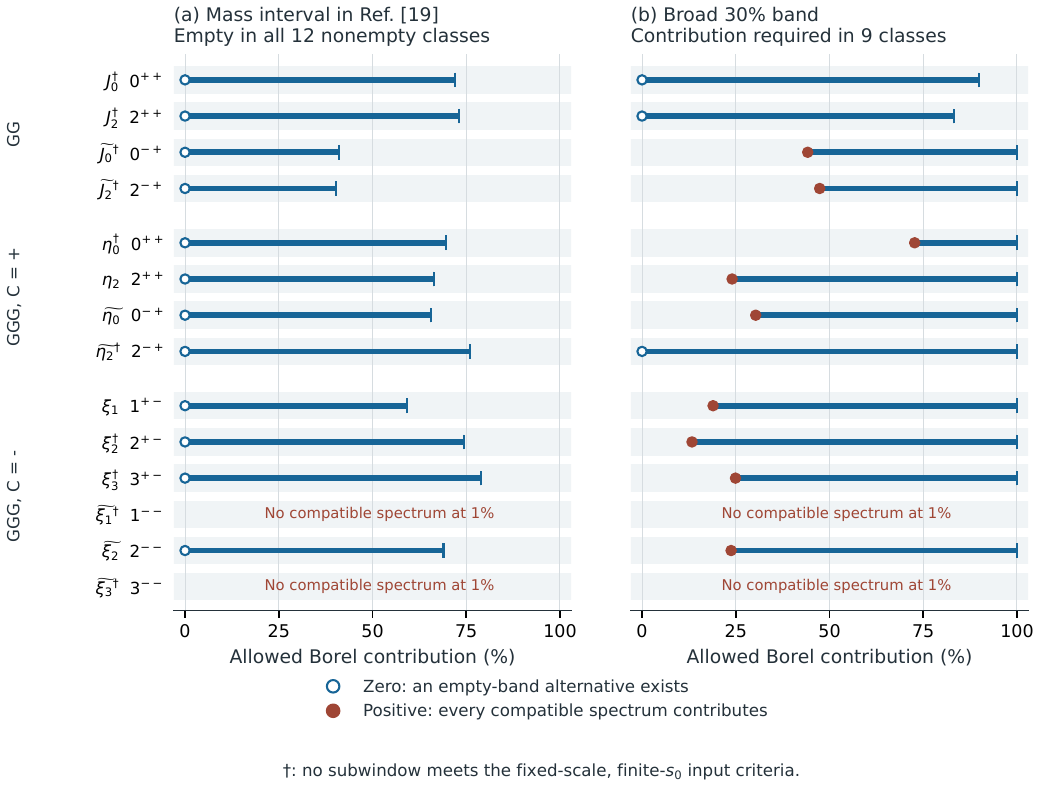}
\caption{The same input can leave a candidate band optional while requiring a broader contribution. Horizontal segments enclose allowed band fractions for the fourteen fixed-scale, full-window models of Table~\ref{tab:chen-comparison}. (a) All twelve nonempty classes admit an empty candidate band. (b) Nine require a positive contribution in the 30\% band. Open left endpoints are attained zeros; filled left endpoints are positive lower bounds. Verified spectra approach the nonzero outer endpoints within 0.03 percentage point before rounding. The two incompatible inputs receive no fraction interval. Daggers mark the separate input-criteria status defined in Table~\ref{tab:chen-comparison}; each current retains its own threshold and Borel window.}\label{fig:fourteen-ranges}
\end{figure*}

\begin{table*}[t]\centering\small
\caption{Mass results in Ref.~\cite{chen2021} and complete-window diagnostics for fourteen fixed-scale inputs. The four separately selected subwindows are tested in Table~\ref{tab:chen-gated}. Each complete-window calculation uses central condensates, the current's threshold and full Borel window from Ref.~\cite{chen2021}, and fixed $\mu^2$ at that window's midpoint. Both moments are constrained within 1\%. Masses and uncertainties are from Table I of Ref.~\cite{chen2021}. The optimal single-pole mismatch $\epsilon_{\rm pole}$, Eq.~\eqref{eq:pole-mismatch}, tests the present fixed-scale input; a value above 1\% excludes that ansatz at our tolerance. The mass intervals serve as external comparison bands.}\label{tab:chen-comparison}
\setlength{\tabcolsep}{4pt}\renewcommand{\arraystretch}{1.18}
\begin{tabular}{lcccrcc}\toprule
\multicolumn{4}{c}{Source current and candidate mass band} & \multicolumn{3}{c}{Present full-window diagnostics}\\
\cmidrule(lr){1-4}\cmidrule(lr){5-7}
Current & $J^{PC}$ & $M_C$ & $\mathcal B_C$ & $\epsilon_{\rm pole}$ & $F_C$ range & $\inf F_{30}$ bracket\\
 & & (GeV) & (GeV) & (\%) & (\%) & (\%)\\\midrule
\addlinespace[3pt]\multicolumn{7}{l}{\textit{Two-gluon currents ($GG$)}}\\[1pt]
$J_0$\textsuperscript{$\dagger$} & $0^{++}$ & $1.78^{+0.14}_{-0.17}$ & $[1.610,1.920]$ & 1.885 & $[0,72.02]$ & $0$\\
$J_2$\textsuperscript{$\dagger$} & $2^{++}$ & $1.86^{+0.14}_{-0.17}$ & $[1.690,2.000]$ & 2.861 & $[0,73.20]$ & $0$\\
$\widetilde{J_0}$\textsuperscript{$\dagger$} & $0^{-+}$ & $2.17^{+0.11}_{-0.11}$ & $[2.060,2.280]$ & 0.592 & $[0,41.08]$ & $[44.19,44.20]$\\
$\widetilde{J_2}$\textsuperscript{$\dagger$} & $2^{-+}$ & $2.24^{+0.11}_{-0.11}$ & $[2.130,2.350]$ & 0.266 & $[0,40.38]$ & $[47.38,47.39]$\\
\addlinespace[3pt]\multicolumn{7}{l}{\textit{Three-gluon currents ($GGG$), $C=+$}}\\[1pt]
$\eta_0$\textsuperscript{$\dagger$} & $0^{++}$ & $4.46^{+0.17}_{-0.19}$ & $[4.270,4.630]$ & 0.144 & $[0,69.69]$ & $[72.72,72.73]$\\
$\eta_2$ & $2^{++}$ & $4.18^{+0.19}_{-0.42}$ & $[3.760,4.370]$ & 10.933 & $[0,66.31]$ & $[24.03,24.04]$\\
$\widetilde{\eta_0}$ & $0^{-+}$ & $4.13^{+0.18}_{-0.36}$ & $[3.770,4.310]$ & 9.198 & $[0,65.58]$ & $[30.31,30.33]$\\
$\widetilde{\eta_2}$\textsuperscript{$\dagger$} & $2^{-+}$ & $4.29^{+0.20}_{-0.22}$ & $[4.070,4.490]$ & 0.745 & $[0,75.94]$ & $0$\\
\addlinespace[3pt]\multicolumn{7}{l}{\textit{Three-gluon currents ($GGG$), $C=-$}}\\[1pt]
$\xi_1$ & $1^{+-}$ & $4.01^{+0.26}_{-0.95}$ & $[3.060,4.270]$ & 27.150 & $[0,59.10]$ & $[18.95,18.97]$\\
$\xi_2$\textsuperscript{$\dagger$} & $2^{+-}$ & $4.42^{+0.24}_{-0.29}$ & $[4.130,4.660]$ & 0.729 & $[0,74.41]$ & $[13.32,13.33]$\\
$\xi_3$\textsuperscript{$\dagger$} & $3^{+-}$ & $4.30^{+0.23}_{-0.26}$ & $[4.040,4.530]$ & 0.771 & $[0,78.90]$ & $[24.92,24.93]$\\
$\widetilde{\xi_1}$\textsuperscript{$\dagger$} & $1^{--}$ & $4.91^{+0.20}_{-0.18}$ & $[4.730,5.110]$ & 2.599 & \textemdash & \textemdash\\
$\widetilde{\xi_2}$ & $2^{--}$ & $4.25^{+0.22}_{-0.33}$ & $[3.920,4.470]$ & 7.591 & $[0,68.95]$ & $[23.75,23.76]$\\
$\widetilde{\xi_3}$\textsuperscript{$\dagger$} & $3^{--}$ & $5.59^{+0.33}_{-0.22}$ & $[5.370,5.920]$ & 6.768 & \textemdash & \textemdash\\\bottomrule\end{tabular}
\par\smallskip\begin{minipage}{\textwidth}\small\raggedright $F_C$ is an outer allowed range for the candidate band defined from the mass interval in Ref.~\cite{chen2021}. The lower endpoint zero is attained. A verified spectrum approaches the rigorous upper bound within 0.03 percentage point before rounding. All twelve nonempty complete-window classes also admit an empty 20\% band; Appendix~\ref{app:channel-atlas} displays those bounds, and Appendix~\ref{app:chen-inputs} identifies their records. The final column instead brackets the lowest fraction in $\mathcal B_{30}=[0.7M_C,1.3M_C]\cap[0,\sqrt{s_0}]$. Certified lower and upper pole bounds are rounded here; masses are optimized over the full low-energy support, including outside the candidate band. Fixed and running midpoint diagnostics are given separately in Table~\ref{tab:prescription-diagnostics}. \textemdash: no nonnegative spectrum at 1\%, hence no fraction bounds. $\dagger$: no subwindow satisfies the fixed-scale, finite-$s_0$ input criteria used here. The four selected-window cases are compared under both derivative prescriptions in Table~\ref{tab:chen-gated}.\end{minipage}\end{table*}
\relax{}

Table~\ref{tab:chen-comparison} compares the mass intervals from Ref.~\cite{chen2021} with the spectral constraints for the surveyed currents. The intervals define external test regions; assessing the source's full mass extraction would additionally require its scale and fitting procedure. The local midpoint ratios $m_{\rm mid}^{\rm F,R}=\sqrt{T_1/T_0}$ are compared separately in Table~\ref{tab:prescription-diagnostics} (Appendix~\ref{app:robustness}).

Seven of the twelve nonempty classes already have an exact nonnegative reference measure from the retained OPE cut and constant term, including the four currents studied further below. For the remaining five nonempty classes, explicit spectra verified over the full Borel window establish feasibility at 1\%. In all twelve, verified spectra can empty $\mathcal B_C$ and the 20\% band. Nine currents nevertheless require a positive contribution in the 30\% band: two from the two-gluon sector, $GG,0^{-+}$ and $GG,2^{-+}$, and seven from the three-gluon sector, $GGG,0^{++},2^{++},0^{-+},1^{+-},2^{+-},3^{+-},2^{--}$. The three allowing even that broader band to be empty are $GG,0^{++}$, $GG,2^{++}$ and $GGG,2^{-+}$. 

In Figure~\ref{fig:fourteen-ranges}, the left endpoint gives the minimum contribution: an attained zero permits an empty band, whereas a positive bound requires every compatible spectrum to contribute. The upper endpoint limits the possible contribution. Appendix~\ref{app:channel-atlas} supplies the spectra and residuals for each current.

Among the nine currents with a positive broad-band bound, $GGG,2^{++},0^{-+},1^{+-},2^{--}$ also have subwindows satisfying the fixed-scale, finite-$s_0$ input criteria. Independent calculations on those subwindows retain positive bounds (Table~\ref{tab:chen-gated}). The other five, $GG,0^{-+},2^{-+}$ and $GGG,0^{++},2^{+-},3^{+-}$, have no such subwindow within their source windows. Their positive bounds remain diagnostics of the specified complete-window inputs. A dagger in Table~\ref{tab:chen-comparison} marks a current with no qualifying subwindow; spectral feasibility is tested separately.

For $GGG,1^{--}$ ($\widetilde{\xi}_1$) and $GGG,3^{--}$ ($\widetilde{\xi}_3$), independent inequalities exclude every nonnegative spectrum under the fixed-scale, full-window constraints at 1\%. Their minimum tolerances for general nonnegative spectra exceed approximately 2.598\% and 6.768\%, respectively; Appendix~\ref{app:chen-inputs} gives the certified brackets, separately from the single-pole test. We leave their band bounds unassigned because the classes are empty. The independent input audit also rejects both complete windows under the fixed-scale, finite-$s_0$ criteria (Table~\ref{tab:prescription-diagnostics}, Appendix~\ref{app:robustness}). These are separate tests of the specified inputs; the spectral exclusion concerns the moment models, rather than the existence of the corresponding physical states.

The fixed-scale convention F defines this classification. Table~\ref{tab:chen-gated} compares F with the running-coupling derivative control R for the four selected currents, at the same windows, thresholds, bands and weights. Their 20\% bands remain optional under F but require a contribution under R. Section~\ref{sec:robustness-controls} examines this dependence on the target derivatives and the separate input-criteria status.

Table~\ref{tab:classification-summary} groups these outcomes into classes A--E, whose physical use is discussed in Section~\ref{sec:classification-meaning}. Sections~\ref{sec:borel-input} and~\ref{sec:tensor-localization} examine spectral alternatives and a broad-band inequality for individual members. Their settings are specified separately from the complete-window protocol.
\relax{}
\relax{}
\subsection{Three-gluon pseudoscalar current: pole and empty-band alternatives}\label{sec:borel-input}
For $\widetilde{\eta}_0$, we compare a pole in a mass band with a smooth spectrum that avoids the band while matching the same moments. This coexistence test uses a different continuum and condensate setting from the undeformed complete-window baseline in Table~\ref{tab:chen-comparison}. The $J^{PC}=0^{-+}$ current of Ref.~\cite{chen2021} is
\begin{equation}\label{eq:pseudoscalar-current}
 \widetilde{\eta}_0=f^{abc}g_s^3
 \widetilde G_a^{\mu\nu}\widetilde G_{b,\nu\rho}
 \widetilde G_c^{\rho}{}_{\mu}.
\end{equation}
Here $G_a^{\mu\nu}$ is the gluon field strength, $\widetilde G_a^{\mu\nu}=\epsilon^{\mu\nu\rho\sigma}G_{a,\rho\sigma}/2$ its dual, and $f^{abc}$ the antisymmetric color structure constant. The current coupling and scalar pole residue are
\begin{equation}\label{eq:pseudoscalar-coupling}
 f_H=\langle0|\widetilde{\eta}_0|H\rangle,\qquad A=|f_H|^2 .
\end{equation}
For this dimension-six current $f_H$ has units $\GeV^5$ and $A$ has units $\GeV^{10}$. Its retained OPE density is
\begin{equation}
 p_\theta(s)=\frac{3\alpha_s^3}{10\pi}s^4+
       \frac{135}{8}\alpha_s^3G s^2+
       \frac{27}{2}\alpha_s^2Gr s .
 \label{eq:new-ope-cut}
\end{equation}
The density has units $\GeV^8$; $G$ and $r$ are defined in Eq.~\eqref{eq:common-condensates}. We evaluate Eq.~\eqref{eq:coupling-law} at fixed $\mu^2=6.305\,\GeV^2$.

We form $O_k(\tau;\theta)=\int_0^\infty s^kp_\theta(s)e^{-\tau s}\ud s$ from the retained OPE and keep the central threshold $s_0=33\,\GeV^2$ adopted in Ref.~\cite{chen2021}. The fixed renormalization scale equals the numerical midpoint of the Borel window $[4.48,8.13]\,\GeV^2$ specified there; only the Borel variable varies during a fit. At the baseline continuum $c=p_\theta$ above $s_0$, the nonnegative measure $p_\theta(s)\ud s$ below $s_0$ reproduces both targets exactly. It provides an analytic reference spectrum against which to test alternatives.

For the coexistence test we prescribe the high-energy model through \mbox{$c_{\beta,\Delta}(s;\theta)=p_\theta(s)[1+\beta b_\Delta(s)]$}, where
\begin{equation}\label{eq:new-continuum}
\begin{aligned}
b_\Delta(s)&=64x^3(1-x)^3,\\
 x&=(s-s_0)/\Delta\in[0,1],\\
 b_\Delta&=0\ \hbox{outside this interval}.
\end{aligned}
\end{equation}
The shape peaks at one and joins zero with two continuous derivatives. Thus $\beta$ is the maximum signed fractional change at the center, $|\beta|$ is the largest local change in magnitude, and $\Delta$ is its width in mass squared. The continuum is nonnegative for $\beta\geq-1$ and returns to the baseline beyond $s_0+\Delta$. We first compare spectra at one fixed member of this model. Section~\ref{sec:continuum-controls} then varies its shape to test input dependence.

For the coexistence example we fix $G=0.060\,\GeV^4$, $r=7.2\,\GeV^2$, $\beta=0.70$ and $\Delta=12\,\GeV^2$. We use $M_B^2\in[5.7,7.2]\,\GeV^2$, with $I=[a,b]=[1/7.2,1/5.7]\,\GeV^{-2}$ and $\tau_*=a$. This noncentral setting satisfies the input criteria. At the central condensates, with all other settings unchanged, the minimum single-pole mismatch is about 1.005\% (Section~\ref{sec:continuum-sensitivity}). The same Borel window and weight are used in the pseudoscalar continuum comparisons of Section~\ref{sec:continuum-controls}. The window was selected from the input criteria before testing spectra; the verification over those parameter families is described there. Other currents retain their own windows and thresholds.

For this current the input criteria of Eq.~\eqref{eq:channel-input-criteria} reduce to
\begin{equation}\label{eq:borel-window-tests}
 \frac{|T_0^{(4)}|}{T_0}\leq5\%,\quad
 \frac{|T_0^{(6)}|}{T_0}\leq10\%,\quad
 \frac{T_0}{O_0}\geq40\%
\end{equation}
Here $T_0^{(4)}$ and $T_0^{(6)}$ are the contributions of the $Gs^2$ and $Grs$ terms after the same continuum subtraction; both have units $\GeV^{10}$. These conditions limit retained condensate terms and keep a substantial share of the input below $s_0$. The 40\% condition refers to the input ratio $T_0/O_0$, whereas $F_{\mathcal B}$ measures a candidate spectrum's contribution in a particular band. Target positivity and spectral feasibility are checked separately.

We chose the coexistence test band $[4.35,4.85]\,\GeV$ around an initial moment-ratio mass near 4.60 GeV, before running the tests. The $\widetilde{\eta}_0$ mass interval reported in Ref.~\cite{chen2021}, $[3.77,4.31]\,\GeV$, is used in the complete-window survey and correlated-input analysis (Sections~\ref{sec:chen-comparison} and \ref{sec:correlated-inputs}). The broader interval $[4.10,5.10]\,\GeV$ tests whether the contribution can be displaced farther away. \mbox{The tensor} example below uses the mass interval reported for it in Ref.~\cite{chen2021}.

\begin{figure*}[t]
\centering\includegraphics[width=\linewidth]{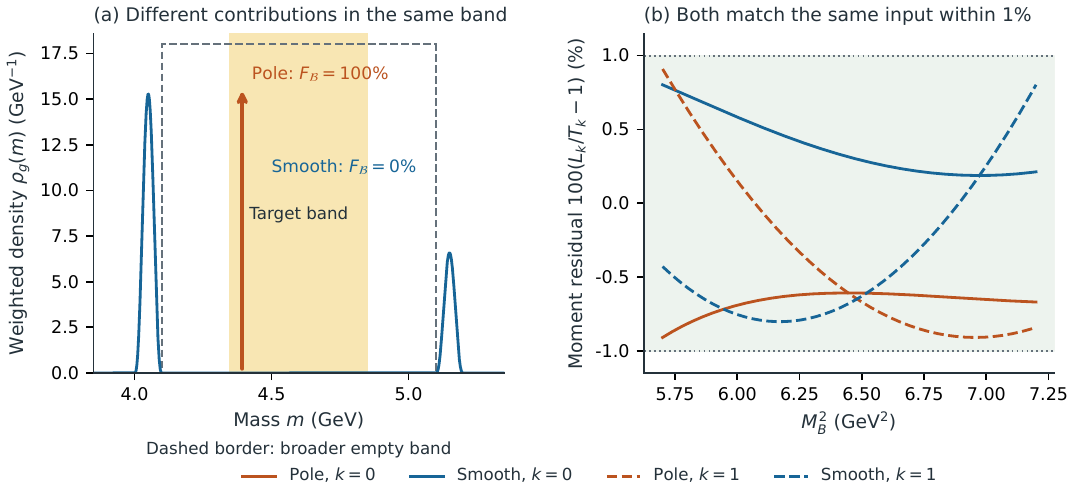}
\caption{A pole-compatible input also permits an empty target band. (a) The pole puts 100\% of its normalized Borel weight in the gold $[4.35,4.85]\,\GeV$ band; the smooth alternative puts zero there and throughout the larger $[4.10,5.10]\,\GeV$ interval outlined in gray, including its edges. The density is $\rho_g(m)=2m e^{-\tau_*m^2}\rho(m^2)/L_0(\tau_*)$; the unit-weight delta at 4.393 GeV has symbolic arrow height. (b) The ordinate is the signed relative moment residual $100\mathcal R_k=100(L_k/T_k-1)$ in percent [Eq.~\eqref{eq:relative-residual}]. Both spectra keep both orders inside the green $\pm1\%$ strip over the full window. The common input is $\Delta=12\,\GeV^2$, $\beta=0.70$, $G=0.060\,\GeV^4$, $r=7.2\,\GeV^2$, $\mu^2=6.305\,\GeV^2$, $s_0=33\,\GeV^2$, $M_B^2\in[5.7,7.2]\,\GeV^2$ and $\tau_*=(7.2\,\GeV^2)^{-1}$. The smooth packets have 100 MeV full support width. Continuous checks establish their compatibility (Appendix~\ref{app:continuum-verification}).}
\label{fig:common-continuum}
\end{figure*}

\subsubsection{Coexisting spectral descriptions}\label{sec:matched-pair}
At the setting above, Figure~\ref{fig:common-continuum} compares the two spectra and their moment residuals. Panel (a) shows their different band contributions; panel (b) plots $100\mathcal R_k=100(L_k/T_k-1)$ for $k=0,1$ [Eq.~\eqref{eq:relative-residual}]. The zero line is exact matching, and the green strip is the allowed $\pm1\%$ discrepancy for each moment. The pole lies near 4.393 GeV, while the smooth spectrum vanishes throughout $[4.10,5.10]\,\GeV$, including its edges. Both satisfy the 1\% constraints over the full window, with certified residual bounds of approximately $0.90957\%$ and $0.93103\%$, respectively. Both also remain feasible in a small parameter neighborhood (Appendix~\ref{app:borel-neighborhood}).

For the target band $[4.35,4.85]\,\GeV$, the two fractions are one and zero. Their mixtures satisfy the same constraints and attain every intermediate fraction, giving the exact allowed range $[0,1]$. In the smooth alternative, contributions below and above the band together reproduce the target's mass weighting. The moment ratio can therefore point to an empty region between them.

The smooth spectrum consists of compact packets on $[4.00,4.10]$ and $[5.10,5.20]\,\GeV$, vanishing at their endpoints. These supports establish the same $[0,1]$ range for every $\tau_*\in I$ and every band $\{m_p\}\subseteq\mathcal B\subseteq[4.10,5.10]\,\GeV$, with $m_p\simeq4.39344\,\GeV$. For each fixed choice of band and $\tau_*$, mixtures of the two spectra attain all intermediate fractions.

The onset of coexistence can also be bounded. Let $\epsilon_{\rm coexist}$ be the infimum tolerance admitting both a pole in $[4.35,4.85]\,\GeV$ and a nonnegative spectrum empty in the open band $(4.10,5.10)\,\GeV$. Exclusion of all poles below one tolerance and the verified pair above it give
\begin{equation}\label{eq:coexistence-tolerance}
0.90\%<\epsilon_{\rm coexist}\leq0.93\%.
\end{equation}
The upper endpoint is a sufficient value from the saved pair, rounded upward. Both spectra remain feasible for every larger allowed tolerance; the bracket locates the onset rather than the full range of coexistence.

\subsection{Three-gluon tensor current: a required broad contribution}\label{sec:tensor-localization}
For the tensor current $\eta_2$, every single pole is excluded at 1\%, yet a broad mass band must contribute. We first use endpoint moment ratios to exclude a pole, then combine the endpoint moments to bound the contribution from a general nonnegative spectrum. The $GGG,2^{++}$ current is
\begin{equation}\label{eq:tensor-current}
\eta_2^{\alpha_1\alpha_2,\beta_1\beta_2}
=f^{abc}\mathcal S\!\left[g_s^3G_a^{\alpha_1\beta_1}
G_b^{\alpha_2\mu}G_{c,\mu}^{\beta_2}
-\{\alpha_2\leftrightarrow\beta_2\}\right].
\end{equation}
The operation $\mathcal S$ symmetrizes and removes traces within the two sets of $\alpha$ and $\beta$ indices simultaneously. Appendix~\ref{app:tensor-input} specifies the positive-parity matrix element and scalar residue.

We use the density in Eq.~\eqref{eq:tensor-ope} and the mass interval $[3.76,4.37]\,\GeV$ reported in Ref.~\cite{chen2021}, corresponding to $4.18^{+0.19}_{-0.42}\,\GeV$. Wider test bands probe coarser mass resolution: $[2.926,5.434]\,\GeV$ equals $[0.7M_C,1.3M_C]$ for $M_C=4.18\,\GeV$, and $[3.34,5.02]\,\GeV$ rounds $[0.8M_C,1.2M_C]$ outward.

We fix $G=0.0635\,\GeV^4$, $r=8.2\,\GeV^2$, $\mu^2=6.905\,\GeV^2$ and $s_0=35\,\GeV^2$, and impose 1\% envelopes on both moments over $M_B^2\in[5.42,7.95]\,\GeV^2$. The baseline continuum is the retained tensor OPE density above $s_0$. The fraction uses $\tau_*=(9\,\GeV^2)^{-1}<a=(7.95\,\GeV^2)^{-1}$, retained from a preliminary wider-window calculation. This choice sets the observable's weight; all OPE constraints remain inside the stated window. Positivity links the normalization to the constrained moment at $a$: the candidate-dependent factor $\chi=T_0(a)/L_0(\tau_*)$ lies within $[0.5923,1.0102]$ for every feasible spectrum. Appendix~\ref{sec:positive-bounds} derives this conservative enclosure and the direct band bound. Appendix~\ref{app:tensor-input} gives the input tests, including the 5\% limit on the higher-order dimension-four term alone.

\noindent\begin{minipage}{\columnwidth}\makeatletter\def\@captype{table}\makeatother\centering\small
\caption{Minimum tensor band fractions at 1\% tolerance. All rows use $GGG,2^{++}$ on $M_B^2\in[5.42,7.95]\,\GeV^2$, weighted at $\tau_*=(9\,\GeV^2)^{-1}$. The fixed continuum is the retained OPE density ($\beta=0$). The family column minimizes jointly over spectra and continuum members with $|\beta|\leq0.75$ at $\Delta=18\,\GeV^2$, keeping $G,r,\mu^2,s_0$ fixed (Section~\ref{sec:continuum-controls}). Fractions use each spectrum's Borel integral below $s_0$. Brackets enclose the infimum between a rigorous lower bound and a feasible value, rounded outward. Every zero is attained by a spectrum at one allowed setting.}\label{tab:tensor-localization}
\begin{tabular}{ccc}\toprule
Mass band ($\GeV$) & \shortstack{Fixed\\continuum (\%)} & \shortstack{Family\\minimum (\%)}\\\midrule
$[3.76,4.37]$ & $0$ & $0$\\
$[3.34,5.02]$ & $0$ & $0$\\
$[2.926,5.434]$ & $[22.2006,22.2009]$ & $0$\\
\bottomrule\end{tabular}
\end{minipage}
\relax{}

All coefficients of $p_T$ in Eq.~\eqref{eq:tensor-ope} are positive. Hence $\ud\rho_{\rm ref}=p_T(s)\ud s$ below $s_0$ is an exact nonnegative reference spectrum. A pole faces a stronger restriction: its mass must be the same at both window endpoints. As $\tau$ increases from $a$ to $b$, the target root-mean-square mass $\sqrt{T_1/T_0}$ decreases from approximately 4.80 to $4.38\,\GeV$. Reconciling these endpoints with a single mass requires a pole tolerance exceeding 4.54\% (Appendix~\ref{app:tensor-pole}). The 1\% class is therefore nonempty but contains no single pole.

For an extended distribution this decrease has a simple origin: the larger $\tau$ suppresses its heavier components more strongly. The endpoint values describe two weightings of the same spectrum, rather than two resonance masses. Pole exclusion establishes that a spread in mass is needed at this input.

The endpoint moments constrain both the overall Borel contribution and its mass weighting. We combine them in a kernel inequality that limits how much contribution can lie outside the broad band. In this tensor example, the lower-bound inequality uses only $a=(7.95\,\GeV^2)^{-1}$ and $b=(5.42\,\GeV^2)^{-1}$; all interior-point multipliers vanish. Every spectrum compatible over the full window obeys these endpoint constraints, so the resulting bound applies to the full spectral class.

Define $x=s/s_0$, with $s_0=35\,\GeV^2$, and $E_t(s)=\exp[-(t-\tau_*)s]$. Against the normalized measure $\ud\eta=e^{-\tau_*s}\ud\rho/L_0(\tau_*)$, $E_t$ integrates to $L_0(t)/L_0(\tau_*)$ and $xE_t$ to $L_1(t)/[s_0L_0(\tau_*)]$. The four kernels are therefore the two constrained moment orders evaluated at the two endpoints. We combine them to bound the negative band indicator:
\begin{equation}\label{eq:tensor-endpoint-kernel}
\begin{aligned}
q(s)={}&v-dE_a(s)+c_0E_b(s)\\
      &+c_a xE_a(s)-c_b xE_b(s)\\
\geq{}&-\mathbf1_{[2.926^2,\,5.434^2]\,\GeV^2}(s)
\quad(0\leq s\leq s_0).
\end{aligned}
\end{equation}
The coefficients are approximately $c_0\simeq5.88$, $c_a\simeq4.41$, $c_b\simeq29.63$ and $v\simeq-0.22$. The normalization correction $d\simeq1.01\times10^{-12}$ is retained. These printed decimals identify the terms; the proof uses the exact stored coefficients on the whole support. Appendix~\ref{app:tensor-certificate} describes an independent check against the archived inequality.

Write $F_{30}$ for $F_{\mathcal B}$ in the broad 30\% band. Integrating Eq.~\eqref{eq:tensor-endpoint-kernel} gives $-F_{30}\leq\int q\,\ud\eta$. The endpoint moment envelopes then bound that integral from above. With $\chi=T_0(a)/L_0(\tau_*)>0$, the two steps are
\begin{equation}\label{eq:tensor-endpoint-data}
\begin{aligned}
-F_{30}&\leq\int q\,\ud\eta\leq v+\chi D\leq v,\\
D={}&-(1-\epsilon)d
 +(1+\epsilon)c_0\frac{T_0(b)}{T_0(a)}\\
 &+(1+\epsilon)c_a\frac{T_1(a)}{s_0T_0(a)}
 -(1-\epsilon)c_b\frac{T_1(b)}{s_0T_0(a)}
 \leq0.
\end{aligned}
\end{equation}
Positive coefficients multiply the upper moment envelope and negative coefficients the lower one. These sign choices give an upper bound on the endpoint-moment combination for every compatible spectrum. At $\epsilon=0.01$, interval evaluation of this data bound with the exact coefficients gives $D\leq0$. It follows that $F_{30}\geq-v>0.22$ for every compatible nonnegative spectrum.

The inequality limits the band contribution because it holds at every allowed mass, before any choice of spectral shape. A three-atom spectrum at approximately $2.93$, $4.07$ and $5.43\,\GeV$ attains $F_{30}<0.22$ while satisfying both moments over the whole window. Its first and third atoms lie strictly outside the closed band at their exact positions. This spectrum shows how closely the lower bound can be approached; its atoms serve as a mathematical construction, without assigning them to glueball states. The endpoint inequality and full-window spectrum bracket the same minimum. Only this particular lower-bound proof reduces to two endpoints; the candidate and other calculations retain their full continuous verification.
\relax{}

The 22.2\% bound requires this share of each compatible candidate spectrum's own low-energy Borel integral, $L_0[\rho](\tau_*)$, to come from the broad band. All resonances and continuous contributions there count toward it. The exact reference spectrum places about 68.1\% in the band; allowing other spectra reduces the requirement to about 22.2\% (Table~\ref{tab:tensor-localization}).

Contracting or expanding the broad band by 25 MeV at each edge gives minima of about 19.6\% and 24.6\%, respectively, at the same input (Table~\ref{tab:tensor-controls}). The contribution remains required under both changes.

\begin{figure*}[t]
\centering\includegraphics[width=\linewidth]{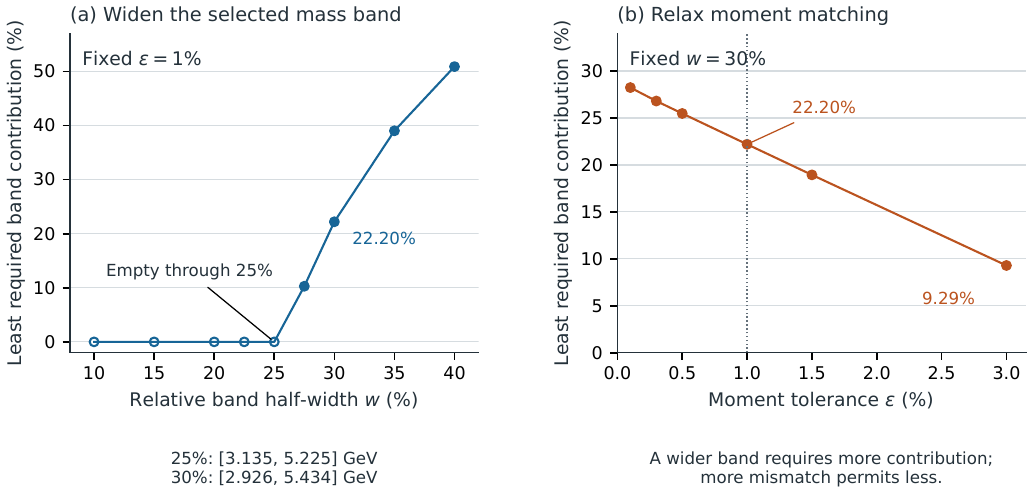}
\caption{A sufficiently broad tensor mass band must contribute, with the required amount set by the allowed moment mismatch. (a) Widening $[(1-w)M_C,(1+w)M_C]$ at fixed 1\% tolerance changes its least allowed contribution from zero to a positive value. Open circles mark attained zeros; filled circles mark positive minima. Two representative mass intervals are printed below the panel. (b) For the 30\% band, relaxing both moment constraints lowers the minimum from 22.20\% at 1\% tolerance to 9.29\% at 3\%. The horizontal percentages describe band half-width or moment tolerance; vertical percentages describe Borel contribution. Both panels use $GGG,2^{++}$ with $M_C=4.18\,\GeV$, $G=0.0635\,\GeV^4$, $r=8.2\,\GeV^2$, $\mu^2=6.905\,\GeV^2$, $s_0=35\,\GeV^2$, $M_B^2\in[5.42,7.95]\,\GeV^2$, $\tau_*=(9\,\GeV^2)^{-1}$ and $\beta=0$. Certified brackets are narrower than 0.0025 percentage point. Connecting lines only guide the eye.}
\label{fig:tensor-resolution}
\end{figure*}

The narrow candidate band and $[3.34,5.02]\,\GeV$ both admit smooth empty-band spectra at 0.1\% tolerance on the wider window $[5.12,9.00]\,\GeV^2$. Those spectra also satisfy 1\% on the common subwindow. Their 100 MeV support-width packets vanish throughout the closed bands, including the edges.

To locate the resolution at which a contribution becomes required, set $\mathcal B_w=[(1-w)M_C,(1+w)M_C]$ with exact closed endpoints. At the baseline continuum and 1\%, a verified spectrum leaves $\mathcal B_{0.25}$ empty, while $w=0.275$ already requires about 10.3\% and $w=0.30$ about 22.2\% (Figure~\ref{fig:tensor-resolution}a). The transition lies between 25\% and 27.5\% relative half-width.

At fixed $w=0.30$, increasing the tolerance reduces the minimum from about 28.2\% at 0.1\% to 22.2\% at 1\% and 9.29\% at 3\% (Figure~\ref{fig:tensor-resolution}b). The exact reference spectrum keeps every class nonempty. Since decreasing $\epsilon$ restricts the class, the 3\% certificate gives a uniform conservative lower bound of 9.29\% throughout $0\leq\epsilon\leq3\%$. The actual minima at smaller tolerances can be larger. At this central tensor input, the endpoint test excludes every single pole throughout this tolerance range.

A sharp band counts weight discontinuously across its edges. The empty 25\% band, $[3.135,5.225]\,\GeV$, extends beyond the candidate interval by 0.625 and 0.855 GeV. Candidate-band emptiness thus survives these finite margins. Relative to the larger band, one saved atom contributing 7.04\% of the spectrum's total low-energy Borel integral at $\tau_*=(9\,\GeV^2)^{-1}$ lies only about 9.57 eV above its upper edge. The exact stored position places it outside the closed band; the small offset specifies support membership rather than physical mass resolution. The achieved margin around the candidate band is therefore more informative than that offset.

The tensor results illustrate the effect of choosing a window and weight. Table~\ref{tab:tensor-localization} gives about 22.2\% on $[5.42,7.95]\,\GeV^2$, weighted at $(9\,\GeV^2)^{-1}$. The full Borel window $[4.77,9.04]\,\GeV^2$ of Ref.~\cite{chen2021}, weighted at its inverse upper endpoint, gives $[24.03,24.04]\%$ in Table~\ref{tab:chen-comparison}. The selected window $[5.11,9.01]\,\GeV^2$, again using its inverse upper endpoint, gives $[23.64,23.66]\%$ in Table~\ref{tab:chen-gated}. The OPE coefficients, threshold, baseline continuum and broad band are the same in all three tests.

The pseudoscalar example admits a pole in the test band and a spectrum that avoids it. The tensor example excludes every single pole, yet still requires a contribution in a broad band. 

Together, the two examples show that pole compatibility and band necessity are independent. For $\widetilde{\eta}_0$, a pole and a smooth empty-band spectrum reproduce the same moments, giving the exact fraction range $[0,1]$. For $\eta_2$, every single pole is excluded, yet every compatible spectrum must supply at least 22.2\% of its own low-energy Borel integral in the broad band. Across the complete-window classification, all twelve nonempty inputs permit empty candidate bands, nine require the 30\% band, and two inputs admit no nonnegative spectrum at 1\%. Section~\ref{sec:input-dependence} tests how far these statements survive input variation.
\relax{}
\section{Dependence on sum-rule inputs and spectral support}\label{sec:input-dependence}
The bounds in Section~\ref{sec:applications} depend on the sum-rule inputs and the allowed spectral support. We first vary the continuum shape, derivative prescription, threshold and support separately, keeping the spectral tests unchanged. We then propagate the condensate and threshold ranges jointly and determine the total moment discrepancy needed to allow an empty band.

\subsection{Continuum shape}\label{sec:continuum-controls}
Changing the high-energy continuum changes the low-energy targets even when the OPE is held fixed. We vary $\beta$ and $\Delta$ in Eq.~\eqref{eq:new-continuum} to test single-pole compatibility and broad-band requirements for the two detailed currents. These sensitivity controls are motivated by the continuum dependence found in solvable models~\cite{lucha2007,lucha2009}; their ranges are not calibrated to physical continuum uncertainties (Section~\ref{sec:physical_controls}).

\subsubsection{Input families}
We first specify the parameter families so that each continuum comparison uses a common input throughout the window. Let $\Theta=[0.060,0.067]\,\GeV^4\times[7.2,9.2]\,\GeV^2$ be the $(G,r)$ rectangle. In the notation of Section~\ref{sec:positive-model}, $\theta=(G,r)$ and $\gamma=(\beta,\Delta)$. Each member specifies one condensate pair, one deformation amplitude and one fixed deformation width. The three input families are
\begin{equation}\label{eq:continuum-parameter-sets}
\begin{aligned}
\mathcal U_{12}&=\Theta\times[-0.75,0.75]\times\{12\,\GeV^2\},\\
\mathcal U_{6}&=\Theta\times[-0.50,0.50]\times\{6\,\GeV^2\},\\
\mathcal U_{18}&=\Theta\times[-0.50,0.50]\times\{18\,\GeV^2\}.
\end{aligned}
\end{equation}
Within each family, $\Delta$ is fixed and a single choice of $(G,r,\beta)$ generates both moments throughout the window. The amplitude scan below restricts $\beta$ to $[-\beta_{\max},\beta_{\max}]$ and optimizes $G,r$ over $\Theta$, keeping $\Delta$, $\mu^2$, $s_0$ and the Borel window fixed. Each setting supplies one common pair of target curves. Increasing the continuum subtracts more from both moments, with an extra factor of $s$ in $T_1$, and can change both the normalization and $T_1/T_0$.

The common window here belongs only to these three pseudoscalar input families and was selected from the input criteria before fitting spectra. The undeformed condensate corners allow $[5.4,8.09]\,\GeV^2$, but enhanced continua fail the retained-term tests at its ends. The narrower interval $M_B^2\in[5.7,7.2]\,\GeV^2$ satisfies the conditions throughout all three families. We retain it for the matched pair and all three pseudoscalar families, with $I=[a,b]=[1/7.2,1/5.7]\,\GeV^{-2}$ and $\tau_*=a$. The correlated-input calculation in Section~\ref{sec:correlated-inputs} uses a separately specified window.

Continuous interval checks cover target positivity and the input criteria of Eq.~\eqref{eq:borel-window-tests} over each full parameter box (Appendix~\ref{app:continuum-verification}). These checks concern the specified moment inputs; a compatible spectrum is verified separately at each setting used for a band conclusion.

\subsubsection{Single-pole compatibility}\label{sec:continuum-sensitivity}
\begin{figure*}[t]
\centering\includegraphics[width=\linewidth]{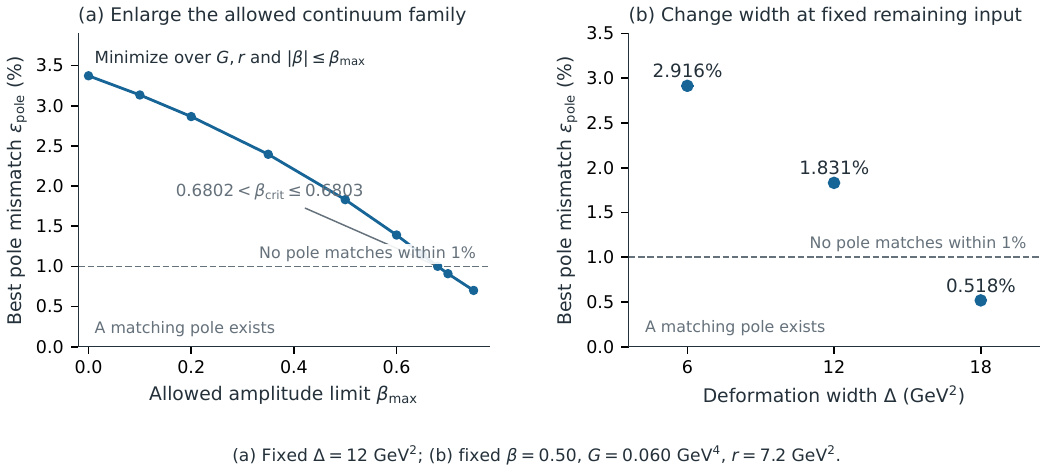}
\caption{Continuum choices can change whether any single pole matches the input within 1\%. The ordinate is the smallest full-window pole mismatch; crossing below the dashed line admits a matching pole. (a) Increasing $\beta_{\max}$ enlarges the allowed family: the minimum ranges over $G\in[0.060,0.067]\,\GeV^4$, $r\in[7.2,9.2]\,\GeV^2$ and $|\beta|\leq\beta_{\max}$ at $\Delta=12\,\GeV^2$. The critical amplitude has a uniform exclusion bound and a feasible endpoint. (b) Only the deformation width changes, with $\beta=0.50$, $G=0.060\,\GeV^4$ and $r=7.2\,\GeV^2$. Independent points show bracket midpoints with their certified lower and upper endpoints, 0.0003 percentage point apart. The printed values are feasible upper endpoints. Both panels keep $\mu^2=6.305\,\GeV^2$, $s_0=33\,\GeV^2$ and $M_B^2\in[5.7,7.2]\,\GeV^2$. The line in (a) guides the eye; (b) reports only the three tested widths. The deformation amplitudes are model controls (Appendix~\ref{app:continuum-verification}).}
\label{fig:continuum-profile}
\end{figure*}

Figure~\ref{fig:continuum-profile} shows how continuum variation changes single-pole compatibility. Panel (a) enlarges the allowed input family, whereas panel (b) changes only the deformation width. A mismatch below the 1\% line admits a compatible pole. At the central condensates, $G=0.0635\,\GeV^4$ and $r=8.2\,\GeV^2$, the width-$12\,\GeV^2$, $\beta=0.70$ continuum gives about 1.005\%, just above 1\%. Raising $\beta$ to 0.75 admits a pole within 1\% at those central inputs.

We also minimize over the common input parameters, a procedure called profiling. It chooses the single parameter set most favorable to a pole across the whole window. On the undeformed continuum, the profiled mismatch is about 3.37\%, compared with 3.47\% at the central input. Allowing $|\beta|\leq\beta_{\max}$ reduces it to about 1.83\% at $\beta_{\max}=0.50$ and 0.909\% at $\beta_{\max}=0.70$.

The 1\% threshold requires about 68\% maximum local enhancement for $\Delta=12\,\GeV^2$. At the verified feasible endpoint $\beta=0.6803$, the continuum's zeroth Borel integral at $\tau_*=(7.2\,\GeV^2)^{-1}$ increases by about 16.0\% relative to the undeformed continuum at the same input. This smaller integrated change averages the deformation with the Borel weight.

The width matters as well. At $\beta=0.50$ and the lower condensate corner, widths 6, 12 and $18\,\GeV^2$ give upper bounds on the optimal mismatch of about 2.92\%, 1.83\% and 0.518\%. Profiling the width-$18\,\GeV^2$ family reaches 1\% at about 41\% local enhancement, corresponding to a 12.7\% increase in the continuum Borel integral. Appendix~\ref{app:continuum-verification} gives the certified brackets and parameter domains. These controls quantify the continuum dependence of a pole description within the declared shape families.

\subsubsection{Broad-band requirements}\label{sec:gap-union}
We now extend the continuum comparison from single-pole fits to band requirements for arbitrary nonnegative spectra. At $\Delta=18\,\GeV^2$, $\beta=0.50$, $G=0.060\,\GeV^4$ and $r=7.2\,\GeV^2$, the moment constraints at 1\% tolerance require a contribution in the broad open band $(4.10,5.10)\,\GeV$. A verified pole establishes that the spectral class is nonempty. A separate inequality excludes every nonnegative spectrum supported on
\begin{equation}\label{eq:gap-support}
 [0,(4.10\,\GeV)^2]\ \cup\ [(5.10\,\GeV)^2,s_0].
\end{equation}
The exclusion applies to arbitrary measures on this support (Appendix~\ref{app:gap-verification}). It establishes a required contribution in the broad band, while leaving its minimum fraction and the contribution in the narrower target band unquantified.

At the same input, a smooth broad-gap spectrum satisfies 1.5\% matching. If $\epsilon_{\rm gap}$ denotes the infimum tolerance permitting a spectrum on the support in Eq.~\eqref{eq:gap-support}, the saved inequality and spectrum give
\begin{equation}\label{eq:gap-tolerance}
1.19\%<\epsilon_{\rm gap}\leq1.48\%.
\end{equation}
The lower endpoint follows from the support inequality and the upper from a verified feasible spectrum. Both keep the input and continuum fixed.

To allow continuum variation, admit any spectrum compatible with at least one setting in a declared family $\mathcal U$ from Eq.~\eqref{eq:continuum-parameter-sets}:
\begin{equation}
 \mathcal A_\epsilon^{\mathcal U}
   =\bigcup_{(\theta,\gamma)\in\mathcal U}\mathcal A_\epsilon(\theta,\gamma).
 \label{eq:positive-union}
\end{equation}
A positive infimum over this union would require a contribution at every admissible setting. Each parameter family includes a $\beta=0$ setting admitting a verified smooth spectrum that matches within 1\% and vanishes throughout $[4.10,5.10]\,\GeV$. For each family, the minimum over its associated union is therefore attained at zero at every $\tau_*\in I$. This covers the broad band $(4.10,5.10)\,\GeV$, the candidate band $[4.35,4.85]\,\GeV$ and every subinterval of $[4.10,5.10]\,\GeV$. The primary family also includes the matched-pair setting, so its associated union retains the exact $[0,1]$ range under the band condition in Section~\ref{sec:matched-pair}. There is therefore no uniform positive band requirement, although particular fixed continuum members still require a contribution.

For the detailed $GGG,2^{++}$ input, continuum suppression can remove the 22.2\% broad-band requirement established in Section~\ref{sec:tensor-localization}. With Eq.~\eqref{eq:new-continuum}, $\Delta=18\,\GeV^2$, $|\beta|\leq0.75$ and fixed condensates, 25\% maximum local suppression lowers the minimum to about 8.47\%; 50\% suppression admits an empty broad band. Table~\ref{tab:tensor-controls} (Appendix~\ref{app:tensor-controls}) gives the numerical brackets. This tensor family differs from the pseudoscalar families. Its continuum changes in moment units use the full zeroth OPE moment as denominator, whereas the pseudoscalar percentages above use the undeformed continuum integral.

At $\beta=-0.75$, nonnegative $C^2$ densities with packet support widths of 5 and 25 MeV also attain zero broad-band fraction at 1\%. These are smooth alternatives within the same continuum family. Appendix~\ref{app:tensor-controls} gives the packet construction and the unresolved searches at 50 and 100 MeV support width.

Table~\ref{tab:chen-inputs} (Appendix~\ref{app:chen-inputs}) extends the comparison to three fixed continuum settings for each current. It identifies which broad bands remain required, which can be empty, and which settings admit no compatible nonnegative spectrum. These per-current comparisons are discrete; the parameter families for the two detailed examples are treated separately above.

\relax{}
\subsection{Derivative prescription, threshold and spectral support}\label{sec:robustness-controls}
We use the four selected currents to compare derivative prescriptions, continuum thresholds and low-mass support cuts. Each comparison keeps the continuum undeformed and holds the other settings fixed. This separates changes to the moment targets from restrictions on the candidate spectra.

\begin{table*}[t]\centering\small
\caption{Paired F/R band minima on the four windows selected with fixed-scale, finite-$s_0$ input criteria. F holds the coupling fixed; R includes its running total derivative. All entries use central condensates, $\beta=0$ and 1\% tolerance. The final three columns give brackets for the minimum fraction.}\label{tab:chen-gated}
\setlength{\tabcolsep}{4pt}\renewcommand{\arraystretch}{1.16}
\begin{tabular}{ccccrccc}\toprule
Channel & $M_B^2$ window & F/R & \shortstack{Finite-limit\\criteria} & $\epsilon_{\rm pole}$ (\%) & $\inf F_C$ (\%) & $\inf F_{20}$ (\%) & $\inf F_{30}$ (\%)\\\midrule
$GGG,\,2^{++}$ & $[5.11,9.01]$ & F & Full & $[9.140,9.141]$ & $0$ & $0$ & $[23.64,23.66]$\\
$GGG,\,2^{++}$ & $[5.11,9.01]$ & R & Partial & $[7.378,7.379]$ & $0$ & $[29.34,29.36]$ & $[66.28,66.30]$\\
\addlinespace[2pt]
$GGG,\,0^{-+}$ & $[4.98,8.11]$ & F & Full & $[6.852,6.853]$ & $0$ & $0$ & $[29.69,29.70]$\\
$GGG,\,0^{-+}$ & $[4.98,8.11]$ & R & Full & $[5.242,5.243]$ & $0$ & $[30.90,30.92]$ & $[68.39,68.40]$\\
\addlinespace[2pt]
$GGG,\,1^{+-}$ & $[3.16,9.03]$ & F & Full & $[27.040,27.041]$ & $0$ & $0$ & $[19.06,19.09]$\\
$GGG,\,1^{+-}$ & $[3.16,9.03]$ & R & Partial & $[22.886,22.887]$ & $0$ & $[26.74,26.80]$ & $[66.03,66.06]$\\
\addlinespace[2pt]
$GGG,\,2^{--}$ & $[7.19,9.44]$ & F & Full & $[3.742,3.743]$ & $0$ & $0$ & $[17.19,17.20]$\\
$GGG,\,2^{--}$ & $[7.19,9.44]$ & R & Partial & $[2.323,2.324]$ & $0$ & $[13.80,13.81]$ & $[60.80,60.81]$\\
\addlinespace[2pt]
\bottomrule\end{tabular}
\par\smallskip\begin{minipage}{\textwidth}\small\raggedright Windows are in GeV$^2$. Full: the adopted OPE-convergence and low-energy-fraction criteria hold throughout the window, using finite upper limits in the convergence ratios. Partial: certified passing and failing subintervals occur. With infinite convergence limits, all eight rows satisfy the criteria throughout these selected windows. Table~\ref{tab:prescription-diagnostics} gives the separate complete-window diagnostics. Each F/R pair has identical $s_0$, window, weight and mass bands; $\tau_*$ is the inverse upper window endpoint and fixed scales are those of Table~\ref{tab:chen-inputs}. Every spectral class has a verified nonnegative witness, irrespective of its input-criteria status. Source masses remain external comparison intervals.\end{minipage}\end{table*}
\relax{}

\begin{figure*}[t]
\centering\includegraphics[width=\linewidth]{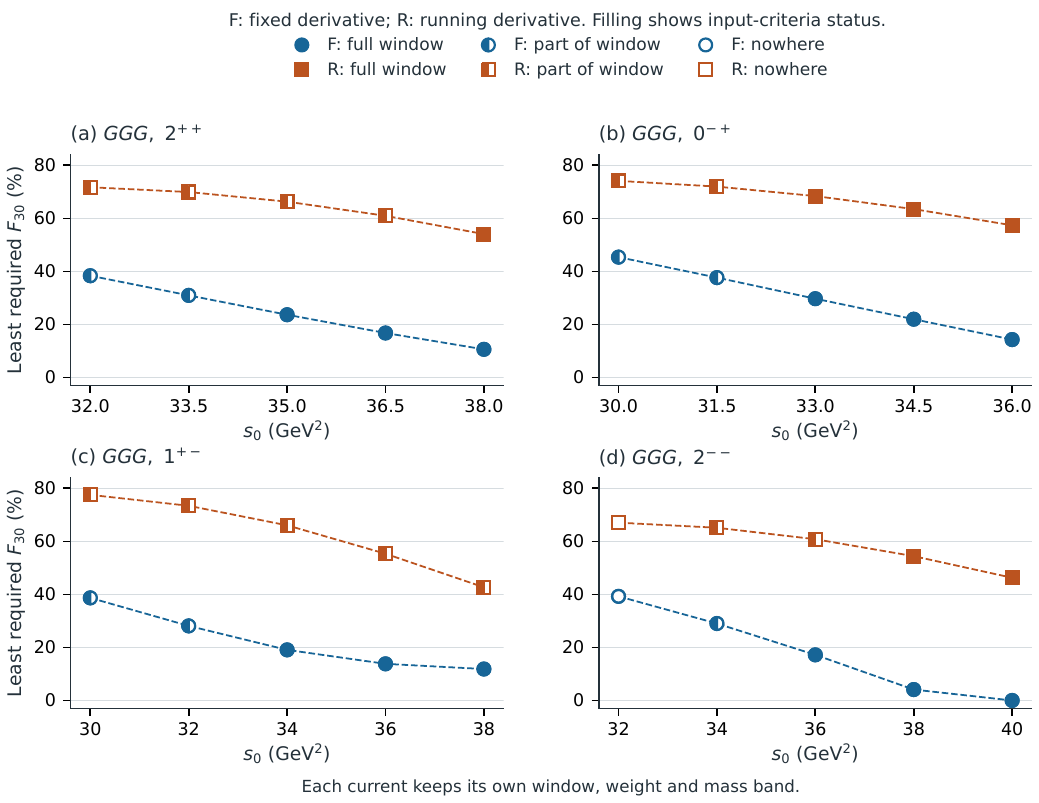}
\caption{Broad-band requirements respond differently to the continuum threshold under fixed and running derivatives. The four panels keep each current's selected window, band and weight from Table~\ref{tab:chen-gated}, central condensates, $\beta=0$ and 1\% tolerance. Blue circles are F and orange squares R. The ordinate gives the minimum contribution to the 30\% band. For both marker shapes, full, half and empty filling mean that the finite-limit input criteria hold throughout, on part of, or nowhere in the window, respectively. Filling indicates input-criteria status, not numerical confidence; even a hollow point carries a verified bound. Numerical brackets are smaller than the symbols. Dashed lines connect tested points to show trends; the running results are pointwise. The fixed-scale embedding proof supplies separate bounds over the continuous threshold range. In the $2^{--}$ panel, the highest threshold admits an empty broad band under F while retaining a positive requirement under R.}\label{fig:threshold-controls}
\end{figure*}

\subsubsection{Moment derivatives at fixed and running coupling}
The F/R comparison tests how the target curves depend on the coupling prescription, with the spectral method and comparison settings unchanged. At the midpoint of a Borel window specified in Ref.~\cite{chen2021}, F and R use the same coupling but different derivatives. R includes the chain rule in $T_1=-\ud T_0/\ud\tau$ with $\mu^2=1/\tau$, acting on both the cut and constant terms. The resulting fourteen midpoint diagnostics agree with the central mass results in Ref.~\cite{chen2021} to within 5 MeV and round to the same two decimal places; F lies about 435--633 MeV higher (Table~\ref{tab:prescription-diagnostics}). For the tensor current the diagnostic falls from about 4.68 to 4.18 GeV, or 10.6\%. The additional derivatives are formally of higher coupling order but substantial at these inputs. This comparison motivates R as a control for the mass results in Ref.~\cite{chen2021}. Midpoint agreement alone cannot identify the source's complete extraction procedure.

There is a second, independent input choice: the upper limit in the OPE-convergence ratios. The baseline window selection uses finite-$s_0$ ratios, as defined in Eq.~\eqref{eq:channel-input-criteria}. Using R with infinite upper limits in these convergence ratios instead gives endpoints close to the tabulated Borel windows (Appendix~\ref{app:robustness}). In this latter test only the convergence ratios use the infinite limit. The spectral targets still use continuum subtraction at $s_0$, and the low-energy fraction always compares finite and full integrals.

The paired bounds in Table~\ref{tab:chen-gated} hold the windows, thresholds, bands and weights identical within each F/R pair. All four windows satisfy the fixed-scale input criteria throughout. Under both prescriptions every spectral class is nonempty, every optimal pole mismatch exceeds 1\%, and every candidate band can be empty. The 20\% bands change from optional under F to required under R; the 30\% minima also increase. These running conclusions concern the four tested currents. The running finite-limit input criteria hold throughout the selected $0^{-+}$ window and on parts of the other three. Infinite-limit criteria hold throughout all four. We keep the windows unchanged so that the spectral comparison isolates the derivative prescription.

Restricting these selected windows to intervals satisfying the R input criteria preserves their empty candidate-band examples, with the targets and Borel weight held fixed. Positive lower bounds need not survive this restriction, because fewer moment constraints allow more spectra. The positive bounds in Table~\ref{tab:chen-gated} refer to the entire selected windows.

For the detailed tensor window, R gives minima of $[0.914,0.916]\%$ in the 20\% band and $[54.08,54.09]\%$ in the 30\% band, with a verified empty $\mathcal B_C$. The band-width transition therefore differs from the fixed-scale 25\%--27.5\% interval in Figure~\ref{fig:tensor-resolution}. Comparisons on the full Borel windows of Ref.~\cite{chen2021} and the detailed tensor pole mismatches appear in Table~\ref{tab:running-full} (Appendix~\ref{app:robustness}).

\subsubsection{Continuum threshold}
To test whether the broad contribution survives a different separation between low and high energy, we vary $s_0$ over the ranges in Ref.~\cite{chen2021} while retaining each selected window, band and weight. Figure~\ref{fig:threshold-controls} compares the resulting minima for F and R. The marker filling reports whether the finite-limit input criteria hold throughout the window, on part of it, or nowhere. This status concerns the OPE inputs; the marker height gives the independently verified spectral bound. Under F, the $2^{++}$, $0^{-+}$ and $1^{+-}$ broad-band minima at the largest thresholds remain above 10.6\%, 14.2\% and 11.8\%, respectively. Adding a nonnegative OPE tail embeds every lower-threshold spectrum into the highest-threshold class. These bounds therefore apply throughout the corresponding threshold ranges (Appendix~\ref{app:threshold-embedding}).

For $2^{--}$ at $s_0=40\,\GeV^2$, the F broad-band minimum is attained at zero, whereas the R minimum lies in $[46.27,46.28]\%$. Both use $M_B^2\in[7.19,9.44]\,\GeV^2$, the same band and weight, and satisfy the finite- and infinite-limit criteria throughout the window. The different band requirements therefore persist at a setting that passes both input tests.

On the detailed tensor window, increasing $s_0$ from 32 to $38\,\GeV^2$ reduces the F minimum from about 35.5\% to 8.55\%. The latter is a conservative bound for the whole interval. R gives about 61.4\% and 43.3\% at the endpoints and positive minima at all five tested thresholds. For R, the plotted points provide individual conclusions; the continuous parameter analysis in Section~\ref{sec:correlated-inputs} uses a separate verification. All these values are normalized fractions. The stored spectra include their absolute band and total integrals so that the change in normalization can be examined.

\begin{table*}[t]
\caption{Minimum Borel-weighted band fractions (percent) for joint input variation in $GGG,0^{-+}$. Lower endpoints bound every allowed parameter member and spectrum; upper endpoints round verified feasible values upward. Every zero is attained outside the closed band. Table~\ref{tab:clearance-control} tests finite edge separation for the R joint-range $\mathcal B_{20}$ zero; the other entries impose no prescribed buffer. Near-edge offsets specify band membership rather than physical mass resolution. All rows use $M_B^2\in[5.75,7.35]\,\GeV^2$, $\tau_*=(7.35\,\GeV^2)^{-1}$, 1\% tolerance and the undeformed continuum. F fixes $\mu^2=6.305\,\GeV^2$; R differentiates with $\mu^2=1/\tau$. The bands are $\mathcal B_C=[3.77,4.31]$, $\mathcal B_{20}=[3.304,4.956]$ and $\mathcal B_{30}=[2.891,5.369]\,\GeV$. Central inputs are $G=0.0635\,\GeV^4$, $H=0.5207\,\GeV^6$ and $s_0=33\,\GeV^2$; ranges are specified in the text. Every class is nonempty. Brackets report numerical enclosures of the fractions.}\label{tab:correlated-bands}
\centering
\begin{tabular}{lcccccc}
\toprule
 & \multicolumn{3}{c}{Fixed prescription F} & \multicolumn{3}{c}{Running prescription R}\\
Input set & $\mathcal B_C$ & $\mathcal B_{20}$ & $\mathcal B_{30}$ & $\mathcal B_C$ & $\mathcal B_{20}$ & $\mathcal B_{30}$\\
\midrule
Central $G,H,s_0$ & $0$ & $0$ & $[27.05,27.06]$ & $0$ & $[14.57,14.58]$ & $[61.06,61.07]$ \\
$G,H$ range; $s_0=33$ & $0$ & $0$ & $[26.74,26.75]$ & $0$ & $[12.29,12.30]$ & $[59.97,59.98]$ \\
$G,H,s_0$ ranges & $0$ & $0$ & $[8.30,8.31]$ & $0$ & $0$ & $[50.21,50.22]$ \\
\bottomrule
\end{tabular}
\end{table*}
\relax{}

\subsubsection{Low-mass support}
These controls test whether empty candidate bands depend on allowing contributions near zero mass. Requiring $m\geq1$ or $m\geq1.5\,\GeV$ while leaving the OPE targets unchanged preserves nonempty spectral classes and empty candidate-band alternatives. This holds under both prescriptions on the four selected windows and the detailed tensor window. Table~\ref{tab:support-controls} (Appendix~\ref{app:threshold-embedding}) gives the bounds and support settings.
\relax{}
\subsection{Joint condensate and threshold variation}\label{sec:correlated-inputs}
For the three-gluon $0^{-+}$ current $\widetilde{\eta}_0$, we now vary the condensates and threshold together, keeping the continuum undeformed. We minimize the band fraction over this joint input set at 1\%, then find the smallest total moment discrepancy that permits an empty band. This extends the separate controls to correlated changes in the targets across the window.

We use the density in Eq.~\eqref{eq:new-ope-cut} with $\beta=0$, $G\in[0.060,0.067]\,\GeV^4$, $H=Gr$, $r\in[7.2,9.2]\,\GeV^2$ and $s_0\in[30,36]\,\GeV^2$. The condensate ranges in Eq.~\eqref{eq:common-condensates} and the threshold range from Ref.~\cite{chen2021} define a deterministic parameter set. One common triple $(G,H,s_0)$ generates both moments throughout the window. Both prescriptions retain the coupling law in Eq.~\eqref{eq:coupling-law} with $\Lambda_{\rm QCD}=300\,\mathrm{MeV}$; independent uncertainty in the coupling parameters is outside this joint propagation. F fixes $\mu^2=6.305\,\GeV^2$ and hence the coupling when differentiating. R evaluates the running coupling at $\mu^2=1/\tau$ and takes the total derivative.

Before optimizing spectra, we select $M_B^2\in[5.75,7.35]\,\GeV^2$ and verify target positivity and the finite-limit input criteria throughout the full parameter region for both prescriptions. Every calculation below, including the central-input control, uses this same window, $\tau_*=(7.35\,\GeV^2)^{-1}$ and support $[0,s_0]$. Changes between rows then measure parameter effects at a common resolution.

\begin{figure*}[t]
\centering\includegraphics[width=\linewidth]{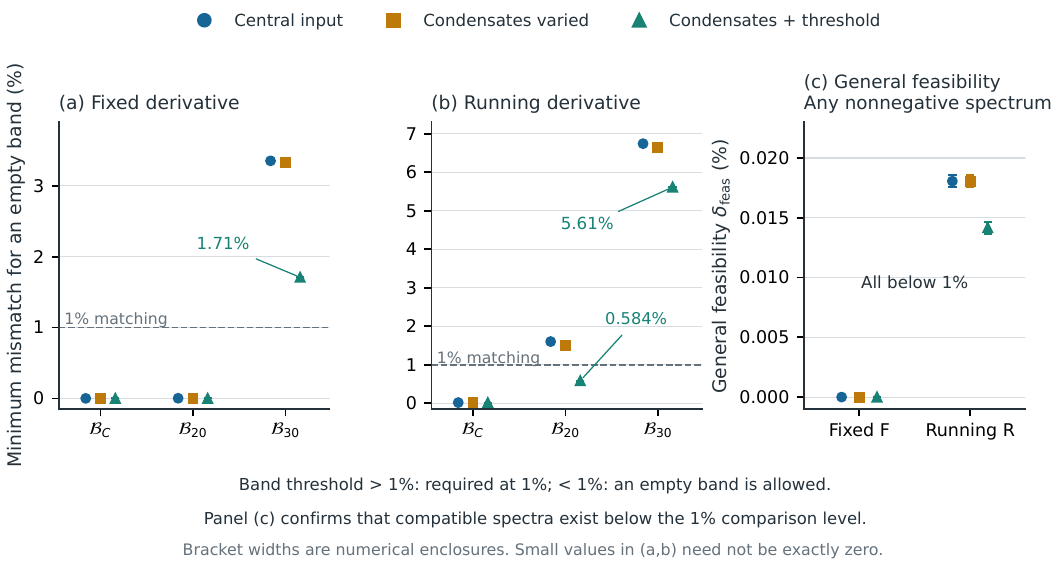}
\caption{Minimum uniform relative discrepancy needed to allow an empty mass band. For $GGG,0^{-+}$, panels (a,b) show the fixed- and running-prescription thresholds $\delta_{\mathcal B}$ for three nested input sets, with the bands and common window of Table~\ref{tab:correlated-bands}. Brackets above the 1\% line preserve the band requirement at that tolerance; brackets below it admit empty-band alternatives. The joint-range 30\% thresholds are about 1.71\% and 5.61\%, while the running 20\% threshold is about 0.584\%. Panel (c), on an enlarged scale, drops the empty-band condition and shows $\delta_{\rm feas}$ for F and R. All general classes are feasible below 1\%. Bracket widths are numerical enclosures, below 0.001 percentage point before outward rounding. F has exact $\delta_{\rm feas}=0$; other near-zero markers need not represent exact zeros. The discrepancy is the maximum relative residual across both moments and the full window, not an extra allowance beyond 1\%; the actual OPE remainder remains to be assessed.}\label{fig:correlated-thresholds}
\end{figure*}

The candidate band $\mathcal B_C$ can be empty in all six settings (Table~\ref{tab:correlated-bands}). Condensate variation alone moderately weakens the wider-band requirements. Adding the threshold range reduces the F 30\%-band minimum from about 27.1\% to 8.30\% and the R minimum from about 61.1\% to 50.2\%. The R 20\%-band requirement, about 14.6\% at central input, vanishes over the joint range. We verify these bounds continuously in $s_0$ for both prescriptions. The positive bounds hold for every compatible spectrum at every input in the family. Each reported zero is attained by a compatible empty-band spectrum at one member; it need not hold at every member. Different entries may use different inputs and spectra.

We next determine the total moment mismatch needed to remove a band requirement. For a parameter set $\mathcal U$, define the empty-band discrepancy threshold by
\begin{equation}\label{eq:correlated-critical}
\begin{aligned}
\delta_{\mathcal B}(\mathcal U)
={}&\inf_{\substack{\theta\in\mathcal U,\,\rho\geq0\\
                   \rho(\mathcal B)=0}}
\ \sup_{\substack{\tau\in I\\ k=0,1}}
\left|\frac{L_k[\rho](\tau)}{T_k(\tau;\theta)}-1\right|.
\end{aligned}
\end{equation}
Here $\theta=(G,H,s_0)$ ranges over the set in the corresponding table row, and $\rho(\mathcal B)=0$ means zero measure on the associated closed interval in $s$. The discrepancy is the maximum relative residual over both moment orders and the full Borel window. Its threshold is defined independently of the matching tolerance $\epsilon$ and compared directly with it; it is not an additional allowance beyond $\epsilon$.

Omitting the empty-band condition defines the general feasibility threshold $\delta_{\rm feas}$. Below it the entire spectral class is empty. For $\epsilon<\delta_{\mathcal B}(\mathcal U)$, no member of $\mathcal U$ admits an empty-band spectrum within that tolerance. Once spectral feasibility is established, the band requirement therefore holds throughout the parameter domain. A certified lower endpoint supports this exclusion; one verified parameter member and empty-band spectrum supplies an upper endpoint. The infimum need not be attained at equality.

Figure~\ref{fig:correlated-thresholds} compares these thresholds with 1\%. In panels (a,b), a bracket above the line excludes an empty band at 1\%, while an upper endpoint below it supplies an empty-band alternative. Panel (c) confirms general spectral compatibility below 1\%, so the positive requirements concern nonempty classes.

Over the full joint range, $\delta_{30}\in[1.709,1.711]\%$ for F and $[5.611,5.613]\%$ for R (Figure~\ref{fig:correlated-thresholds}). The thresholds are about 1.71 and 5.61 times the 1\% tolerance. Under this discrepancy measure, the F requirement is therefore much closer to being removed. For R, $\delta_{20}\in[0.583,0.584]\%$ already admits an empty-band spectrum below 1\%. Its general-feasibility and candidate-band thresholds both lie in $[0.013,0.015]\%$ at the computed resolution. For F, $\delta_{\rm feas}=0$ exactly, and the candidate-band thresholds have upper bounds below 0.001\%. Thus feasible spectra are available well below the tolerance used for the positive band bounds.

Each verified spectrum supplies an absolute residual $R_0(\tau)=L_0(\tau)-T_0(\tau;\theta)$ with $R_1=-\partial_\tau R_0$ across the \mbox{window}. The dimensionless residual in Eq.~\eqref{eq:relative-residual} is $\mathcal R_k=R_k/T_k$. Appendix~\ref{app:correlated-inputs} gives the continuous parameter verification and source audit.

\begin{table*}[t]
\centering
\caption{Finite outward shifts of the running-prescription joint-input $\mathcal B_{20}$ at 1\%. Both expanded edges are included. ``Empty'' has a continuously verified spectrum at one allowed parameter member; ``excluded'' rules out empty-band spectra over the entire joint domain. The general spectral class remains nonempty. For the 10 and 25 MeV expansions, the verified witnesses remain at least 11.95 and 1.46 MeV outside the expanded closed band, respectively.}\label{tab:clearance-control}
\setlength{\tabcolsep}{2pt}\begin{tabular}{ccc}\toprule
Shift per edge & Expanded closed band & Empty-band test\\
(MeV) & (GeV) & \\\midrule
10 & $[3.294,4.966]$ & Empty\\
25 & $[3.279,4.981]$ & Empty\\
50 & $[3.254,5.006]$ & Excluded\\\bottomrule
\end{tabular}
\end{table*}
\relax{}

For the R 20\% band, we also test whether emptiness survives a finite outward shift of its edges. The original witness for an empty $\mathcal B_{20}$ places $62.81\%$ of its weight about $10.1$ eV above the upper edge at $4.956\,\GeV$. We require the enlarged closed band $[3.304-h,4.956+h]\,\GeV$ to be empty, with $h=0.010,0.025,0.050\,\GeV$. All other inputs, including the joint parameter domain and 1\% tolerance, remain unchanged.

The 10 and 25 MeV expansions admit verified empty-band spectra (Table~\ref{tab:clearance-control}). Both use $G=0.067\,\GeV^4$, $H=0.6164\,\GeV^6$ and $s_0=36\,\GeV^2$; their nearest contributions lie at least 21.95 and 26.46 MeV from the original band, respectively. A separate inequality excludes every empty-band spectrum for the 50 MeV expansion throughout the entire joint parameter domain. Thus the \mbox{original} zero survives a finite outward shift, while a slightly broader region becomes required. These tests concern band-edge separation; their atomic spectra impose no minimum smooth support width. Appendix~\ref{app:clearance-control} gives the continuous checks.

The input controls give different limits on the broad-band requirements. The joint condensate and threshold ranges preserve the $\widetilde{\eta}_0$ 30\%-band contribution under both prescriptions at 1\%, whereas the specified continuum deformations can remove the requirements in the detailed examples. The discrepancy thresholds quantify how far the joint-input targets remain from an empty-band spectrum. Section~\ref{sec:scope} discusses how to use these results in hadronic-model tests and input-error assessment.
\relax{}
\section{Implications for spectral interpretation}\label{sec:scope}
A spectral model specifies both where the current couples and how much each mass region contributes. The results above constrain these allocations. We discuss what the five classes permit, how a model can be tested against a band bound, and what input accuracy is needed for a physical interpretation.
\subsection{Band contributions and spectral alternatives}\label{sec:classification-meaning}
Table~\ref{tab:classification-summary} organizes the survey results by what they allow and exclude. The classes distinguish two tests at the same input, rather than ranking currents by physical credibility.

\begin{table*}[t]
\caption{Five diagnostic classes for the fourteen complete-window inputs at 1\% matching, central condensates, fixed scale and the undeformed continuum. A--D have nonempty spectral classes; E admits no nonnegative spectrum. Empty means an attained zero for the indicated band, and required means a positive lower bound for every compatible spectrum. All A--D inputs also admit an empty 20\% band. The pole test covers the full allowed mass domain. A qualifying subwindow satisfies the fixed-scale, finite-$s_0$ input criteria; its existence does not qualify the complete window. The four subwindow calculations in Table~\ref{tab:chen-gated} independently retain positive 30\%-band bounds. Table~\ref{tab:chen-comparison} gives the complete-window values.}\label{tab:classification-summary}
\centering
\setlength{\tabcolsep}{4pt}
\begin{tabular}{cclllcl}
\toprule
Class & Count & \shortstack[l]{Single-pole\\mismatch} & \shortstack[l]{Candidate\\band} & 30\% band & \shortstack{Qualifying\\subwindow} & Currents \\
\midrule
A & 5 & $\epsilon_{\rm pole}\leq1\%$ & Empty & Required & No & $GG:0^{-+},2^{-+}$; $GGG:0^{++},2^{+-},3^{+-}$\\
B & 1 & $\epsilon_{\rm pole}\leq1\%$ & Empty & Empty & No & $GGG:2^{-+}$\\
C & 4 & $\epsilon_{\rm pole}>1\%$ & Empty & Required & Yes & $GGG:2^{++},0^{-+},1^{+-},2^{--}$\\
D & 2 & $\epsilon_{\rm pole}>1\%$ & Empty & Empty & No & $GG:0^{++},2^{++}$\\
E & 2 & $\epsilon_{\rm pole}>1\%$ & --- & --- & No & $GGG:1^{--},3^{--}$\\
\bottomrule
\end{tabular}
\end{table*}
\relax{}

Classes A and C require a 30\%-band contribution while allowing the narrower candidate band to be empty. In A, a single pole can match the input; in C, every single pole is excluded at 1\%. The common result bounds the total contribution in the broad band, without fixing the number of contributing states. A successful pole fit is one possible allocation, and a failed pole fit need not remove the information about a broader region.

Classes B and D also allow the 30\% band to be empty, but still constrain how much weight the candidate region can carry. Their candidate-band upper bounds remain below unity: about 75.9\% for B and 72.0\% and 73.2\% for the two D inputs. D additionally excludes all single poles. Class E has no nonnegative spectrum at 1\%, so no mass interpretation can satisfy these moment constraints without changing the input or tolerance.

The qualifying-subwindow column records the separate input audit. The four C currents retain positive bounds in independent selected-window calculations (Table~\ref{tab:chen-gated}); the five A currents remain complete-window diagnostics. Section~\ref{sec:physical_controls} discusses the further accuracy assessment needed for physical use.

\subsection{Testing hadronic spectral models}\label{sec:physical_strength}
To test a hadronic model, add its Borel-weighted pole residues and continuous contribution in the band, then divide by its own full low-energy Borel integral. The model must use the same current, input, window and weight as the bound. For the detailed $2^{++}$ example $\eta_2$ (Section~\ref{sec:tensor-localization}), at least 22.2\% of that integral must come from $[2.926,5.434]\,\GeV$ at 1\% tolerance. A model assigning only 10\% to this band is therefore excluded. A 25\% assignment passes this necessary test; its two moments must still satisfy the full-window constraints.

This band test applies to multiple poles and continuous low-energy contributions as well as to a single pole. The $\epsilon_{\rm pole}$ test applies specifically to one low-energy pole after subtraction of the common high-energy continuum. Models with additional low-energy contributions must be checked through their full spectral moments.

The current weights each state's contribution by its coupling. If an isolated state is assumed to exhaust the band, Section~\ref{sec:band-fraction} connects the fraction to its squared coupling, mass and total Borel integral. That additional assumption is needed for a decay-constant interpretation. Gluonic currents can couple to mixed configurations~\cite{liu2019review,jiang2023,chen2023review}; a large band fraction measures dominance in this sum rule, rather than gluonic purity of a state.

For $GG,0^{-+}$, the mass interval in Ref.~\cite{chen2021} defines the candidate band $[2.060,2.280]\,\GeV$. Its central value, $2.17\,\GeV$, lies closest to the nominal $X(2370)$ mass among the pseudoscalar estimates considered here. At the central fixed-scale complete-window input, this band can be empty, whereas the 30\% band $[1.519,2.821]\,\GeV$ has a minimum of about 44.2\% of each compatible spectrum's own low-energy Borel integral. The latter region includes the nominal $X(2370)$ mass.

The same input also admits a spectrum empty throughout the 20\% band $[1.736,2.604]\,\GeV$, which includes $X(2370)$ (Table~\ref{tab:chen-comparison} and Appendix~\ref{app:channel-atlas}). Thus the required broad contribution can avoid that entire narrower region. The pole test optimizes over the full support, rather than fixing the mass at either $2.17\,\GeV$ or the $X(2370)$ value; its saved pole lies near $2.603\,\GeV$. These results neither assign the broad contribution to $X(2370)$ nor exclude a physical state there. This row has no qualifying subwindow under the fixed-scale, finite-$s_0$ criteria, so the comparison remains a diagnostic of its specified complete-window input.

\subsection{Input accuracy and QCD uncertainties}\label{sec:physical_controls}\label{sec:phenomenology}
The thresholds give an accuracy target for using the surviving broad-band requirements in a physical analysis. For $\widetilde{\eta}_0$, an empty 30\% band requires a total relative moment discrepancy of at least about 1.71\% under F or 5.61\% under R over the joint parameter range (Figure~\ref{fig:correlated-thresholds}). An error assessment in the same conventions must be compared with these thresholds for both moments over the full window. The smaller F threshold demands tighter control of the remaining input error.

R changes the target derivative and the mass weighting to be reproduced. Its larger threshold measures the distance to an empty-band spectrum for that input; it does not establish that R is more accurate in QCD. F and R do not bracket a complete RG result.

The retained terms indicate the accuracy assessment still needed. Over the same joint domain and common window, the audit encloses $T_0^{(4)}/T_0$ and $T_0^{(6)}/T_0$ within $[2.00,4.28]\%$ and $[3.45,10.00]\%$ for F, and $[1.99,4.39]\%$ and $[3.71,9.97]\%$ for R. These conservative envelopes describe terms already included in the targets, rather than attained extrema or estimates of omitted terms. Passing the input criteria therefore does not itself establish percent-level accuracy. The retained terms cannot simply be added to the 1\% matching tolerance.

This accuracy assessment must include omitted OPE contributions and the high-energy continuum in the same conventions~\cite{leinweber1997,gubler2019,narison2023status,narison2018correlations}. At fixed input parameters, the verified absolute residuals obey $R_1=-\partial_\tau R_0$; relating them to omitted terms requires a calculation for both moments. Direct instantons and topological charge screening in spin-zero two-gluon sum rules~\cite{forkel2005} illustrate the current dependence of such estimates. A complete scale analysis also includes operator evolution and mixing. The present thresholds give quantitative targets against which those estimates can be compared, while the continuum controls separately expose sensitivity to the high-energy model~\cite{lucha2007,lucha2009}. Together, the five classes, the model test and the accuracy thresholds provide a conditional but quantitative companion to a pole-based mass extraction.

\relax{}

\section{Conclusions}
The glueball sum-rule inputs studied here can require a contribution in a broad mass band while allowing a narrower candidate band to be empty. At central fixed-scale inputs, undeformed continua and 1\% matching, twelve of the fourteen inputs admit nonnegative spectra. Each allows an empty candidate band and, separately, an empty 20\% band; nine require a contribution in the 30\% band. These comparisons use the candidate mass intervals of Ref.~\cite{chen2021}, each current's complete Borel window and threshold, and a pole search over the full allowed support.

Among the nine inputs requiring the 30\% band, $GG:0^{-+},2^{-+}$ and $GGG:0^{++},2^{+-},3^{+-}$ admit a single pole (class A in Table~\ref{tab:classification-summary}), whereas $GGG:2^{++},0^{-+},1^{+-},2^{--}$ exclude every single pole (C). The four C currents retain positive bounds in independent calculations on subwindows satisfying the fixed-scale, finite-$s_0$ input criteria; the five A currents have no such subwindow and remain complete-window diagnostics. The 30\% band can be empty for $GGG,2^{-+}$ (B), which admits a pole, and for $GG:0^{++},2^{++}$ (D), which exclude all single poles. For $GGG:1^{--},3^{--}$ (E), the 1\% constraints exclude every nonnegative spectrum, so a mass interpretation requires reassessing those inputs.

At the deformed-continuum setting for $\widetilde{\eta}_0$ ($GGG,0^{-+}$), a pole in $[4.35,4.85]\,\GeV$ and a smooth spectrum empty in that band both match the moments within 1\%, giving the exact fraction range $[0,1]$. In the detailed $\eta_2$ ($GGG,2^{++}$) example with an undeformed continuum, every single pole is excluded at 1\%, yet every compatible spectrum must supply at least 22.2\% of its own low-energy Borel integral in $[2.926,5.434]\,\GeV$. A spectrum compatible throughout the window nearly saturates this bound.

Input variation separates the two diagnostics. Empty candidate-band alternatives survive the tested changes of derivative prescription, low-mass support, continuum shape and threshold, so candidate-band emptiness is not an artifact of one input choice. Broader-band requirements are more fragile: they can weaken or disappear under continuum suppression, change qualitatively between fixed- and running-coupling derivatives, or be removed at the highest tested threshold in some channels. For $\widetilde{\eta}_0$ with an undeformed continuum, however, the 30\% band remains required throughout the joint condensate and threshold ranges. Removing this requirement would need a total relative moment discrepancy of order 1.7\% under F or 5.6\% under R. These thresholds are not physical error estimates, but they specify the input accuracy that a QCD uncertainty assessment must reach before the broad-band requirement can be used or discarded.

The $GG,0^{-+}$ mass result in Ref.~\cite{chen2021} is closest to the nominal $X(2370)$ mass among the pseudoscalar currents considered here. At its central fixed-scale input, the 20\% band contains $X(2370)$ but can be empty; only the 30\% band among the three tested intervals requires a contribution, with a minimum of about 44.2\% of each compatible spectrum's own low-energy Borel integral. This current has no qualifying subwindow. The result constrains a broad contribution at the complete-window input, without determining the existence or gluonic content of $X(2370)$. Turning this conditional spectral diagnostic into a statement about the physical nature of X(2370) would require an assessment of omitted OPE terms and continuum uncertainty in the same conventions, together with lattice glueball spectra, production and decay measurements, and mixing analyses. More theoretical and experimental progress is needed to determine the physical nature of $X(2370)$.

\relax{}

\section*{Data availability}
The input data, candidate spectra, bound coefficients and verification code supporting this work are maintained in the currently private GitHub repository \url{https://github.com/yyywinky580-sketch/glueball-sumrule-bounds}. The manuscript-specific reproducibility package is available from the contact author upon reasonable request. It includes a result-to-evidence index, instructions for independent verification and the individual-current diagnostic figures. Appendix~\ref{sec:verification} describes the verification methods.
\section*{Acknowledgments}
ChatGPT and Codex (OpenAI) assisted with code development, numerical checks, and manuscript drafting and revision. The saved inputs, verification code and calculation records document the resulting computational evidence. The authors take full responsibility for the content and integrity of the work.

\appendix

\section{Input conventions and tensor controls}\label{sec:input}
\subsection{Known-spectrum input construction}\label{app:known-input}

In the mass variable $m=\sqrt{s}$, each packet has probability density $B_h(m-m_c)=35[1-((m-m_c)/h)^2]_+^3/(32h)$, where $h=0.10\,\GeV$ is the half support width and $[x]_+=\max(x,0)$. Each single packet therefore has normalized measure $B_h(m-m_c)\ud m$. The two-packet spectrum uses an equal mixture of these measures for the full Borel-weighted measure $\ud\eta=e^{-as}\ud\rho/L_0(a)$. Setting $L_0^{\rm true}(a)=1$ fixes arbitrary common units.

\begin{samepage}
The common known continuum is
\begin{equation}
\begin{aligned}
 c(s)&=s^4/J,\qquad s>s_0,\\
 J&=\int_{s_0}^{\infty}s^4e^{-as}\,\ud s.
\end{aligned}
\end{equation}
\end{samepage}
Its zeroth Borel moment at $a$ is one. Exact subtraction recovers the targets in Eq.~\eqref{eq:mock-offset}, whose offset is supplied entirely by the prescribed admixture.

\subsubsection{Changing the Borel weight parameter}\label{app:weight-transfer}
For the known-spectrum tests in Section~\ref{sec:known-truth}, the weight parameter $\tau_*$ can change while the moment constraints stay fixed. Keep the input, tolerance, Borel window, support and mass band unchanged, and let $F_t$ denote a candidate's fraction with weight $e^{-ts}$. If every candidate has $F_a\geq\ell$, with $0\leq\ell\leq1$, then
\begin{equation}
 F_{\tau_*}\geq
 \frac{\ell}{\ell+(1-\ell)e^{(\tau_*-a)m_+^2}},
 \qquad a\leq\tau_*\leq b.
 \label{eq:borel-weight-transfer}
\end{equation}
Reweighting multiplies the band integral by at least $e^{-(\tau_*-a)m_+^2}$, while the complementary integral cannot increase. For the two $\epsilon=0.1\%$ cases, the conservative values $\ell=0.7082$ and $0.6904$ give lower bounds above 50.6\% and 48.5\% throughout $I$, without reoptimizing. At $\epsilon=1\%$, both inputs retain the exact range $[0,1]$ for every $\tau_*\in I$, since the verified zero and one fractions follow from support.

\subsection{Tensor input and band-fraction tests}\label{app:tensor-input}
For $GGG,2^{++}$, the retained OPE spectral density of Ref.~\cite{chen2021} is
\begin{equation}\label{eq:tensor-ope}
\begin{aligned}
 p_T(s)={}&\frac{2\alpha_s^3}{315\pi}s^4+\frac{\alpha_s^2G_2}{15}s^2\\
 &+\frac{53\alpha_s^3G_2}{320\pi}s^2+\frac{\alpha_s^2Gr}{3}s,
 \qquad G_2=4\pi G.
\end{aligned}
\end{equation}
The scalar invariant is the positive-parity spin-2 coefficient of the $\eta_2\eta_2^\dagger$ correlator selected in Ref.~\cite{chen2021}, with its normalization fixed by Eq.~\eqref{eq:tensor-ope}. For the current in Eq.~\eqref{eq:tensor-current}, write $A=(\alpha_1\alpha_2,\beta_1\beta_2)$. Its positive-parity matrix element is
\begin{equation}\label{eq:tensor-matrix-element}
\begin{aligned}
\langle0|\eta_2^A|T(p,\lambda)\rangle
 &=i f'_T V^A(p,\lambda),\\
V^A&=\mathcal S\!\left[
 \epsilon^{\alpha_1\beta_1\mu_1\nu_1}
 \epsilon^{\alpha_2\beta_2\mu_2\nu_2}\right.\\
 &\hspace{2.8em}\left.{}\times p_{\nu_1}p_{\nu_2}
 \varepsilon_{\mu_1\mu_2}(p,\lambda)\right].
\end{aligned}
\end{equation}
Here $\varepsilon_{\mu_1\mu_2}$ is the transverse, symmetric, traceless spin-2 polarization tensor and $\mathcal S$ is the same operation as in the current definition. The two momentum factors distinguish $f'_T$ from the effective scalar coupling. The positive-parity projection of the polarization sum therefore has the form
\begin{equation}\label{eq:tensor-residue}
\begin{aligned}
 \ud\rho_{T,\mathrm{pole}}(s)&=A_T\delta(s-m_T^2)\ud s,\\
 A_T&=C_T m_T^4|f'_T|^2\equiv|g_T|^2\geq0.
\end{aligned}
\end{equation}
Here $C_T>0$ is the dimensionless normalization of that projected polarization sum, including the source's symmetrization convention. We define the effective scalar coupling by $g_T=\sqrt{C_T}\,m_T^2 f'_T$, absorbing that fixed normalization rather than assigning $C_T=1$. The source coupling $f'_T$ has units $\GeV^3$; $g_T$ and $A_T$ have units $\GeV^5$ and $\GeV^{10}$. The nonnegative measure includes only this projected component. We use Eq.~\eqref{eq:tensor-ope} directly, with the kinematic factor included in the effective coupling $g_T$. The common normalization cancels in $F_{\mathcal B}$.

We fix $\alpha_s$ from Eq.~\eqref{eq:coupling-law} at $\mu^2=6.905\,\GeV^2$, the midpoint of the Borel window in Ref.~\cite{chen2021}. The full OPE moments use $G=0.0635\,\GeV^4$, $r=8.2\,\GeV^2$ and $s_0=35\,\GeV^2$. The continuum retains all four terms through $p_T(1+\beta b_{18})$, with $\Delta=18\,\GeV^2$.

On the detailed tensor window $[5.42,7.95]\,\GeV^2$, interval checks establish positive $T_0,T_1$, a higher-order dimension-four contribution of at most 5\%, a dimension-six contribution of at most 10\%, and $T_0/O_0\geq40\%$, for all $|\beta|\leq0.75$. The 5\% numerator is the continuum-subtracted third term of Eq.~\eqref{eq:tensor-ope}, proportional to $\alpha_s^3G_2$; it is not the sum of both dimension-four terms. This differs from the pseudoscalar test in Eq.~\eqref{eq:borel-window-tests}.

Besides the broad-band bound and feasible spectrum, the tensor records include a $\beta=-0.75$ empty-band measure and two smooth empty-band spectra at $\beta=0$. The latter satisfy 0.1\% envelopes on $M_B^2\in[5.12,9.00]\,\GeV^2$, hence also 1\% on the common subwindow. All three rows of Table~\ref{tab:tensor-localization} retain $\tau_*=(9\,\GeV^2)^{-1}$.

\subsubsection{Single-pole necessary condition}\label{app:tensor-pole}
Set $r_i=T_1(\tau_i)/T_0(\tau_i)>0$ at the two endpoints of the detailed tensor window, and let $r_{\min}$ and $r_{\max}$ be the smaller and larger of these ratios. The indexed $r_i$ are moment ratios, distinct from the condensate ratio $r$. If a pole satisfies both relative moment envelopes at $0\leq\epsilon<1$, then, with $\zeta=\frac{1+\epsilon}{1-\epsilon}$,
\begin{equation}\label{eq:tensor-pole-condition}
 \frac{r_i}{\zeta}\leq m_p^2\leq\zeta r_i,
 \qquad
 \epsilon\geq\frac{\sqrt{r_{\max}/r_{\min}}-1}
 {\sqrt{r_{\max}/r_{\min}}+1}.
\end{equation}
The second inequality follows by intersecting the two mass intervals. At $\beta=0$, the endpoint values of $\sqrt{r_i}$ are approximately $4.80181995$ and $4.38432599\,\GeV$; interval evaluation gives the conservative bound $\epsilon>4.5448\%$. Requiring the pole to lie below $4.37\,\GeV$ strengthens it to $\epsilon>9.3954\%$ using the first endpoint alone. These endpoint conditions give necessary lower bounds on the pole tolerance; the optimized values use the full-window calculation.

\subsubsection{Endpoint certificate and feasible spectrum}\label{app:tensor-certificate}
The independent tensor check verifies the four nonzero endpoint multipliers against Eq.~\eqref{eq:tensor-endpoint-kernel}, retaining the $10^{-12}$ normalization repair. It checks the kernel on continuous spectral support and the three-atom spectrum over the whole Borel window, using the exact stored coefficients. The exact rational coefficients, normalization repair and verification record are included in the reproducibility package and linked to the original tensor certificate through its evidence index. The stored interval for the data inequality lies in the outward-rounded enclosure $D\in[-5.381867,-5.381865]\times10^{-11}$, whose upper endpoint is strictly negative.

The three-atom witness has exact squared masses $s_1=8.5614759\,\GeV^2$, $s_2=(7931/480)\,\GeV^2$ and \mbox{$s_3=29.5283561\,\GeV^2$}. Its outer atoms lie approximately 17.09 and 9.20 eV outside the lower and upper closed-band edges, respectively. These positions specify membership for the feasible witness, which approaches the minimum; the all-spectrum lower bound follows from the independent kernel inequality. The small magnitude of $D$ records how close the fixed-input data inequality is to equality.

\subsubsection{Boundary, continuum and packet controls}\label{app:tensor-controls}
Table~\ref{tab:tensor-controls} changes either the closed mass band at fixed input or $\beta$ at fixed band. The input checks cover all amplitudes used here through $|\beta|\leq0.75$.

\begin{table}[!htbp]\centering\small
\caption{Tensor broad-band minima at separate fixed settings and 1\% tolerance. Inputs, window and weight are those of Table~\ref{tab:tensor-localization}. The first two rows vary the continuum amplitude at $\Delta=18\,\GeV^2$. The last two move each band edge by 25 MeV at the baseline continuum. Brackets enclose the infimum and are rounded outward; the zero is attained. Here $\beta$ is the peak local fractional deformation; Appendix~\ref{app:tensor-controls} expresses its effect on the integrated moment.}\label{tab:tensor-controls}
\begin{tabular}{ccc}\toprule
$\beta$ & Mass band ($\GeV$) & Minimum (\%)\\\midrule
$-0.25$ & $[2.926,5.434]$ & $[8.4731,8.4760]$\\
$-0.50$ & $[2.926,5.434]$ & $0$\\
$0$ & $[2.951,5.409]$ & $[19.6293,19.6297]$\\
$0$ & $[2.901,5.459]$ & $[24.6194,24.6198]$\\
\bottomrule\end{tabular}
\end{table}

\relax{}

To express the continuum changes in moment units, at each Borel point we divide the absolute change in the continuum's zeroth Borel integral by the full zeroth OPE moment, and then take the maximum over the detailed tensor window. This maximum lies in $[3.58,3.64]\%$ and $[7.16,7.29]\%$ for $\beta=-0.25$ and $-0.50$, respectively. At $\beta=-0.75$ it lies in $[10.7450,10.9290]\%$. These intervals enclose maxima over the window, not pointwise ranges. The denominator is the full OPE moment, whereas the pseudoscalar percentages in Section~\ref{sec:continuum-sensitivity} use the undeformed continuum integral.

At $\beta=-0.75$, we use the $C^2$ mass profile of Appendix~\ref{app:known-input} with full widths 5, 25, 50 and 100 MeV, support above $1\,\GeV$ and an empty closed broad band. The density uses the fixed $\tau_*$ weighting, with normalization fitted jointly to both moments. The 5 and 25 MeV spectra pass continuous verification. Searches at all prescribed refinements remain unresolved for 50 and 100 MeV, leaving existence in those packet classes open.

\subsubsection{Band-width and tolerance scans}\label{app:tensor-resolution}
The scans fix the tensor baseline of Table~\ref{tab:tensor-localization}. One varies the band $[(1-w)M_C,(1+w)M_C]$ from $w=0.10$ to $0.40$ at 1\%; the other fixes $w=0.30$ and varies $\epsilon$ from 0.1\% to 3\%. All bands lie within $[0,\sqrt{s_0}]$. The exact 20\% band is $[3.344,5.016]\,\GeV$, distinct from the rounded band in Table~\ref{tab:tensor-localization}. The complete scan settings are retained with the figure data and verification records.

At each setting, a feasible nonnegative atomic measure bounds the infimum from above and a continuous weak-dual inequality bounds it from below. Zeros are attained by full-window feasible spectra with empty closed bands. The largest bracket width is below 0.0025 percentage point. Enlarging a nested band cannot lower the infimum, whereas relaxing the tolerance cannot raise it. Each tolerance has its own dual coefficients and data inequality, with $\chi>0$.
\relax{}
\relax{}
\section{Derivation of the spectral bounds}\label{app:bounds}
The bounds use nonnegativity of the candidate low-energy measure and the stated moment envelopes. The first derivation bounds a band fraction; the second explains what a moment ratio says about spectral spread.
\subsection{Continuous bounds for nonnegative Borel spectra}\label{sec:positive-bounds}

Normalize the low-energy Borel weight by setting
$\ud\eta=e^{-\tau_*s}\ud\rho/L_0(\tau_*)$ and $\chi=T_0(a)/L_0(\tau_*)>0$.
Then $\eta$ has total weight one and band weight $F_{\mathcal B}$. If $\tau_*<a$, nonnegativity on $[0,s_0]$ gives
\begin{equation*}
 e^{-(a-\tau_*)s_0}L_0(\tau_*)\leq L_0(a)\leq L_0(\tau_*).
\end{equation*}
Combining this with the constraint at $a$ yields
\begin{equation*}
 \frac{e^{-(a-\tau_*)s_0}}{1+\epsilon}\leq\chi\leq\frac{1}{1-\epsilon}.
\end{equation*}
For the detailed tensor input at $\epsilon=1\%$, this gives the outward-rounded enclosure $[0.5923,1.0102]$. The denominator is therefore controlled through positivity even when its weight lies outside the matching window. The factor $T_0(a)$ sets the input normalization independently of the observable weight $\tau_*$; the moment kernels are evaluated within the constrained window. Selecting a subset of Borel points relaxes the constraints, so its bounds remain valid for the full window. Let $i$ label the selected pairs $(k_i,\tau_i)$ of moment order and Borel parameter, and define
\begin{equation}
\begin{aligned}
 H_i(s)&=(s/s_0)^{k_i}e^{-(\tau_i-\tau_*)s},\\
 d_i&=\frac{T_{k_i}(\tau_i)}{s_0^{k_i}T_0(a)}.
\end{aligned}
\end{equation}
Equation~\eqref{eq:positive-class} implies
$(1-\epsilon)d_i\chi\leq\int H_i\,\ud\eta\leq(1+\epsilon)d_i\chi$.

\begin{samepage}
For either sign $\sigma=\pm1$, if real coefficients $p_i,v$ obey
\begin{equation}\label{eq:positive-dual}
\begin{aligned}
v+\sum_i p_iH_i(s)&\geq \sigma\mathbf1_{[m_-^2,m_+^2]}(s)\\
 &\quad(0\leq s\leq s_0),\\
 \sum_i p_id_i+\epsilon\sum_i|p_i|d_i&\leq0,
\end{aligned}
\end{equation}
\par\end{samepage}
integration yields
\begin{equation*}
\begin{aligned}
\sigma F_{\mathcal B}
&\leq v+\chi\left(\sum_i p_id_i+\epsilon\sum_i|p_i|d_i\right)\\
&\leq v.
\end{aligned}
\end{equation*}
The factor $\chi$ remains candidate dependent; its positivity makes the second coefficient inequality sufficient.
Thus $\sigma=1$ gives $F_{\mathcal B}\leq v$, and $\sigma=-1$ gives $F_{\mathcal B}\geq-v$.
This is weak duality: any coefficients $p_i,v$ satisfying the inequalities bound every feasible spectrum, regardless of optimality.

Taylor expansions with enclosed remainders and polynomial coefficient bounds in a Bernstein basis verify the first inequality at every $s$, including both band boundaries. In this basis, the coefficient extrema bound a polynomial throughout its interval. Separate interval subdivision checks candidate spectra throughout the Borel window. Both verifications cover complete intervals.
\relax{}

\subsection{Borel moment ratios and spectral spread}\label{app:moment-ratio}
For a nonzero nonnegative measure on $[0,s_0]$, differentiating $L_k(\tau)=\int s^ke^{-\tau s}\ud\rho$ under the integral gives
\begin{equation}
\begin{aligned}
 -\partial_\tau\log L_0&=\frac{L_1}{L_0}=\langle s\rangle_\tau,\\
 \partial_\tau\frac{L_1}{L_0}&=-\left[\frac{L_2}{L_0}-\left(\frac{L_1}{L_0}\right)^2\right].
\end{aligned}
\end{equation}
The bracket is the variance under $e^{-\tau s}\ud\rho/L_0(\tau)$. Zero variance forces one support point. At finite matching tolerance, however, several spectra may lie within the same moment envelopes. The identities describe candidate nonnegative measures; existence for the OPE targets is checked separately. The ambiguity studied here concerns finite tolerance, whereas an exactly known Laplace transform on an open interval uniquely determines its measure.
\relax{}
\section{Numerical verification}\label{sec:verification}
Floating-point optimization proposes spectra and bounding inequalities; independent interval checks establish their validity between grid points. The checks cover the full Borel window and spectral support, including closed-band edges, and continuous parameter ranges wherever claimed. They use exact stored coefficients, arbitrary-precision Arb arithmetic, and Taylor and Bernstein enclosures. Quadrature cross-checks the implementation at diagnostic points. The reproducibility package supplies the records and replay commands; its verification-details document collects arithmetic precision, search grids, coverage counts and corruption tests.

\subsection{Nonnegative Borel spectra and known-spectrum tests}\label{app:positive-evidence}
Packet moments use exact polynomial integrals with an exponential Taylor remainder. On each Borel subinterval, endpoint log residuals $\log(L_k/T_k)$ and a second-derivative bound enclose the residual throughout. The latter uses the finite-support variance of a nonnegative measure. Subtracted OPE targets require separate interval bounds because they need not be moments of such a measure (Appendix~\ref{app:target-positivity}). All twelve known-spectrum cases have verified attainable endpoints and outer bounds, with the generating spectra checked separately.

\subsection{Continuous parameter profiles and continuum controls}\label{app:continuum-verification}
The common window $M_B^2\in[5.7,7.2]\,\GeV^2$ satisfies the retained-term tests for all three continuum families. Feasible poles give upper endpoints for the minimum pole mismatch; interval inequalities exclude every mass and positive amplitude below the lower endpoints.

For profiles over the input and continuum parameters, write $\zeta_\epsilon=(1+\epsilon)/(1-\epsilon)$
and $b=\sup I$. The two moments at $b$ imply
$s_p\leq\zeta_\epsilon T_1(b)/T_0(b)$. At a selected $t\leq b$, pole feasibility requires
\begin{equation}
 \log\frac{T_1(t)}{T_1(b)}
 -(b-t)\zeta_\epsilon\frac{T_1(b)}{T_0(b)}-\log\zeta_\epsilon\leq0.
\end{equation}
Interval derivative signs over exact subdivisions of the $(G,H,\beta)$
domain, with $H=Gr$, locate the most favorable corner for this exclusion test.
The rectangle $G\in[0.060,0.067]\,\GeV^4$,
$H\in[0.432,0.6164]\,\GeV^6$ contains the declared input trapezoid.
The selected point $t$ is allowed anywhere in the window, including its interior.

The window tests cover all input values and $\beta\in[-0.75,0.75]$
for width $12\,\GeV^2$, and separately $\beta\in[-0.5,0.5]$
for widths 6 and $18\,\GeV^2$.
After multiplying by positive denominators, the Borel-window inequalities are
affine in each of $G,r,\beta$ separately; all parameter corners are checked over
continuous Borel intervals. Widths are discrete controls rather than
a verified continuum of width values.

The verified critical amplitudes are
$0.6802<\beta_{\rm crit}^{(12)}\leq0.6803$
for the primary family and
$0.4076<\beta_{\rm crit}^{(18)}\leq0.4078$
for the wider control. Each critical amplitude is the infimum of $\beta_{\max}$ allowing a 1\% pole fit after the stated parameter optimization. They quantify the onset of a compatible pole within the stated continuum family.

\subsection{Positive targets throughout the continuum families}\label{app:target-positivity}
The three synthetic inputs have $T_0,T_1>0$ because Eq.~\eqref{eq:mock-offset} combines nonzero nonnegative measures at positive mass. For the pseudoscalar three-gluon input, the continuum correction can have either sign; positivity is checked on the full boxes in Eq.~\eqref{eq:continuum-parameter-sets}.

A common subtraction preserves $T_1=-\partial_\tau T_0$, with the algebraic representation
\begin{equation*}
\begin{aligned}
 \rho_{\rm eff}(s)&=p_\theta(s)-\mathbf1_{s>s_0}c(s),\\
 T_k(\tau)&=\int_0^\infty s^k e^{-\tau s}\rho_{\rm eff}(s)\,\ud s.
\end{aligned}
\end{equation*}
Above $s_0$, $\rho_{\rm eff}=-\beta p_\theta b_\Delta$ can be negative, unlike the nonnegative candidate hadronic spectrum. This signed effective density represents the target moments algebraically. Feasibility requires a separate nonnegative measure below $s_0$, to which the identities in Appendix~\ref{app:moment-ratio} apply.

At fixed $\tau$, $T_0$ is affine in each of $G,r,\beta$ separately. Continuous enclosures at all eight corners establish $T_0\geq0.4O_0>0$ and the retained-term tests. Multiplying those tests by positive $T_0$ gives the same separate affine dependence, extending them to the whole box. In particular, the dimension-six test $|T_0^{(6)}|\leq0.1T_0$ is the pair $\pm T_0^{(6)}\leq0.1T_0$; each inequality has this separate affine dependence.

For $T_1$, define the continuum-subtracted monomial integral
\begin{equation}
\begin{aligned}
Q_n(\tau)={}&\int_0^{s_0}s^ne^{-\tau s}\ud s\\
&-\beta\int_{s_0}^{s_0+\Delta}s^nb_\Delta(s)e^{-\tau s}\ud s.
\end{aligned}
\end{equation}
The preceding corner argument for the retained-term tests needs no prior sign assumption on $Q_1$. Since $0\leq b_\Delta\leq1$ and $\beta\leq3/4$, for $n=2,3$,
\begin{equation}\label{eq:target-monomial-positivity}
Q_n(\tau)\geq\frac{n!}{\tau^{n+1}}
\left[1-\frac74 e^{-x}\sum_{j=0}^{n}\frac{x^j}{j!}\right]>0.
\end{equation}
For the continuum families considered here, $s_0=33\,\GeV^2$ and $\tau\geq(7.2\,\GeV^2)^{-1}$, so $x=s_0\tau\geq55/12$. The function $e^{-x}\sum_{j=0}^{n}x^j/j!$ decreases with $x$. At $x_0=55/12$, the fifth-order Taylor sum, itself below $e^{x_0}$, exceeds $(7/4)\sum_{j=0}^{3}x_0^j/j!$ by $66537959/5971968>0$.

Equation~\eqref{eq:new-ope-cut} gives $T_1=c_4Q_5+c_2GQ_3+c_1GrQ_2$, with $c_4,c_2,c_1>0$. Thus $T_1$ increases with $G,r$ and decreases with $\beta$, placing each family's minimum at its lower condensate corner and largest $\beta$. The pole certificates establish positivity of the complete $T_1$, including $c_4Q_5$, throughout $\tau$ at those three corners; no separate assumption on $Q_5$ is used.

Table~\ref{tab:target-positivity} gives the smallest continuously verified enclosure endpoints, rounded down. They are conservative lower bounds, not attained minima; the exact partitions and arithmetic settings are retained with the certificates.

\par\medskip\noindent\begin{minipage}{\columnwidth}
\makeatletter\def\@captype{table}\makeatother
\centering
\renewcommand{\arraystretch}{1.10}
\caption{Uniform positive lower bounds for the three-gluon targets on the full parameter sets $\mathcal U_\Delta$ and $\tau\in[5/36,10/57]\,\GeV^{-2}$. The $T_0$ and $T_1$ columns have units $\GeV^{10}$ and $\GeV^{12}$, respectively.}\label{tab:target-positivity}
\normalsize
\begin{tabular}{rrr}
\hline
$\Delta\,[\GeV^2]$ & $T_0$ lower bound & $T_1$ lower bound\\
\hline
12 & 176.621 & 3468.508\\
6 & 193.053 & 3934.471\\
18 & 181.454 & 3575.109\\
\hline
\end{tabular}
\end{minipage}\par\medskip
\relax{}

\subsection{Broad-gap exclusion and the continuum union}\label{app:gap-verification}
For the fixed $\Delta=18\,\GeV^2$, $\beta=0.50$ example at $G=0.060\,\GeV^4$, $r=7.2\,\GeV^2$, a dual kernel formed from both moment kernels at the saved Borel points is nonnegative over
$[0,4.10^2]\cup[5.10^2,33]$ in $\GeV^2$,
while its data upper bound is strictly negative at $\epsilon=1\%$
(approximately $-0.00191336$ in the saved normalization).
A nonnegative measure on this support would integrate the continuously nonnegative kernel to a nonnegative value, contradicting the data bound. This excludes every broad-gap spectrum on that support.

At 1.5\%, an explicit nonnegative $C^2$ empty-band spectrum is continuously feasible.
At this input, the saved multiplier gives $C_d+\epsilon P_d<0$, where $C_d=\sum_jp_jd_j$ and $P_d=\sum_j|p_j|d_j>0$. The dual sums bound the threshold from below; the verified witness residual bounds it from above. Their bracket need not identify an attained threshold.
The three-gluon gap witnesses use packets of 100 MeV full support width in $m=\sqrt{s}$, restricted to $m\geq1.5\,\GeV$. These are construction choices for the witnesses; the exclusion inequality covers support down to $s=0$.
The $\beta=0$ witness belongs to the declared continuum union and has
exactly zero fraction in both the broad gap and the narrower target.
The primary $\beta=0.70$ pole/gap pair attains zero and one at identical
inputs, establishing the union statements in Section~\ref{sec:gap-union}.

\subsection{A neighborhood of the matched pole and empty-band spectra}\label{app:borel-neighborhood}
Fix both low-energy spectra of Section~\ref{sec:matched-pair}, including the pole residue and packet amplitudes. Let $T_k$ and $T'_k$ denote the targets before and after changing the OPE and continuum parameters. The fixed spectra have continuous certificates giving the common residual bound $e<0.009310303072$ for $k=0,1$ throughout $I$. If $|T'_k/T_k-1|\leq d<1$, then
\begin{equation}\label{eq:borel-neighborhood-error}
 \left|\frac{L_k}{T'_k}-1\right|
 \leq\frac{e+d}{1-d}.
\end{equation}
Parameter-derivative bounds follow from
$\int_0^\infty s^n e^{-\tau s}\ud s\leq n!/a^{n+1}$,
disjoint low-energy and continuum-change supports, and $0\leq b_\Delta\leq1$.
With the positive target bounds in Appendix~\ref{app:target-positivity}, they give~\mbox{$d<0.000463617418$} throughout
\begin{equation}\label{eq:borel-neighborhood-box}
\begin{aligned}
G&\in[0.060,0.060001]\,\GeV^4,\\
r&\in[7.2,7.2001]\,\GeV^2,\\
\beta&\in[0.69995,0.70005],
\end{aligned}
\end{equation}
at fixed $\Delta=12\,\GeV^2$, $\mu^2=6.305\,\GeV^2$ and $s_0=33\,\GeV^2$. Equation~\eqref{eq:borel-neighborhood-error} then bounds both residuals by $0.977846\%$ throughout the Borel window, preserving feasibility and zero/one fractions. This conservative box gives a verified neighborhood within the declared input ranges.
\par
\relax{}
\relax{}
\section{Fourteen-current inputs and checks}\label{app:chen-inputs}
\subsection{Interpolating operators}\label{app:current-definitions}
We use the color tensors $f^{abc}$ (antisymmetric) and $d^{abc}$ (symmetric), and $\alpha_s=g_s^2/(4\pi)$. All fields in a current are evaluated at the same spacetime point. With color indices summed, the field strength and its dual are
\begin{equation}\label{eq:field-definitions}
\begin{aligned}
G^a_{\mu\nu}&=\partial_\mu A^a_\nu-\partial_\nu A^a_\mu
 +g_sf^{abc}A^b_\mu A^c_\nu,\\
\widetilde G^a_{\mu\nu}&=\tfrac12\epsilon_{\mu\nu\rho\sigma}G^{a,\rho\sigma}.
\end{aligned}
\end{equation}
We follow the dual and projection conventions of Ref.~\cite{chen2021}. The operation $\mathcal S$ symmetrizes and subtracts traces in the two Lorentz-index sets $\{\alpha_i\}$ and $\{\beta_i\}$ simultaneously. The four two-gluon currents are
\begin{equation}\label{eq:gg-currents}
\begin{aligned}
J_0&=g_s^2G_a^{\mu\nu}G_{a,\mu\nu},&
\widetilde J_0&=g_s^2G_a^{\mu\nu}\widetilde G_{a,\mu\nu},\\
J_2^{\alpha_1\alpha_2,\beta_1\beta_2}
 &=\mathcal S[g_s^2G_a^{\alpha_1\beta_1}G_a^{\alpha_2\beta_2}],\\
\widetilde J_2^{\alpha_1\alpha_2,\beta_1\beta_2}
 &=\mathcal S[g_s^2G_a^{\alpha_1\beta_1}\widetilde G_a^{\alpha_2\beta_2}].
\end{aligned}
\end{equation}
They supply the $0^{++},0^{-+},2^{++},2^{-+}$ projections, respectively.

For the three-gluon $C=+$ sector, $\widetilde\eta_0$ and $\eta_2$ are defined in Eqs.~\eqref{eq:pseudoscalar-current} and \eqref{eq:tensor-current}. The remaining two currents are
\begin{equation}\label{eq:ggg-plus-currents}
\begin{aligned}
\eta_0&=f^{abc}g_s^3G_a^{\mu\nu}G_{b,\nu\rho}G_c^\rho{}_\mu,\\
\widetilde\eta_2^{\alpha_1\alpha_2,\beta_1\beta_2}
 &=f^{abc}\mathcal S[g_s^3\widetilde G_a^{\alpha_1\beta_1}
 \widetilde G_b^{\alpha_2\mu}\widetilde G_{c,\mu}^{\beta_2}
 -\{\alpha_2\leftrightarrow\beta_2\}].
\end{aligned}
\end{equation}
These provide the $0^{++}$ and $2^{-+}$ projections.
For $C=-$, the six currents are
\begin{equation}\label{eq:ggg-minus-currents}
\begin{aligned}
\xi_1^{\alpha\beta}
 &=d^{abc}g_s^3G_a^{\mu\nu}G_{b,\mu\nu}G_c^{\alpha\beta},\\
\widetilde\xi_1^{\alpha\beta}
 &=d^{abc}g_s^3G_a^{\mu\nu}G_{b,\mu\nu}\widetilde G_c^{\alpha\beta},\\
\xi_2^{\alpha_1\alpha_2,\beta_1\beta_2}
 &=d^{abc}\mathcal S[g_s^3\widetilde G_a^{\alpha_1\beta_1}
 G_b^{\alpha_2\mu}\widetilde G_{c,\mu}^{\beta_2}
 -\{\alpha_2\leftrightarrow\beta_2\}],\\
\widetilde\xi_2^{\alpha_1\alpha_2,\beta_1\beta_2}
 &=d^{abc}\mathcal S[g_s^3G_a^{\alpha_1\beta_1}
 \widetilde G_b^{\alpha_2\mu}G_{c,\mu}^{\beta_2}
 -\{\alpha_2\leftrightarrow\beta_2\}],\\
\xi_3^{\alpha_1\alpha_2\alpha_3,\beta_1\beta_2\beta_3}
 &=d^{abc}\mathcal S[g_s^3G_a^{\alpha_1\beta_1}
 G_b^{\alpha_2\beta_2}G_c^{\alpha_3\beta_3}],\\
\widetilde\xi_3^{\alpha_1\alpha_2\alpha_3,\beta_1\beta_2\beta_3}
 &=d^{abc}\mathcal S[g_s^3\widetilde G_a^{\alpha_1\beta_1}
 \widetilde G_b^{\alpha_2\beta_2}\widetilde G_c^{\alpha_3\beta_3}].
\end{aligned}
\end{equation}
The untilded currents supply $1^{+-},2^{+-},3^{+-}$ and their tilded partners supply $1^{--},2^{--},3^{--}$. Higher-spin currents may couple to both parities; the retained invariant selects the indicated projection. For spin $j$, its matrix element has the form
\begin{equation}\label{eq:general-spin-projection}
\begin{aligned}
\langle0|J^{\alpha_1\cdots\alpha_j,\beta_1\cdots\beta_j}|X\rangle
 &=if_X\,\mathcal S\!\left[
 \prod_{i=1}^{j}\epsilon^{\alpha_i\beta_i\mu_i\nu_i}p_{\nu_i}
 \right]\varepsilon_{\mu_1\cdots\mu_j}.
\end{aligned}
\end{equation}
The two index sets and trace subtraction are essential to this convention. Equation~\eqref{eq:tensor-matrix-element} spells out the selected tensor case. The effective scalar residue includes the projected polarization normalization; no unspecified factor is set to unity.

\subsection{Retained OPE and input criteria}\label{app:ope-coefficients}
Table~\ref{tab:ope-coefficients} gives every retained coefficient of Ref.~\cite{chen2021} in a common notation. For a two-gluon current,
\begin{equation}\label{eq:gg-ope-template}
 p_J(s)=a_J\alpha_s^2s^2+b_J\alpha_s^2G_2,
 \qquad C_J=d_J\pi\alpha_s H .
\end{equation}
For a three-gluon current,
\begin{equation}\label{eq:ggg-ope-template}
\begin{aligned}
 p_J(s)&=\frac{a_J\alpha_s^3}{\pi}s^4
 +b_J\alpha_s^2G_2s^2
 +\frac{c_J\alpha_s^3G_2}{\pi}s^2
 +d_J\alpha_s^2Hs,\\
 C_J&=0 .
\end{aligned}
\end{equation}
Here $C_J$ is the nonintegrated contribution to the zeroth moment, not a term in the cut density. The retained factorized dimension-eight terms vanish in this input calculation; this does not set the full higher-order remainder to zero.

\begin{table*}[t]
\centering
\renewcommand{\arraystretch}{1.4}
\caption{Retained OPE coefficients for the fourteen currents, in Eqs.~\eqref{eq:gg-ope-template} and \eqref{eq:ggg-ope-template}. $G_2=4\pi G$ and $H=Gr$. In the two-gluon block, $d_J$ multiplies the nonintegrated term and $c_J$ is unused (shown as zero); in the three-gluon block, $d_J$ multiplies the dimension-six cut term and $C_J=0$. All signs and coupling powers are retained.}\label{tab:ope-coefficients}
\begin{tabular}{ccrrrr}
\toprule Current & $J^{PC}$ & $a_J$ & $b_J$ & $c_J$ & $d_J$\\\midrule
\multicolumn{6}{l}{Two-gluon currents: Eq.~\eqref{eq:gg-ope-template}}\\
$J_0$ & $0^{++}$ & $32$ & $60$ & $0$ & $24$\\
$J_2$ & $2^{++}$ & $\tfrac{2}{15}$ & $-\tfrac{5}{24}$ & $0$ & $\tfrac{1}{3}$\\
$\widetilde J_0$ & $0^{-+}$ & $32$ & $0$ & $0$ & $-40$\\
$\widetilde J_2$ & $2^{-+}$ & $\tfrac{2}{5}$ & $\tfrac{1}{12}$ & $0$ & $-\tfrac{1}{2}$\\
\midrule
\multicolumn{6}{l}{Three-gluon currents: Eq.~\eqref{eq:ggg-ope-template}}\\
$\eta_0$ & $0^{++}$ & $\tfrac{3}{10}$ & $0$ & $\tfrac{135}{32}$ & $-\tfrac{81}{2}$\\
$\eta_2$ & $2^{++}$ & $\tfrac{2}{315}$ & $\tfrac{1}{15}$ & $\tfrac{53}{320}$ & $\tfrac{1}{3}$\\
$\widetilde\eta_0$ & $0^{-+}$ & $\tfrac{3}{10}$ & $0$ & $\tfrac{135}{32}$ & $\tfrac{27}{2}$\\
$\widetilde\eta_2$ & $2^{-+}$ & $\tfrac{2}{315}$ & $-\tfrac{1}{15}$ & $\tfrac{57}{320}$ & $\tfrac{5}{4}$\\
$\xi_1$ & $1^{+-}$ & $\tfrac{4}{81}$ & $\tfrac{10}{9}$ & $\tfrac{35}{36}$ & $\tfrac{5}{27}$\\
$\xi_2$ & $2^{+-}$ & $\tfrac{1}{324}$ & $-\tfrac{5}{108}$ & $\tfrac{15}{128}$ & $\tfrac{65}{216}$\\
$\xi_3$ & $3^{+-}$ & $\tfrac{5}{2016}$ & $\tfrac{1}{16}$ & $-\tfrac{59}{512}$ & $-\tfrac{1}{2}$\\
$\widetilde\xi_1$ & $1^{--}$ & $\tfrac{4}{81}$ & $-\tfrac{10}{9}$ & $\tfrac{25}{36}$ & $\tfrac{35}{27}$\\
$\widetilde\xi_2$ & $2^{--}$ & $\tfrac{1}{324}$ & $\tfrac{5}{108}$ & $\tfrac{15}{128}$ & $\tfrac{5}{24}$\\
$\widetilde\xi_3$ & $3^{--}$ & $\tfrac{5}{2016}$ & $-\tfrac{1}{16}$ & $-\tfrac{49}{1536}$ & $-\tfrac{11}{432}$\\
\bottomrule
\end{tabular}
\end{table*}
\relax{}

For each current, the target is
\begin{equation}\label{eq:survey-target}
T_{J,k}(\tau)=\int_0^{s_0}s^k p_J(s)e^{-\tau s}\ud s
 +\delta_{k0}C_J-\beta B_{J,k}(\tau),
\end{equation}
where $B_{J,k}=\int_{s_0}^{s_0+\Delta}s^kp_J(s)b_\Delta(s)e^{-\tau s}\ud s$ and $b_\Delta$ is Eq.~\eqref{eq:new-continuum}. Differentiation is at fixed $\mu$. The constant contributes only to $T_{J,0}$. The continuum $p_J(1+\beta b_\Delta)$ is nonnegative above $s_0$ for all tabulated settings. This is verified by expanding $p_J(s_0+t)$ in nonnegative powers of $t$ and checking its coefficients, together with $\beta\geq-1$. Positivity below threshold is checked separately.

The common parameters are $G=0.0635\,\GeV^4$, $r=8.2\,\GeV^2$, $\langle g_s^2GG\rangle=4\pi G$ and $\langle g_s^3G^3\rangle=Gr$. The coupling is evaluated with Eq.~\eqref{eq:coupling-law} at the fixed $\mu^2$ in Table~\ref{tab:chen-inputs}. The mass intervals reported in Ref.~\cite{chen2021} serve as comparison bands. The baseline $T_1$ holds the coupling fixed; Appendix~\ref{app:robustness} derives the running counterpart and compares their midpoint ratios.

At $\beta=0$, Eq.~\eqref{eq:survey-target} gives an exact reference measure whenever $p_J\geq0$ on $[0,s_0]$ and $C_J\geq0$:
\begin{equation}\label{eq:analytic-reference}
 \ud\rho_{\rm ref}(s)=p_J(s)\ud s+C_J\delta_0(\ud s).
\end{equation}
Here $\delta_0$ denotes a unit point measure at zero, representing the retained constant term rather than a physical massless state. At the central inputs, nine cuts are nonnegative, but only seven also have a nonnegative constant: $GG,0^{++}$ and $GGG,2^{++},0^{-+},2^{-+},1^{+-},2^{+-},2^{--}$. The $GG,0^{-+}$ and $GG,2^{-+}$ cuts are nonnegative but their constants are negative, so this construction does not apply. Polynomial coefficient signs or an analytic cubic minimum, evaluated with interval arithmetic, establish these statements. All four fixed-scale rows represented in Table~\ref{tab:chen-gated} have an exact reference measure; three other such currents fail the fixed-scale, finite-$s_0$ input criteria. These exact reference measures establish nonemptiness. Input-criteria status and minimum band fractions require their respective separate checks.

\begin{table*}[t]\centering\small
\caption{Inputs and minimum 30\%-band fractions at three fixed continuum settings and 1\% tolerance. The local OPE templates and coefficients are given in Eqs.~\eqref{eq:gg-ope-template}--\eqref{eq:ggg-ope-template} and Table~\ref{tab:ope-coefficients}. Each bracket in the last three columns encloses one minimum.}\label{tab:chen-inputs}
\setlength{\tabcolsep}{5pt}\renewcommand{\arraystretch}{1.18}
\begin{tabular}{ccrrcccc}\toprule
\multicolumn{5}{c}{Full-window input; dimensional columns in GeV$^2$} & \multicolumn{3}{c}{Minimum broad-band fraction (\%)}\\
\cmidrule(lr){1-5}\cmidrule(lr){6-8}
Current & $J^{PC}$ & $s_0$ & $\mu^2$ & $M_B^2$ window & Suppressed & Baseline & Enhanced\\
 & & & & & $\beta=-0.5$ & $\beta=0$ & $\beta=0.5$\\\midrule
\addlinespace[3pt]\multicolumn{8}{l}{\textit{Two-gluon currents ($GG$)}}\\[1pt]
$J_0$\textsuperscript{$\dagger$} & $0^{++}$ & 9 & 3.945 & $[3.70,4.19]$ & $0$ & $0$ & $0$\\
$J_2$\textsuperscript{$\dagger$} & $2^{++}$ & 10 & 4.295 & $[3.99,4.60]$ & $0$ & $0$ & $0$\\
$\widetilde{J_0}$\textsuperscript{$\dagger$} & $0^{-+}$ & 9 & 3.490 & $[3.28,3.70]$ & $0$ & $[44.19,44.20]$ & \textemdash\\
$\widetilde{J_2}$\textsuperscript{$\dagger$} & $2^{-+}$ & 10 & 3.735 & $[3.27,4.20]$ & $0$ & $[47.38,47.39]$ & \textemdash\\
\addlinespace[3pt]\multicolumn{8}{l}{\textit{Three-gluon currents ($GGG$), $C=+$}}\\[1pt]
$\eta_0$\textsuperscript{$\dagger$} & $0^{++}$ & 33 & 7.430 & $[7.25,7.61]$ & $[78.05,78.06]$ & $[72.72,72.73]$ & $[98.14,98.15]$\\
$\eta_2$ & $2^{++}$ & 35 & 6.905 & $[4.77,9.04]$ & $[17.34,17.35]$ & $[24.03,24.04]$ & \textemdash\\
$\widetilde{\eta_0}$ & $0^{-+}$ & 33 & 6.305 & $[4.48,8.13]$ & $[17.74,17.77]$ & $[30.31,30.33]$ & \textemdash\\
$\widetilde{\eta_2}$\textsuperscript{$\dagger$} & $2^{-+}$ & 35 & 8.315 & $[8.10,8.53]$ & $0$ & $0$ & $[76.57,76.58]$\\
\addlinespace[3pt]\multicolumn{8}{l}{\textit{Three-gluon currents ($GGG$), $C=-$}}\\[1pt]
$\xi_1$ & $1^{+-}$ & 34 & 6.125 & $[3.16,9.09]$ & $[22.83,22.87]$ & $[18.95,18.97]$ & \textemdash\\
$\xi_2$\textsuperscript{$\dagger$} & $2^{+-}$ & 35 & 7.810 & $[7.53,8.09]$ & $0$ & $[13.32,13.33]$ & $[98.35,98.36]$\\
$\xi_3$\textsuperscript{$\dagger$} & $3^{+-}$ & 33 & 8.045 & $[7.69,8.40]$ & $0$ & $[24.92,24.93]$ & \textemdash\\
$\widetilde{\xi_1}$\textsuperscript{$\dagger$} & $1^{--}$ & 34 & 6.290 & $[5.81,6.77]$ & \textemdash & \textemdash & \textemdash\\
$\widetilde{\xi_2}$ & $2^{--}$ & 36 & 7.635 & $[5.80,9.47]$ & $[6.73,6.78]$ & $[23.75,23.76]$ & \textemdash\\
$\widetilde{\xi_3}$\textsuperscript{$\dagger$} & $3^{--}$ & 38 & 6.685 & $[6.15,7.22]$ & \textemdash & \textemdash & \textemdash\\\bottomrule\end{tabular}
\par\smallskip\begin{minipage}{\textwidth}\small\raggedright \textit{Conventions.} $\Delta=18s_0/35$ is rounded to nine significant digits; $\tau_*$ is the inverse upper Borel-window endpoint. Zero is attained by an empty-band spectrum. The two target moments remain positive in all 42 tabulated settings. A dash means that no nonnegative spectrum satisfies the 1\% constraints. $\dagger$ identifies baseline failure of the input criteria as in Table~\ref{tab:chen-comparison}. These criteria are checked separately from spectral feasibility for every continuum setting in the survey.\end{minipage}\end{table*}
\relax{}

For the baseline finite-$s_0$ input criteria, write $T_{J,0}^{\rm hc}$ for the retained higher-coupling-order condensate contribution and $T_{J,0}^{(6)}$ for the dimension-six contribution. Then
\begin{equation}\label{eq:channel-input-criteria}
 \frac{|T_{J,0}^{\rm hc}|}{T_{J,0}}\leq5\%,\qquad
 \frac{|T_{J,0}^{(6)}|}{T_{J,0}}\leq10\%,\qquad
 \frac{T_{J,0}}{O_{J,0}}\geq40\%.
\end{equation}
Here $O_{J,0}$ uses the same retained OPE integrated to infinity, including $C_J$. For $GG$, the first numerator contains the $b_J\alpha_s^2G_2$ term and the second contains $C_J$. For $GGG$, they contain the $c_J\alpha_s^3G_2s^2/\pi$ and $d_J\alpha_s^2Hs$ terms, respectively, after the same continuum subtraction as the target. Terms in each specified numerator are summed before taking the absolute value. In particular, the tensor 5\% numerator contains only its $\alpha_s^3G_2$ contribution, not both dimension-four terms. Equation~\eqref{eq:borel-window-tests} is the pseudoscalar specialization. The low-energy fraction is an input criterion; it does not by itself establish single-pole dominance. Target positivity and spectral existence are tested separately.

Continuous term-by-term enclosures establish target positivity and test the input criteria. Ten complete-window inputs fail at least one fixed-scale, finite-limit criterion at every point; four have qualifying subintervals. Those intervals are rounded inward to 0.01 GeV$^2$ and verified throughout before fitting spectra (Table~\ref{tab:chen-gated}). R reuses these windows; its criteria are classified separately in Appendix~\ref{app:robustness}. The fixed scale remains the original midpoint.

For $GGG,1^{--}$ and $GGG,3^{--}$, the minimum tolerance for a general nonnegative spectrum lies in $[2.59846,2.59868]\%$ and $[6.76815,6.76837]\%$, respectively. Independent weak-dual exclusions and feasible spectra give these outward-rounded brackets; they are not inferred from the single-pole bounds. The twelve nonempty complete-window inputs have separate, continuously verified empty-20\%-band spectra. These are distinct from the selected-window calculations.

The continuum-control width is $18s_0/35$, rounded to nine significant digits in the saved inputs. A band extending beyond $\sqrt{s_0}$ uses its exact intersection in $s$, including an atom at $s_0$. Rounding the mass cap down would incorrectly place that atom outside the band. The certificates and replay preserve the exact intersection.
\relax{}
\section{Input-prescription and support controls}\label{app:robustness}
\subsection{Running-coupling target and verification}
The running prescription retains the printed current normalization and central condensates, sets $\beta=0$, and evaluates the same continuum subtraction with $\alpha_s(1/\tau)$. Define $\ell=-\log(0.09\,\GeV^2\tau)$ and $D=-\partial_\tau$. For a term proportional to $\alpha_s^n$, the derivatives used in the verification are
\begin{equation}\label{eq:running-jet}
\begin{aligned}
D\alpha_s^n&=-\frac{n\alpha_s^n}{\tau\ell},\\
D^2\alpha_s^n&=\frac{n\alpha_s^n}{\tau^2}
 \left[\frac{n+1}{\ell^2}-\frac1\ell\right],\\
D^3\alpha_s^n&=-\frac{n\alpha_s^n}{\tau^3}
 \left[\frac2\ell-\frac{3(n+1)}{\ell^2}
       +\frac{(n+1)(n+2)}{\ell^3}\right].
\end{aligned}
\end{equation}
Let $I_p(\tau,S)=\int_0^S s^pe^{-\tau s}\ud s$, where $S$ is a generic integration limit. The low-energy targets use $S=s_0$. Each target term is differentiated by
\begin{equation}\label{eq:running-leibniz}
D^k(\alpha_s^n I_p)
=\sum_{j=0}^k \binom{k}{j}(D^j\alpha_s^n)I_{p+k-j}.
\end{equation}
The same derivatives act on any coupling-dependent nonintegrated term. Only $k=0,1$ are constrained; $k=2,3$ enter the continuous residual bounds. Interval enclosures combine these coefficient derivatives with monotone endpoint enclosures of $I_p$.

The candidate spectrum remains one nonnegative measure, with the same Borel kernels throughout each window. The exact reference construction of Eq.~\eqref{eq:analytic-reference} applies to the fixed-scale baseline. For R, we first verify target positivity and then establish feasibility or exclusion with a separate spectrum or inequality.

\subsection{Convergence conventions}
For each prescription, the higher-order coupling and dimension-six ratios are evaluated with both the finite $s_0$ and infinite integration limits. The numerator and denominator of each convergence ratio use the same limit, whereas the low-energy fraction remains the finite-to-full zeroth-moment ratio. The tensor higher-order numerator is the specified $\alpha_s^3\langle g_s^2GG\rangle$ term alone; the other currents retain their own printed term combinations. The limits remain 5\%, 10\% and 40\%, respectively.

\begin{table*}[t]\centering\small
\caption{Midpoint mass diagnostics and continuous input-criteria classifications on the fourteen Borel windows printed in Ref.~\cite{chen2021}. Mass columns are in GeV. F/R denote fixed/running coupling; the final columns specify the integration limit in the convergence ratios.}\label{tab:prescription-diagnostics}
\setlength{\tabcolsep}{5pt}\renewcommand{\arraystretch}{1.16}
\begin{tabular}{cccccccc}\toprule
Channel & Source $M_C$ & $m_{\rm mid}^{\rm F}$ & $m_{\rm mid}^{\rm R}$ & F, $s_0$ & F, $\infty$ & R, $s_0$ & R, $\infty$\\\midrule
$GG,\,0^{++}$ & $1.78$ & $2.273$ & $1.776$ & N & A & N & A\\
$GG,\,2^{++}$ & $1.86$ & $2.339$ & $1.859$ & N & P & N & P\\
$GG,\,0^{-+}$ & $2.17$ & $2.608$ & $2.170$ & N & P & N & P\\
$GG,\,2^{-+}$ & $2.24$ & $2.675$ & $2.237$ & N & P & N & P\\
$GGG,\,0^{++}$ & $4.46$ & $5.030$ & $4.461$ & N & P & N & P\\
$GGG,\,2^{++}$ & $4.18$ & $4.676$ & $4.181$ & P & P & P & A\\
$GGG,\,0^{-+}$ & $4.13$ & $4.631$ & $4.134$ & P & P & P & P\\
$GGG,\,2^{-+}$ & $4.29$ & $4.896$ & $4.289$ & N & P & N & A\\
$GGG,\,1^{+-}$ & $4.01$ & $4.462$ & $4.014$ & P & P & P & A\\
$GGG,\,2^{+-}$ & $4.42$ & $5.034$ & $4.424$ & N & P & N & P\\
$GGG,\,3^{+-}$ & $4.30$ & $4.832$ & $4.303$ & N & P & N & P\\
$GGG,\,1^{--}$ & $4.91$ & $5.526$ & $4.907$ & N & P & N & P\\
$GGG,\,2^{--}$ & $4.25$ & $4.767$ & $4.249$ & P & P & P & P\\
$GGG,\,3^{--}$ & $5.59$ & $6.223$ & $5.594$ & N & P & N & P\\
\bottomrule\end{tabular}
\par\smallskip\begin{minipage}{\textwidth}\small\raggedright A: every point satisfies all criteria. P: a nonzero accepted interval is certified, but the whole printed window is not accepted. N: every point violates at least one criterion. U: no conclusion established. Narrow unresolved boundary intervals are retained in the certificates. The low-energy fraction always uses the finite-to-infinite ratio. The midpoint ratios do not optimize a common pole over a window.\end{minipage}\end{table*}
\relax{}
\begin{table*}[t]\centering\small
\caption{Prescription comparison on the full Borel windows specified in Ref.~\cite{chen2021} and the detailed tensor window. Central inputs, the baseline continuum and 1\% tolerance are fixed. All classes are nonempty.}\label{tab:running-full}
\setlength{\tabcolsep}{5pt}\renewcommand{\arraystretch}{1.16}
\begin{tabular}{ccccrccc}\toprule
Channel & $M_B^2$ window & $1/\tau_*$ & Prescription & $\epsilon_{\rm pole}$ (\%) & $\inf F_C$ (\%) & $\inf F_{20}$ (\%) & $\inf F_{30}$ (\%)\\\midrule
$GGG,\,2^{++}$ & $[4.77,9.04]$ & $9.04$ & F & $[10.932,10.933]$ & $0$ & $0$ & $[24.03,24.04]$\\
$GGG,\,2^{++}$ & $[4.77,9.04]$ & $9.04$ & R & $[9.096,9.097]$ & $0$ & $[34.75,34.76]$ & $[68.74,68.75]$\\
$GGG,\,0^{-+}$ & $[4.48,8.13]$ & $8.13$ & F & $[9.197,9.199]$ & $0$ & $0$ & $[30.31,30.33]$\\
$GGG,\,0^{-+}$ & $[4.48,8.13]$ & $8.13$ & R & $[7.528,7.529]$ & $0$ & $[43.17,43.19]$ & $[72.40,72.41]$\\
$GGG,\,1^{+-}$ & $[3.16,9.09]$ & $9.09$ & F & $[27.150,27.151]$ & $0$ & $0$ & $[18.95,18.97]$\\
$GGG,\,1^{+-}$ & $[3.16,9.09]$ & $9.09$ & R & $[22.937,22.938]$ & $0$ & $[28.08,28.10]$ & $[66.63,66.66]$\\
$GGG,\,2^{--}$ & $[5.80,9.47]$ & $9.47$ & F & $[7.590,7.591]$ & $0$ & $0$ & $[23.75,23.76]$\\
$GGG,\,2^{--}$ & $[5.80,9.47]$ & $9.47$ & R & $[5.800,5.801]$ & $0$ & $[17.28,17.29]$ & $[62.15,62.17]$\\
$GGG,\,2^{++}$ (detail) & $[5.42,7.95]$ & $9.00$ & F & $[6.709,6.710]$ & $0$ & $0$ & $[22.20,22.21]$\\
$GGG,\,2^{++}$ (detail) & $[5.42,7.95]$ & $9.00$ & R & $[5.474,5.475]$ & $0$ & $[0.91,0.92]$ & $[54.08,54.09]$\\
\bottomrule\end{tabular}
\par\smallskip\begin{minipage}{\textwidth}\small\raggedright The dimensional columns are in GeV$^2$. Each fraction entry brackets the minimum. The detailed tensor weight is outside its constrained window and specifies only the observable; no OPE constraint is imposed there. Fixed full-window rows retain their archived survey certificates. All displayed decimal endpoints are rounded outward.\end{minipage}\end{table*}
\relax{}

The baseline criteria in Eq.~\eqref{eq:channel-input-criteria} use finite limits. The infinite-limit running calculation gives endpoints close to those in Ref.~\cite{chen2021}, with deviations consistent with their printed precision. Table~\ref{tab:prescription-diagnostics} classifies the printed windows continuously without shifting their endpoints; partial satisfaction remains marked as such.

\subsection{Threshold embedding and restricted support}\label{app:threshold-embedding}
For the four central fixed-scale inputs, $p_J\geq0$, $C_J\geq0$, and the upper edge of $\mathcal B_{30}$ lies below the smallest threshold in the scan. Fix the window, weight, coupling, condensates, band and tolerance, and write $S\leq H$ for lower and higher values of $s_0$ in this argument. Any feasible measure at $S$ extends to
\begin{equation}\label{eq:threshold-embedding}
 \ud\rho_H=\ud\rho_S+\mathbf1_{(S,H]}p_J(s)\ud s.
\end{equation}
The target and candidate moments acquire the same nonnegative integral, so their absolute residuals stay fixed while the relative-error allowances increase. The added measure lies outside the band: its numerator is unchanged and its denominator increases. Therefore $\inf_{\mathcal A_H}F_{30}\leq\inf_{\mathcal A_S}F_{30}$. A certified lower bound at the largest threshold is uniform over the whole stated threshold interval. The positive reference density establishes nonemptiness at every threshold. This argument concerns the fixed input models; the input criteria are checked separately and need not hold over the entire threshold range. The four-current running-threshold results instead use separate certificates at each tested point.

Raising the candidate support to $[m_{\min}^2,s_0]$ leaves the OPE target unchanged and selects a subset of the original measures. Existing lower bounds remain valid on this subset, but a nonempty subset and an attainable zero require new witnesses. The cuts at 1 and 1.5 GeV define the chosen support controls. The control calculation applies the support restriction to both the primal witnesses and the continuous dual inequalities.

\begin{table*}[t]\centering\small
\caption{Removing low-mass support at fixed input. Each entry brackets a minimum Borel-weighted band fraction in percent at 1\% tolerance. The four currents use the selected windows of Table~\ref{tab:chen-gated}; the last pair uses the detailed tensor window.}\label{tab:support-controls}
\setlength{\tabcolsep}{5pt}\renewcommand{\arraystretch}{1.16}
\begin{tabular}{cccccc}\toprule
Channel & Prescription & $m_{\min}=1$: $\inf F_C$ & $m_{\min}=1$: $\inf F_{20}$ & $m_{\min}=1.5$: $\inf F_C$ & $m_{\min}=1.5$: $\inf F_{20}$\\\midrule
$GGG,\,2^{++}$ & F & $0$ & $0$ & $0$ & $0$\\
$GGG,\,2^{++}$ & R & $0$ & $[30.04,30.06]$ & $0$ & $[30.83,30.86]$\\
$GGG,\,0^{-+}$ & F & $0$ & $0$ & $0$ & $0$\\
$GGG,\,0^{-+}$ & R & $0$ & $[30.91,30.93]$ & $0$ & $[30.91,30.93]$\\
$GGG,\,1^{+-}$ & F & $0$ & $0$ & $0$ & $0$\\
$GGG,\,1^{+-}$ & R & $0$ & $[26.74,26.80]$ & $0$ & $[28.95,28.99]$\\
$GGG,\,2^{--}$ & F & $0$ & $0$ & $0$ & $0$\\
$GGG,\,2^{--}$ & R & $0$ & $[13.80,13.81]$ & $0$ & $[13.80,13.81]$\\
$GGG,\,2^{++}$ (detail) & F & $0$ & $0$ & $0$ & $0$\\
$GGG,\,2^{++}$ (detail) & R & $0$ & $[0.91,0.92]$ & $0$ & $[0.91,0.92]$\\
\bottomrule\end{tabular}
\par\smallskip\begin{minipage}{\textwidth}\small\raggedright Support cuts are in GeV. F/R have the meaning of Table~\ref{tab:chen-gated}. The OPE targets, windows, thresholds and weights are unchanged when the candidate support is raised to $[m_{\min}^2,s_0]$. Every restricted class has a verified feasible spectrum; every zero is attained. These cuts are sensitivity choices, not universal hadronic thresholds.\end{minipage}\end{table*}
\relax{}

Brackets retain their verified widths, including those that miss the precision target. Stored absolute band and total Borel integrals describe individual feasible spectra, not separately optimized extrema.
\relax{}
\section{Correlated-input propagation and compatibility thresholds}\label{app:correlated-inputs}
\subsection{Affine parameters and the common window}
For the pseudoscalar density in Eq.~\eqref{eq:new-ope-cut}, write $T_k=P_k+GU_k+HV_k$, with $P_0=3\alpha_s^3I_4/(10\pi)$, $U_0=135\alpha_s^3I_2/8$ and $V_0=27\alpha_s^2I_1/2$, where $I_j=\int_0^{s_0}s^j e^{-\tau s}\ud s$. Fixed or running derivatives preserve the affine dependence through order three. This current has no nonintegrated term. The condensate polygon is $0.060\leq G\leq0.067$, $7.2G\leq H\leq9.2G$ in the corresponding GeV units. Its four vertices suffice to check affine positivity and linear-fractional input criteria with positive denominators.

Before spectral optimization, interval checks cover $I\times[30,36]\,\GeV^2$ with an exact rectangular partition. At every condensate vertex they establish positive $T_0,T_1$, dimension-four and dimension-six fractions below 5\% and 10\%, and a finite/full zeroth-moment ratio above 40\%. The saved partition specifies every rectangle and its bounds.

Choose $T_{\rm ref}>0$ at the central input and $\tau_*$ for each prescription. With the normalized measure $\eta$ of Appendix~\ref{sec:positive-bounds}, set $\chi=T_{\rm ref}/L_0(\tau_*)$, $g=\chi G$ and $h=\chi H$. Every constraint uses the same combination $\chi P_k+gU_k+hV_k$, with $0.060\chi\leq g\leq0.067\chi$ and $7.2g\leq h\leq9.2g$. The implementation uses equivalent shared nonnegative weights on the polygon vertices, summing to $\chi$.

\subsection{Uniform bounds and feasible spectra}
The kernels use the fixed normalization $S_{\rm n}=36\,\GeV^2$ in $(s/S_{\rm n})^k$. For a threshold interval $[S_-,S_+]$, we verify the kernel inequality on $[0,S_+]$ and its data inequality at every condensate vertex throughout $s_0\in[S_-,S_+]$. Interval bounds on the $s_0$ derivative control variation between endpoints. Exact interval coverage gives a bound for the union. For F, all three bands lie below the smallest threshold, so the nonnegative-tail embedding in Appendix~\ref{app:robustness} extends the upper-threshold certificate to the full range. It preserves empty bands and reduces relative residuals, hence also covers compatibility thresholds.

Verification uses the proposed spectrum's exact parameter triple, checking both moments over the full window, closed-band membership, allowed support and normalization.

A threshold lower bound follows from a nonnegative dual kernel on the allowed support and a strictly negative data upper bound. A full-window feasible spectrum provides the upper bound. Before outward rounding for display, the verified brackets enclose every fraction minimum within 0.01 percentage point and every discrepancy threshold within 0.001 percentage point; bisection locates the latter. Each positive 1\% result has a separately verified nonempty class.

\subsection{Independent checks and input sources}
Independent high-precision differentiation of incomplete-gamma expressions checks both prescriptions through third order. Collapsing the condensate polygon reproduces the fixed-input certificates after the kernel-normalization change; nested input sets obey the expected ordering within certified brackets. The verification-details document locates these cross-checks and the recorded rejection tests.

Ref.~\cite{chen2021} obtains vanishing selected factorized dimension-eight terms but includes only one of a larger set of higher-coupling-order diagrams. The remaining terms require assessment. Condensates fitted to a truncated expansion can absorb higher-dimensional contributions~\cite{narison2018correlations}; additional independent errors or coefficients from other currents require matching to avoid double counting or convention changes. The sourced input ranges and computed thresholds thus leave a physical remainder to be assessed, including uncalibrated continuum-shape uncertainty.

\subsection{Finite separation from the joint-input band edges}\label{app:clearance-control}
The three outward-expansion tests retain the R joint-input domain, window, weight and tolerance of Section~\ref{sec:correlated-inputs}. Requiring zero measure in $[3.304-h,4.956+h]\,\GeV$ treats both expanded edges as part of the band. The 10 and 25 MeV witnesses use the upper condensate corner and $s_0=36\,\GeV^2$. Their full-window relative residuals are below $0.999995\%$ and $0.999997\%$, respectively, and their band integrals vanish exactly.

At 50 MeV, nine threshold subintervals cover $[30,36]\,\GeV^2$ without gaps. On each subinterval a kernel inequality is nonnegative outside the expanded band while its data inequality is strictly excluding at every condensate-polygon vertex. The affine parameter dependence extends it across the polygon. This excludes empty-band spectra for the whole family, not only at the witness parameters of the smaller expansions. The original feasible spectrum confirms that the unrestricted spectral class remains nonempty. The result requires a contribution in the expanded band without determining a new minimum fraction there.

The 50 MeV exclusion follows from these continuous inequalities, independently of the unsuccessful spectrum searches retained in the evidence. None of the three tests imposes a minimum packet width.
\relax{}
\clearpage
\onecolumngrid
\noindent\begin{minipage}{\linewidth}
\section{Diagnostic plots for each current}\label{app:channel-atlas}
\makeatletter\def\@captype{figure}\makeatother
\centering\includegraphics[width=\linewidth]{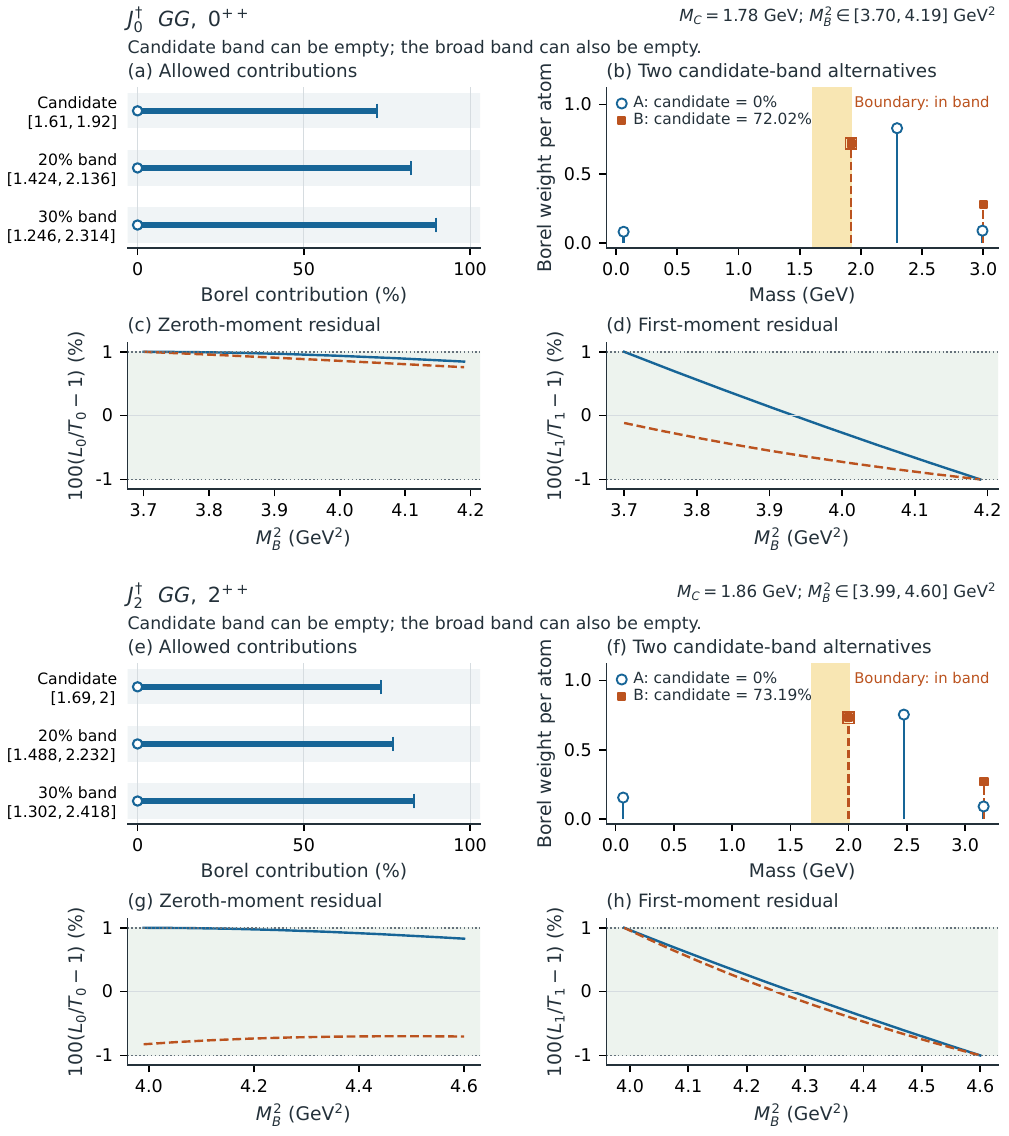}
\caption{Allowed contributions and explicit alternatives for $GG,0^{++}$ and $GG,2^{++}$. (a,e) Fraction ranges: open left endpoints are attained zeros; filled endpoints require a positive contribution. (b,f) A empties the gold candidate band; B nearly maximizes its fraction. Heights are normalized Borel weights. (c,d,g,h) The corresponding moment residuals stay within 1\% over each complete window. Across the atlas, A/B target the candidate band only; wider-band endpoints can use other spectra. Current-specific inputs and dagger status are in Tables~\ref{tab:chen-comparison} and~\ref{tab:chen-inputs}.}\label{fig:atlas-1}
\end{minipage}
\clearpage
\begin{figure}[!ht]\centering\includegraphics[width=\linewidth]{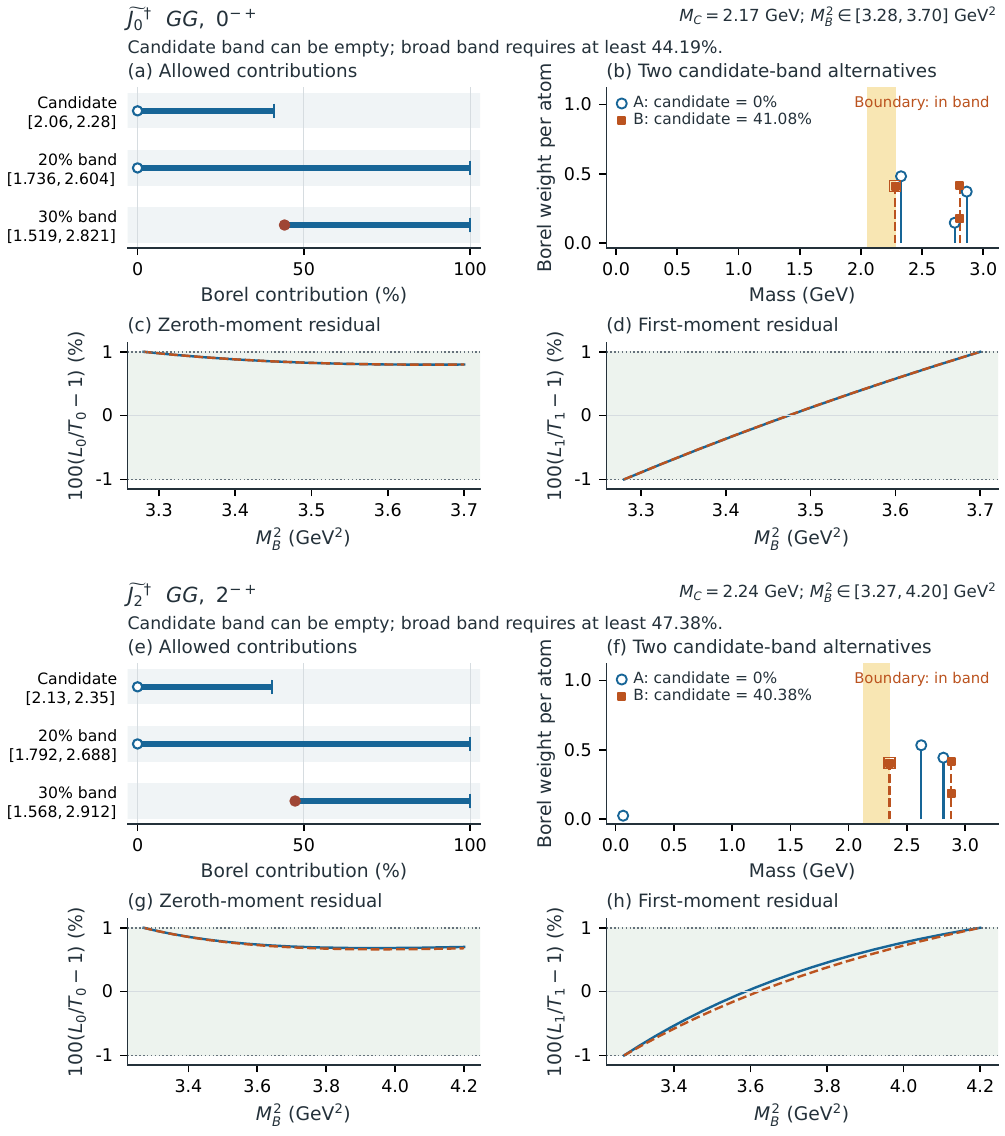}
\caption{Allowed contributions and explicit alternatives for $GG,0^{-+}$ and $GG,2^{-+}$. (a,e) Fraction ranges: open left endpoints are attained zeros; filled endpoints require a positive contribution. (b,f) A empties the gold candidate band; B nearly maximizes its fraction. Heights are normalized Borel weights. (c,d,g,h) The corresponding moment residuals stay within 1\% over each complete window. Colors identify the same spectra throughout. Current-specific inputs and dagger status are in Tables~\ref{tab:chen-comparison} and~\ref{tab:chen-inputs}.}\label{fig:atlas-2}\end{figure}
\clearpage
\begin{figure}[!ht]\centering\includegraphics[width=\linewidth]{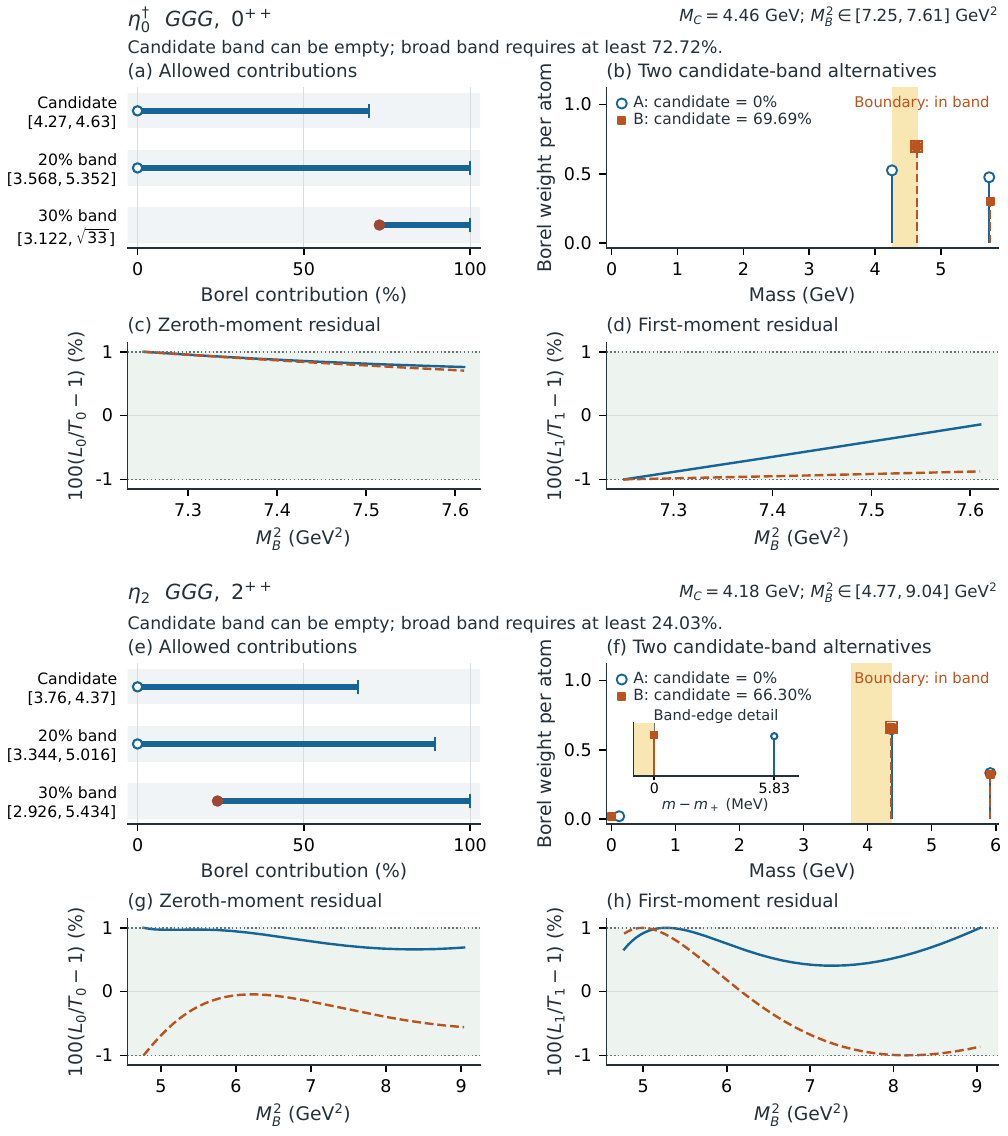}
\caption{Allowed contributions and explicit alternatives for $GGG,0^{++}$ and $GGG,2^{++}$. (a,e) Fraction ranges: open left endpoints are attained zeros; filled endpoints require a positive contribution. (b,f) A empties the gold candidate band; B nearly maximizes its fraction. Heights are normalized Borel weights. (c,d,g,h) The corresponding moment residuals stay within 1\% over each complete window. Colors identify the same spectra throughout. Current-specific inputs and dagger status are in Tables~\ref{tab:chen-comparison} and~\ref{tab:chen-inputs}. The tensor A spectrum empties its candidate band; B contributes about 66.30\% there. Both satisfy 1\% matching. The inset in (f) resolves the upper-edge atom: A is about 5.83 MeV outside, while B is on the included boundary.}\label{fig:atlas-3}\end{figure}
\clearpage
\begin{figure}[!ht]\centering\includegraphics[width=\linewidth]{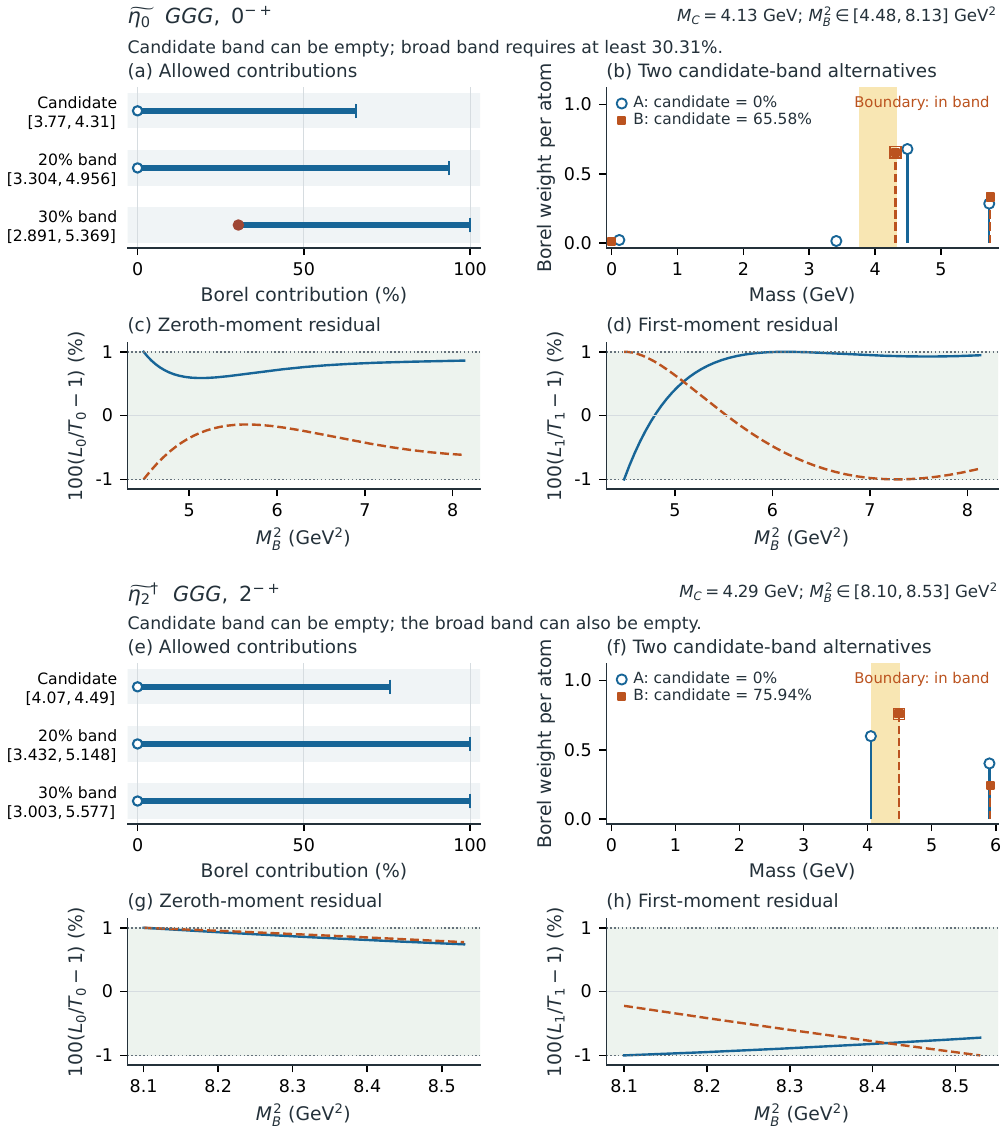}
\caption{Allowed contributions and explicit alternatives for $GGG,0^{-+}$ and $GGG,2^{-+}$. (a,e) Fraction ranges: open left endpoints are attained zeros; filled endpoints require a positive contribution. (b,f) A empties the gold candidate band; B nearly maximizes its fraction. Heights are normalized Borel weights. (c,d,g,h) The corresponding moment residuals stay within 1\% over each complete window. Colors identify the same spectra throughout. Current-specific inputs and dagger status are in Tables~\ref{tab:chen-comparison} and~\ref{tab:chen-inputs}.}\label{fig:atlas-4}\end{figure}
\clearpage
\begin{figure}[!ht]\centering\includegraphics[width=\linewidth]{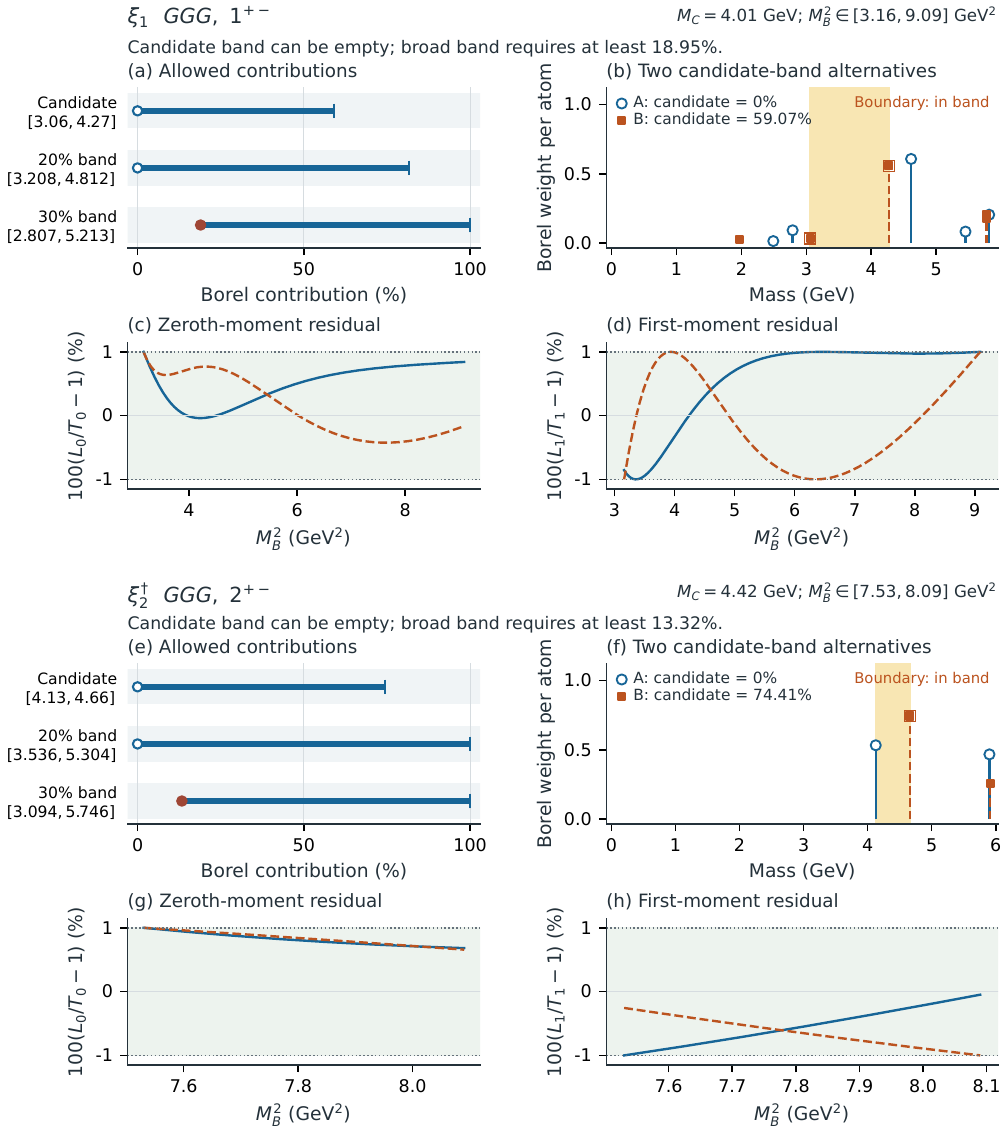}
\caption{Allowed contributions and explicit alternatives for $GGG,1^{+-}$ and $GGG,2^{+-}$. (a,e) Fraction ranges: open left endpoints are attained zeros; filled endpoints require a positive contribution. (b,f) A empties the gold candidate band; B nearly maximizes its fraction. Heights are normalized Borel weights. (c,d,g,h) The corresponding moment residuals stay within 1\% over each complete window. Colors identify the same spectra throughout. Current-specific inputs and dagger status are in Tables~\ref{tab:chen-comparison} and~\ref{tab:chen-inputs}.}\label{fig:atlas-5}\end{figure}
\clearpage
\begin{figure}[!ht]\centering\includegraphics[width=\linewidth]{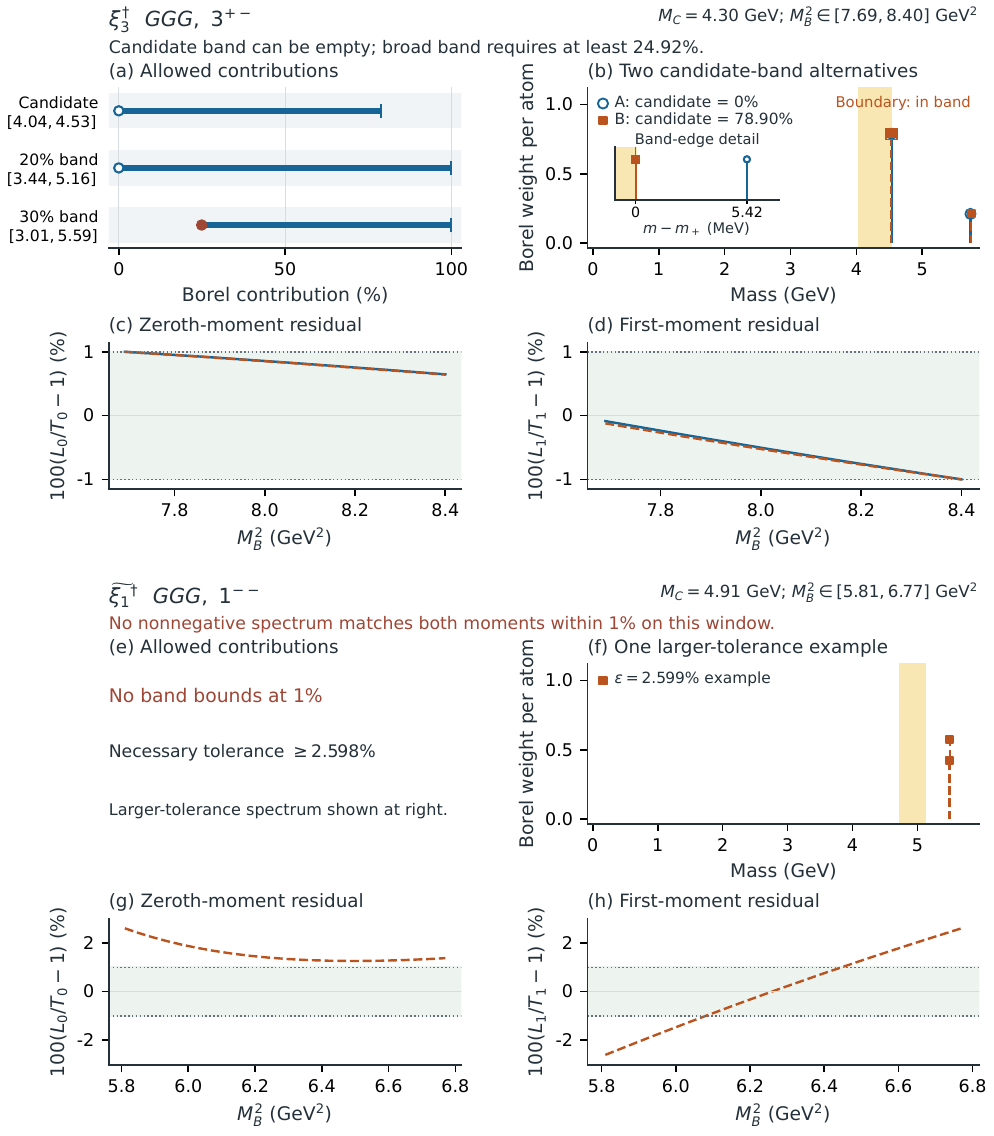}
\caption{Contrasting compatibility for $GGG,3^{+-}$ and $GGG,1^{--}$. Upper row (a--d): A empties the candidate band, B nearly maximizes its fraction, and both satisfy 1\%; the left panel also bounds the wider bands. Lower row (e--h): no nonnegative spectrum is compatible at 1\%, so no band bounds are assigned. The single spectrum in (f) satisfies 2.599\% tolerance; (g,h) show its residuals. It is not a band extremum. Gold selects the closed candidate band; green marks $\pm1\%$. Inputs and dagger: Tables~\ref{tab:chen-comparison} and~\ref{tab:chen-inputs}. The inset resolves A outside the band and B on its included edge.}\label{fig:atlas-6}\end{figure}
\clearpage
\begin{figure}[!ht]\centering\includegraphics[width=\linewidth]{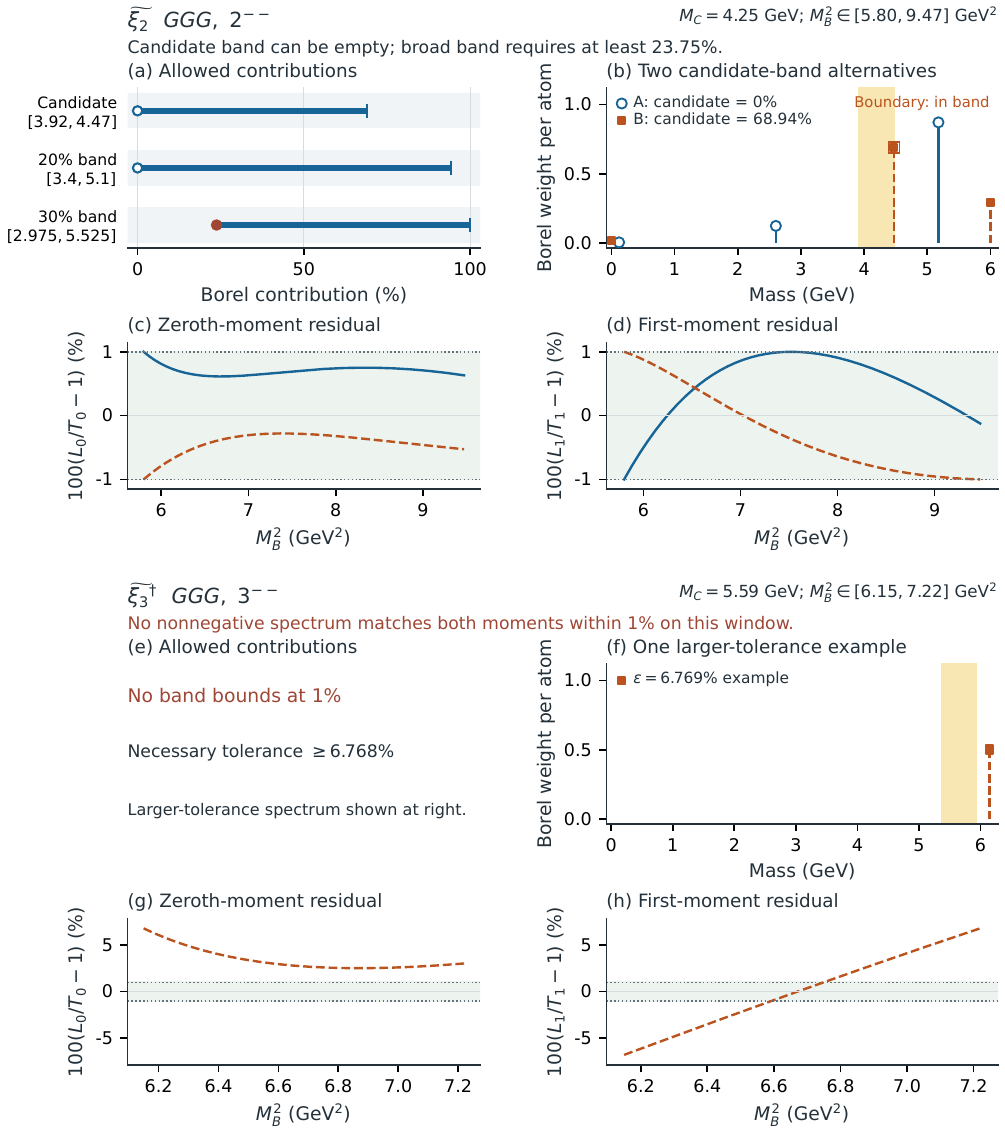}
\caption{Contrasting compatibility for $GGG,2^{--}$ and $GGG,3^{--}$. Upper row (a--d): A empties the candidate band, B nearly maximizes its fraction, and both satisfy 1\%; the left panel also bounds the wider bands. Lower row (e--h): no nonnegative spectrum is compatible at 1\%, so no band bounds are assigned. The single spectrum in (f) satisfies 6.769\% tolerance; (g,h) show its residuals. It is not a band extremum. Gold selects the closed candidate band; green marks $\pm1\%$. Inputs and dagger: Tables~\ref{tab:chen-comparison} and~\ref{tab:chen-inputs}.}\label{fig:atlas-7}\end{figure}
\clearpage

\twocolumngrid
\relax{}

\end{document}